\documentclass[english,11pt]{article}
\usepackage[a4paper,bottom=4.2cm,top=1cm,head=3cm,width=18cm,dvipdfm]{geometry}
\evensidemargin=\oddsidemargin

\usepackage{braket}
\usepackage[T1]{fontenc}
\usepackage[latin9]{inputenc}
\usepackage{amstext}
\usepackage{babel}
\usepackage{verbatim}
\usepackage{amsfonts}
\usepackage{mathrsfs}
\usepackage{amsmath}
\usepackage{amssymb}
\usepackage{bm}
\usepackage{extarrows}
\usepackage{slashed}
\usepackage{isodateo}
\usepackage{xcolor}
\usepackage{graphicx}
\usepackage[low-sup]{subdepth}
\usepackage{isodateo}
\usepackage[CJKbookmarks=true, bookmarksnumbered=true,bookmarksopen=true]{hyperref}
\hypersetup{colorlinks,%
	linkcolor=[rgb]{0,0,1}, %
	citecolor=[rgb]{0,0,1}}

\hypersetup{
	colorlinks=true,
	linkcolor=blue,
	filecolor=blue,
	urlcolor=blue,
	citecolor=blue,
}

\numberwithin{equation}{section}

\makeatletter
\@ifundefined{textcolor}{}
{%
	\definecolor{BLACK}{gray}{0}
	\definecolor{WHITE}{gray}{1}
	\definecolor{RED}{rgb}{1,0,0}
	\definecolor{GREEN}{rgb}{0,1,0}
	\definecolor{BLUE}{rgb}{0,0,1}
	\definecolor{CYAN}{cmyk}{1,0,0,0}
	\definecolor{MAGENTA}{cmyk}{0,1,0,0}
	\definecolor{YELLOW}{cmyk}{0,0,1,0}
}
\newcommand{\fr}[2]{\mbox{$\frac{\,{#1}\,}{#2}$}}
\renewcommand{\rm}{\mathrm}
\def\bge{\begin{equation}}
	\def\ede{\end{equation}}
\def\bga{\begin{aligned}}
	\def\eda{\end{aligned}}
\newcommand{\beq}{\begin{equation}}
	\newcommand{\eeq}{\end{equation}}
\newcommand{\bq}{\begin{equation}}
	\newcommand{\eq}{\end{equation}}
\newcommand{\ba}{\begin{array}}
	\newcommand{\ea}{\end{array}}
\newcommand{\beqa}{\begin{eqnarray}}
	\newcommand{\eeqa}{\end{eqnarray}}
\newcommand{\beqs}{\begin{subequations}}
	\newcommand{\eeqs}{\end{subequations}}

\def\to{\rightarrow}
\def\ito{\!\rightarrow\!}

\def\over{\overline}
\def\nn{\nonumber}

\def\({\left(}
\def\){\right)}

\def\deg{\circ}

\def\End{\end{document}}

\def\d{\text{d}}
\def\ii{{\tt i}}
\def\over{\overline}

\def\ga{\gamma}

\def\ep{\epsilon}

\renewcommand{\rm}{\mathrm}
\def\bge{\begin{equation}}
\def\ede{\end{equation}}
\def\bga{\begin{aligned}}
\def\eda{\end{aligned}}

\def\nn{\nonumber}

\def\({\left(}
\def\){\right)}
\def\[{\left[}
\def\]{\right]}

\def\deg{\circ}
\def\End{\end{document}}

\newcommand{\gammaslash}[1]{#1 \!\!\!/}
\DeclareMathOperator\dif{d\!}

\def\TT{\mathcal{T}}

\def\over{\overline}

\def\ga{\gamma}

\def\ep{\epsilon}

\def\si{\sigma}

\def\to{\rightarrow}

\def\ii{\mathrm{i}}

\def\cut{\Lambda}

\newcommand{\mO}{\mathcal{O}}

\def\HC{\rm{h.c.}}
\def\vs{\vspace*{1mm}}

\def\hs{\hspace*{0.3mm}}

\def\hsm{\hspace*{-0.3mm}}
\def\hsmx{\hspace*{-0.5mm}}

\newlength{\halfpagewidth}
\divide\halfpagewidth by 2

\def\End{\end{document}}
\renewcommand{\thefootnote}{\fnsymbol{footnote}}
\makeatother

\begin{document}

\thispagestyle{empty}

\title{}
\author{}


\vspace*{6mm}

\begin{center}
{\Large\bf\boldmath Probing Neutral Triple Gauge Couplings via $ZZ$ Production 
\\[2mm]
at $e^+e^-$ Colliders with Machine Learning}
\vspace*{10mm}

{\bf John Ellis}$\hs^{a,}$\footnote{john.ellis@cern.ch},~ 
{\bf Hong-Jian He}$\hs^{b,c,}$\footnote{hjhe@sjtu.edu.cn},~ 
{\bf Rui-Qing Xiao}$\hs^{d,}$\footnote{xiaoruiqing@pku.edu.cn},~ 
{\bf Shi-Ping Zeng}$\hs^{b,}$\footnote{spzeng@sjtu.edu.cn}

\vspace*{3mm}
{\small
$^a$\,Tsung-Dao Lee Institute, Shanghai Jiao Tong University, Shanghai, China;\\
Department of Physics, King's College London, Strand, London WC2R 2LS, UK;\\
Theoretical Physics Department, CERN, CH-1211 Geneva 23, Switzerland
\\[1mm]
$^b$\,Tsung-Dao Lee Institute \& School of Physics and Astronomy, \\
Shanghai Jiao Tong University, Shanghai, China
\\[1mm]
$^c$\,Department of Physics, Tsinghua University, Beijing, China;\\
Center for High Energy Physics, Peking University, Beijing, China
\\[1mm]
$^d$\,School of Physics, Peking University, Beijing, China
}


\end{center}

\vspace*{20mm}

\begin{abstract}
Neutral triple gauge couplings (nTGCs) first arise from the dimension-8 operators of the Standard Model Effective Field Theory (SMEFT), rather than the dimension-4 SM Lagrangian and dimension-6 SMEFT operators, opening up a unique window for probing new physics at the dimension-8 level.\ In this work, we formulate the nTGC form factors of $ZZV^*$ ($V\!\!=\!Z,\gamma$) that are compatible with the spontaneous breaking of the SU(2)$\otimes$U(1) electroweak gauge symmetry and consistently match the dimension-8 nTGC operators in the broken phase.\  We study the sensitivities for probing both the $ZZV^*$ form factors and the corresponding new physics scales through $ZZ$ production (with visible/invisible fermionic $Z$ decays) at high energy $e^+e^-$ colliders including CEPC, FCC-ee, LCF, ILC and CLIC.\ In particular, we identify the dimension-8 operator that contributes to the pure triple $Z$ boson coupling $ZZZ^*$ alone, but not the mixed $ZZ\gamma^*$ coupling.\  We further study the correlations between probes of the $ZZZ^*$ and $ZZ\gamma^*$ couplings.\  Using machine learning, we show that angular distributions of the final-state fermions can play key roles in suppressing the SM backgrounds.\ The sensitivities can be further improved by using polarized $e^\mp$ beams, and we find that the optimal sensitivity bounds on the nTGC correlations are given by the mixed setting including both the unpolarized operation and the follow-up polarized operation.\ We demonstrate that machine learning is advantageous for handling the 4-body final states from $ZZ$ decays and improves significantly the sensitivity reaches of probes of nTGCs in $e^+e^-$ collisions.\ We find that nTGC new physics scales can be probed up to the multi-TeV scale at the proposed $e^+e^-$ colliders.
\\[4mm]
KCL-PH-TH/2025-28, CERN-TH-2025-122  
\\[2mm]
\href{https://doi.org/10.1103/w9wn-fxkc}{Phys.\ Phys.\ D\,113 (2026) 075005}
[{\hs}arXiv:2506.21433{\hs}{]}.

\end{abstract}

\newpage
\baselineskip 18pt
\tableofcontents

\vspace*{5mm}

\setcounter{footnote}{0}
\renewcommand{\thefootnote}{\arabic{footnote}}

\section{\hspace*{-2.5mm}Introduction}
\label{sec:1}

Studies of neutral triple gauge couplings (nTGCs)
open a unique window on new physics beyond the Standard Model (BSM),\
since they are absent both in the Standard Model (SM) and
in the SM Effective Field Theory (SMEFT)\,\cite{SMEFT-Rev} at the dimension-6 level.\
However, nTGCs are generated by dimension-8 SMEFT operators,
and offer a rare opportunity to probe directly BSM physics at the dimension-8 level,
without obscuration by dimension-4 SM contributions and dimension-6 SMEFT effects.

In a series of works\,\cite{Ellis:2025ghl}-\cite{Ellis:2019zex},
we have explored the potential sensitivities of probes of the nTGCs
at $e^+ e^-$ and $pp$ colliders through the reaction $f\bar{f}\!\ito\! Z\gamma\hs$,
and further studied the UV completion of the nTGCs\,\cite{nTGC-UV}\cite{nTGC-UVx}.\
In the course of these studies, we have revisited the conventional nTGC form factor 
formalism\,\cite{Gounaris:1999kf}\cite{Degrande:2013kka},
which respects only the residual U(1)$_{\rm{em}}^{}$ gauge symmetry of QED.\ 
In our formulation of nTGC form factors
we have emphasized\,\cite{Ellis:2025ghl}-\cite{newformfactor}
the importance of incorporating the full electroweak gauge group SU(2)$\otimes$U(1) of the SM 
with spontaneous symmetry breaking.
In particular, we have shown that disregarding the full electroweak gauge group 
in the conventional nTGC formalism\,\cite{Gounaris:1999kf}
gives unphysically large predictions for the nTGC sensitivity bounds that are stronger than
the correct bounds by about two orders of magnitude in the cases of the LHC probes in initial
CMS and ATLAS experimental analyses\,\cite{CMS2016nTGC-FF}\cite{Atlas2018nTGC-FF}.\ 
However, the ATLAS collaboration\,\cite{ATLAS2025-ZA} has recently performed a new LHC measurement of nTGCs 
by using the consistent nTGC form factor formulas\,\cite{Ellis:2023ucy}\cite{newformfactor}
and confirmed our theoretical predictions for the prospective experimental sensitivities.\ 
The lesson is that, although the contributions of dimension-8 operators to collider processes  
generally have higher energy-power dependences than those of dimension-6 operators\,\cite{dim6}\cite{dim6x}, 
it is important to ensure that high-energy behaviors of the dimension-8 contributions are
consistent with the full electroweak gauge group SU(2)$\otimes$U(1) 
with spontaneous symmetry breaking, because it can
impose nontrivial energy cancellations for the physical scattering amplitudes that arise from
BRST identities\,\cite{Ellis:2025ghl}-\cite{newformfactor}.

Previous studies have mainly focused on analyses of $Z\ga$ 
production\,\cite{Ellis:2025ghl}-\cite{Ellis:2019zex}\cite{nTGC-other}.\footnote{%
In passing, we note that a recent preprint\,\cite{XeeZZ} studied $ZZ$ production at $e^+e^-$ colliders,
but differs substantially from our work.\ It did not study the $ZZZ^*$ coupling because it considered 
only the four operators
$\mO_{\hsmx\tilde{B}W}^{},\hs\mO_{\!B\tilde{W}}^{},\hs 
\mO_{\!\tilde{W}W}^{},\hs \mO_{\hsm\!\tilde{B}B}^{}$ that do not contribute to the $ZZZ^*$ vertex, 
as shown in our Table\,\ref{tab:1}.\ Moreover, the operators $\mO_{\!\tilde{W}W}^{}$ and $\mO_{\hsm\!\tilde{B}B}^{}$ 
have zero contribution to the $ZZ\gamma^*$ vertex as well, whereas $\mO_{\!B\tilde{W}}^{}$ gives the same
$ZZ\gamma^*$ contribution as $\mO_{\hsmx\tilde{B}W}^{}$ (up to a trivial overall minus sign as shown in our Table\,\ref{tab:1}).\
Hence, Ref.\,\cite{ XeeZZ} effectively studied only the $ZZ\gamma^*$ coupling generated by $\mO_{\hsmx\tilde{B}W}^{}$,
and could not study any correlations between different nTGCs as we do.\ Finally, it did not
study the probe of any related nTGC form factors, which is a major subject of our work.}\   
However, these cannot probe the pure $Z$ boson nTGC $ZZZ^*$.\
For this purpose, we study in the present work a different reaction process, $e^+ e^- \!\!\ito\hsm ZZ$
with on-shell final-state $Z$ bosons,
which can probe $ZZZ^*$ and $ZZ\gamma^*$ nTGCs.\
The final-state $Z$ bosons decay dominantly through the charged-fermion channels and invisible channels\footnote{%
We note that at $e^+ e^-$ colliders, in contrast to $pp$ colliders, it is possible to make a kinematic selection of invisible decays of on-shell $Z$ bosons 
($Z\ito \nu\bar{\nu}$) in the mixed final states 
$ZZ \ito \ell\bar{\ell}/q\bar{q} + \nu\bar{\nu}\hs$.}$\!$,
$Z\ito q\bar{q},\ell\bar{\ell},\nu\bar{\nu}\hs$, 
whose angular distributions can be used to suppress the SM backgrounds
and enhance the sensitivities of probes of the nTGCs.\
We demonstrate in our analysis how the sensitivities of probes of nTGCs can be significantly enhanced by using machine learning to optimize the discriminatory power of the 4-body final states from $ZZ$ decays.\ 
We show that using polarized electron and positron ($e^\mp$) beams 
can further increase the sensitivity reaches of probes of nTGCs.\
Moreover, we find that the optimal sensitivity bounds on the nTGC correlations are given by the mixed setup including both the unpolarized operation of $e^-e^+$ collisions
and the follow-up polarized operation.


This paper is organized as follows.\
In Section\,\ref{sec:2.1} we analyze how the nTGC couplings arise from dimension-8 SMEFT operators, 
and establish a consistent formulation of the nTGC form factors that incorporates the 
full electroweak gauge group SU(2)$\otimes$U(1) with spontaneous symmetry breaking.\   
Then, in Section\,\ref{sec:2.2} we derive the unitarity constraints on nTGCs that are significantly weaker
than the sensitivity reaches for probes of the nTGCs at high-energy $e^+e^-$ colliders, as we show later.\
In Section\,\ref{sec:3} we derive the SM and nTGC contributions to the scattering amplitudes and cross sections
for $ZZ$ production channels, and analyze the differences between their angular distributions.\
Section\,\ref{sec:4} demonstrates how these may be used to probe $ZZZ^*$ and $ZZ\gamma^*$ nTGCs 
at high-energy $e^+ e^-$ colliders.\
We analyze the visible fermionic decay channels
$Z \!\ito \ell\bar{\ell},q\bar{q}\,$ in Section\,\ref{sec:4.1} and the invisible decay channels $Z \!\ito\! \nu\bar{\nu}$ 
in Section\,\ref{sec:4.2}.\
We apply machine learning algorithms to enhance 
sensitivity reaches of probes of nTGCs in Section\,\ref{sec:4.3}.\ Then, we analyze the improved sensitivities with polarized electron/positron beams in Section\,\ref{sec:4.4}.\
We analyze correlations between the nTGC parameters in Section\,\ref{sec:4.5}.\
Comparisons with the previous analyses of probes of the nTGCs via the $Z\gamma$ production channel 
are presented in Section\,\ref{sec:4.6}.\
Finally, we conclude in Section\,\ref{sec:5}.

\section{\hspace*{-2.5mm}\boldmath{${ZZV^*}\!$} Couplings from Operators and Form Factors}
\label{sec:2}

We present the formulation of the nTGC couplings of ${ZZV^*}$ in terms of the
gauge-invariant dimension-8 effective operators and the corresponding form factors in Section\,\ref{sec:2.1}.\
Then,  we analyze the unitarity constraints on the nTGCs in Section\,\ref{sec:2.2}.\
We will show that these unitarity bounds are much weaker than our collider constraints, 
and thus do not affect our collider analyses in the next Section.\

\subsection{\hspace*{-2.5mm}nTGC Operators and Form Factors for \boldmath{${ZZV^*}\!$} Couplings}
\label{sec:2.1}

In previous papers\,\cite{Ellis:2020ljj}\cite{Ellis:2019zex}
we studied possible probes of neutral triple gauge couplings (nTGCs) via on-shell $Z\ga$ production.\
In this Section, we will analyze the nTGC contributions to the production process $\,e^-e^+\ito ZZ\,$
through the vertices $ZZV^*$ (with $V\!=\!Z,\gamma$).\
We work in the general framework of the Standard Model Effective Field Theory (SMEFT),
in which nTGCs are first generated by dimension-8 operators\,\cite{Degrande:2013kka}.\
The general dimension-8 SMEFT Lagrangian can be written as
%
\beqa
\Delta\mathcal{L}(\text{dim-8})
\,=\, \sum_{j} \!\frac{c_j}{\,\tilde{\cut}^4\,}\mathcal{O}_j^{}
= \sum_{j} \!\frac{\,\text{sign}(c_j^{})\,}{\,\cut_j^4\,}\mathcal{O}_j^{} 
= \sum_{j} \!\frac{1}{\,[\cut_j^4]\,}\mathcal{O}_j^{} 
\,,
\label{cj}
\eeqa
where the dimensionless coefficients $\,c_j^{}$ may be $O(1)$,
with signs $\,\text{sign}(c_j^{})=\pm\hs$.\ We define the corresponding cutoff scale 
$\,\cut_j^{} \!\equiv\hsm \tilde{\cut}/|c_j^{}|^{1/4}\,$
and introduce the shorthand notation $[\cut_j^4]\hsm\equiv\hsm\rm{sign}(c_j^{})\cut_j^4$.\ 

A CP-conserving $ZZV^*$ vertex can be parameterized as follows\,\cite{Degrande:2013kka}:
\begin{align}
\text{i}\hs e\hs\Gamma^{\alpha\beta\mu}_{ZZV^*}(q_1^{},q_2^{},q_3^{})
= \frac{\,e(q_3^2\!-\!M_V^2)\,}{M_Z^2}f_5^V
\epsilon^{\mu\alpha\beta\rho}(q_1^{}\!-\!q_2^{})_{\rho}^{} \, ,
\end{align}
and a CP-conserving $Z\gamma V^*$ vertex can be parameterized as follows\,\cite{newformfactor}:
\begin{align}
\text{i}\hs e\hs\Gamma^{\alpha\beta\mu}_{Z\gamma V^*} (q_1^{},q_2^{},q_3^{})
= \frac{\,e(q_3^2\!-\!M_V^2)\,}{M_Z^2}\!\!
\left[\hsm\!\(\!h_3^V\!+\!\frac{h_4^V}{\,2M_Z^2\,}q_3^2\!\)\!q_{2\nu}^{}\epsilon^{\alpha\beta\mu\nu}
\!+\!\frac{h_4^V}{\,M_Z^2\,}q_2^\alpha q_{3\nu}q_{2\sigma}\epsilon^{\beta\mu\nu\sigma}\right] \! .
\end{align}
The $ZZV^*$ nTGC vertices can arise only from dimension-8
operators with two SM Higgs-doublet fields.\ 
As shown in Ref.\,\cite{nTGC-UV}, we note that there are 7 independent CP-conserving nTGC operators of dimension-8
with two SM Higgs-doublet fields after taking into account the equivalence relations arising from integration by parts.\
We choose the following operator basis for our nTGC analysis:
\begin{subequations}
\label{eq:O8-ZZV*}
\begin{align}
\mathcal{O}_{\tilde{W}W} &=\, {\rm i}H^{\dagger}\tilde{W}_{\mu\nu}W^{\nu\rho}\{\!D_{\rho},D^{\mu}\}H+\HC \hs ,
\\
\mathcal{O}_{\tilde{W}W}^{\,\prime} &=\, {\rm i}H^{\dagger}\tilde{W}_{\mu\nu}\big(\!D_{\rho}W^{\nu\rho}\big)D^{\mu}H+ \HC \hs ,
\\
\mathcal{O}_{\tilde{B}B} &=\, {\rm i}H^{\dagger}\tilde{B}_{\mu\nu}B^{\nu\rho}\{\!D_{\rho},D^{\mu}\}H+\HC \hs ,
\\
\mathcal{O}_{\tilde{B}B}^{\,\prime} &=\, {\rm i}H^{\dagger}\tilde{B}_{\mu\nu}\big(\!D_{\rho}B^{\nu\rho}\big)D^{\mu}H+\HC \hs ,
\\
\mathcal{O}_{\tilde{B}W} &=\, {\rm i}H^{\dagger}\tilde{B}_{\mu\nu}W^{\nu\rho}\{\!D_{\rho},D^{\mu}\}H+\HC \hs ,
\\
\mathcal{O}_{\tilde{B}W}^{\,\prime} &=\, {\rm i}H^{\dagger}\tilde{B}_{\mu\nu}\big(\! D_{\rho}W^{\nu\rho}\big)D^{\mu}H+\HC \hs ,
\\
\mathcal{O}_{\tilde{W}B} &=\, {\rm i}H^{\dagger}\tilde{W}_{\mu\nu}B^{\nu\rho}\{\! D_{\rho},D^{\mu}\}H+\HC \hs ,
\end{align}
\end{subequations}
where $H$ denotes the SM Higgs doublet and we define the dual field strengths
$\tilde{B}_{\mu\nu}^{}\!=\!\epsilon_{\mu\nu\alpha\beta}^{}B^{\alpha\beta}$ and
$\tilde{W}_{\mu\nu}^{}\!=\!\epsilon_{\mu\nu\alpha\beta}^{}W^{\alpha\beta}$,
where $W_{\mu\nu}\!=\!W^a_{\mu\nu}\frac{\,\sigma^a}{2}$ and $\sigma^a$ denotes the Pauli matrices.\
The above operators are all Hermitian and their Wilson coefficients $c_j^{}$ are assumed to be real, so that they conserve CP.\
As discussed in the text below Eq.(2.8) of Ref.\,\cite{nTGC-UV}, 
the operators of Eq.\eqref{eq:O8-ZZV*} are linear combinations of the operators
${\mathcal{O}}_{W^{2}\phi^{2}D^{2}}^{(9)}$,
${\mathcal{O}}_{W^{2}\phi^{2}D^{2}}^{(17)}$, 
${\mathcal{O}}_{WB\phi^{2}D^{2}}^{(14)}$,
${\mathcal{O}}_{WB\phi^{2}D^{2}}^{(15)}$, 
${\mathcal{O}}_{WB\phi^{2}D^{2}}^{(18)}$,
${\mathcal{O}}_{B^{2}\phi^{2}D^{2}}^{(10)}$, and 
${\mathcal{O}}_{B^{2}\phi^{2}D^{2}}^{(12)}$
that are listed in Ref.\cite{Chala:2021cgt} (cf.\ its Table\,2) 
for the dimension-8 SMEFT analysis, up to non-nTGC terms.\ 
These operators are not given in the basis of Refs.\cite{Murphy,JHY},  
but they can be derived from operators of the type $(D\phi)^2 F^2$ 
\cite{Murphy,JHY} via integration by parts.\

The contributions of the operators \eqref{eq:O8-ZZV*} 
to $ZZZ^*$ and $ZZ\gamma^*$ vertices are summarized in Table\,\ref{tab:1},
where $\braket{H^0}\!=\!v/\hsm\sqrt{2\,}$ denotes the SM Higgs vacuum expectation value,
$\theta_W^{}$ is the weak mixing angle, and we use the notations
$(s_W^{},\hs c_W^{})\!=\!(\sin\theta_W^{},\hs \cos\theta_W^{})$.\

We see that the operators $\mathcal{O}_{\tilde{B}W}^{}$ and $\mathcal{O}_{\tilde{W}B}^{}$
both contribute to the $ZZ\gamma^*$ coupling, but do not contribute to the $ZZZ^*$ coupling.\
Moreover, their contributions to the $ZZ\gamma^*$ coupling differ only by an overall minus sign,
so we need only consider the operator $\mathcal{O}_{\tilde{B}W}^{}$ for our study of the $ZZ\gamma^*$ nTGC.\
In summary, we have 4 remaining relevant operators 
($\mO_{\hsmx\tilde{W}W}^{\prime},\hs\mO_{\!\tilde{B}B}^{\prime},\hs 
\mO_{\!\tilde{B}W}^{},\hs \mO_{\hsm\!\tilde{B}W}'$), among which $\mO_{\!\tilde{B}W}^{}$ contributes to
the $ZZ\gamma^*$ coupling alone and the other three operators contribute to both the
$ZZZ^*$ and $ZZ\gamma^*$ couplings.\ 
\begin{table}[t] 			
\begin{center}
\tabcolsep 1pt
\begin{tabular}{c||c|c}
\hline\hline
& & \\[-4mm]
~~Operators~~  &  $ZZZ^*$  &  $ZZ\gamma^*$ \\[0.4mm]
\hline\hline 
& &\\[-4mm]
${\mO}_{\tilde{W}W}$  &  0  &  0
\\[0.3mm]
\hline
& &\\[-4mm]
	${\mO}_{\hsmx\tilde{W}W}^{\,\prime}$
	&  $\frac{e\hs v^2}{\,4\tan\theta_W^{}\,}\!\big(p_3^2\!-\!M_Z^2\big)\!
	\big(p_1^\rho \!-\! p_2^\rho\big)\epsilon_{\alpha\beta\mu\rho}$
	&  $\frac{\,e\hs v^2}{4}p_3^2\big(p_1^\rho \!-\! p_2^\rho\big)\epsilon_{\alpha\beta\mu\rho}^{}$
	\\
	& &\\[-4mm]
	\hline
	& &\\[-4.3mm]
	$\mathcal{O}_{\hsm\!\tilde{B}B}^{}$  &  0  &  0\\[0.7mm]
	\hline
	& &\\[-4mm]
	${\mO}_{\hsm\!\tilde{B}B}^{\,\prime}$
	&  ~$ev^2\hsm\tan\hsm\theta_W^{}\big(p_3^2\!-\!M_Z^2\big)\!\big(p_1^\rho \!-\! p_2^\rho\big)\epsilon_{\alpha\beta\mu\rho}^{}$~
	&  $-ev^2p_3^2 \big(p_1^\rho \!-\! p_2^\rho\big)\epsilon_{\alpha\beta\mu\rho}^{}$
	\\[0.6mm]
	\hline
	& &\\[-4.2mm]
	${\mO}_{\hsm\!\tilde{B}W}^{}$  &  0
	&  $-\frac{e\hs v^2}{\,2s_W^{}c_W^{}\,}p_3^2\big(p_1^\rho \!-\! p_2^\rho\big)\epsilon_{\alpha\beta\mu\rho}^{}$
	\\[1.4mm]
	\hline
	& &\\[-4.3mm]
	${\mO}_{\hsm\!\tilde{B}W}^{\,\prime}$
	&  $\frac{\,e\hs v^2\,}{2}\big(p_3^2\!-\!M_Z^2\big)\!\big(p_1^\rho\!-\!p_2^\rho\big)\epsilon_{\alpha\beta\mu\rho}^{}$
	&  ~$\frac{\,e\hs v^2\,}{2}\hsm\tan \hsm\theta_W^{} p_3^2\big(p_1^\rho\!-\!p_2^\rho\big)\epsilon_{\alpha\beta\mu\rho}^{}$~
	\\[0.8mm]
	\hline
	& &\\[-4.3mm]
	${\mO}_{\hsm\tilde{W}B}^{}$  &  0
	& $\frac{e\hs v^2}{\,2s_W^{}c_W^{}\,}p_3^2\big(p_1^\rho\!-\!p_2^\rho\big)\epsilon_{\alpha\beta\mu\rho}^{}$
	\\[1.4mm]
	\hline\hline
\end{tabular}
\caption{\small {Summary of contributions by the relevant CP-conserving dimension-8 operators 
		to the nTGC vertices $ZZZ^*$ and $ZZ\gamma^*$.}}
\label{tab:1}
\end{center}
\end{table}

We note in Table\,\ref{tab:1} that three dimension-8 operators 
($\mO_{\hsmx\tilde{W}W}^{\,\prime},\hs\mO_{\!\tilde{B}B}^{\,\prime},\hs 
\mO_{\hsm\!\tilde{B}W}^{\,\prime}$) contribute to the $ZZZ^*$ vertex, as follows:
\beq 
\label{f5Z-all}
f_5^Z \,= v^2M_Z^2\!\(\!
\frac{1}{\,4\tan\theta_W[\Lambda_{\tilde{W}W}'^{\,4}]\,} +
\frac{\tan\theta_W}{\,[\Lambda_{\tilde{B}B}'^{\,4}]\,} +
\frac{1}{\,2[\Lambda_{\tilde{B}W}'^{\,4}]\,}
\!\) \!.
\eeq 
In order to study the $ZZZ^*$ and $ZZ\gamma^*$ couplings separately,
we can construct an operator that contributes to the $ZZZ^*$ coupling alone:
\beq 
\label{eq:O3Z}
{\mO}_{3Z}^{} \,\equiv\, {\mO}_{\tilde{W}W}^{\prime}+\frac{1}{\hs 4\hs} {\mO}_{\tilde{B}B}^{\prime} \, ,
\eeq 
which yields the following $ZZZ^*$ coupling:
\beq
\frac{\,e\hs v^2}{~4\hs s_W^{}c_W^{}~} (p_3^2\!-\!M_Z^2)
(p_1^\rho \!-\!p_2^\rho) \hs\epsilon_{\alpha\beta\mu\rho}^{} \,.
\eeq
Its contribution to the $ZZZ^*$ form factor $f_5^Z$ is as follows:
\begin{align}
	\label{f5z}
	f_5^Z \,=\frac{~v^2M_Z^2~}{\,4\hs s_W^{}c_W^{}[\Lambda_{3Z}^4]\,} \, .
\end{align} 

We also see in Table\,\ref{tab:1} that there are four dimension-8 operators 
($\mO_{\hsmx\tilde{W}W}^{\,\prime},\hs\mO_{\!\tilde{B}B}^{\,\prime},\hs 
\mO_{\!\tilde{B}W}^{},\hs \mO_{\hsm\!\tilde{B}W}^{\,\prime}$) that
contribute to the  $ZZ\gamma^*$ vertex, whose contributions to the $ZZ\gamma^*$ form factor are as follows: 
\begin{align}
\label{f5a}
f_5^\gamma \,= v^2M_Z^2\!\(\!
\frac{1}{\,4[\Lambda_{\tilde{W}W}'^{\,4}]\,} +
\frac{1}{\,[\Lambda_{\tilde{B}B}'^{\,4}]\,} +
\frac{1}{\,2\hs s_W^{}c_W^{}[\Lambda_{\tilde{B}W}^4]\,}+
\frac{\tan\theta_W}{\,2[\Lambda_{\tilde{B}W}'^{\,4}]\,}
\!\) \!.
\end{align}
Among the above four operators, the operator $\mO_{\!\tilde{B}W}^{}$ contributes to the form factor $f_5^\gamma$
only, but not to $f_5^Z$.\ This is in contrast with the operator ${\mO}_{3Z}^{}$,
which contributes to the form factor $f_5^Z$ only, instead of $f_5^\gamma$.\
As discussed in Ref.\,\cite{newformfactor}, the operator ${\mO}_{\!\tilde{B}W}^{}$ also contributes
to the $Z\gamma Z^*$ vertex, and the corresponding form factor $h_3^Z$ is given as follows:
\begin{align}
\label{eq:h3Z}
h_3^Z \,= \frac{~v^2M_Z^2~}{~2\hs s_W^{}c_W^{}[\Lambda_{\tilde{B}W}^4]~} \,.
\end{align}
Inspecting Eqs.\eqref{f5a} and \eqref{eq:h3Z}, we deduce the following relation:
\begin{align}
f_5^\gamma=h_3^Z \,,
\hspace*{12mm} (\text{for}~\mathcal{O}_{\tilde{B}W}^{}) {\hs}.
\end{align}

As seen in Eq.\eqref{eq:O8-ZZV*}, spontaneous electroweak symmetry breaking (EWSB) and the Higgs vacuum expectation value (VEV) 
are essential for generating $ZZV^*$ couplings.\
We also note that these vertices could only generate on-shell $Z$ pairs with one $Z$ polarized transversely and the other $Z$
polarized longitudinally, since the kinematic structure with the antisymmetric Levi-Civita tensor 
causes the production of $Z$ pairs with the same polarizations to vanish.\ Due to the antisymmetric structure, 
the pure gauge operators $\mO_{\!G\pm}^{}$ that we introduced in Ref.\,\cite{Ellis:2020ljj} 
make vanishing contributions to the $ZZV^*$ couplings.

\vspace*{1.5mm}

Working in the electroweak broken phase of the SM in which
$\rm{SU(2)}_{W}^{}\hsm\otimes\hsm \rm{U(1)}_Y^{}
 \hsm\ito \rm{U(1)}_{\rm{em}}^{}$,
we expand the operators \eqref{eq:O8-ZZV*} 
and derive the relevant neutral triple gauge vertices (nTGVs) as follows:
{\small 
\begin{subequations}
\begin{align}
\hspace*{-5mm}	
\mathcal{O}_{\tilde{W}W} & \rightarrow -\frac{e{\hs}v^2}{8s_W^{}c_W^{}}\!
\(\hsm s_W^2\tilde{A}_{\mu\nu}A^{\nu\rho} \!+\!s_W^{}c_W^{}\tilde{A}_{\mu\nu}^{}Z^{\nu\rho}
\!+\!s_W^{}c_W^{}\tilde{Z}_{\mu\nu}^{}A^{\nu\rho}\!+\!c_W^2\tilde{Z}_{\mu\nu}^{}Z^{\nu\rho}\hsm\)\!
(\partial^\mu Z_\rho \!+\!\partial_\rho Z^\mu),
\\
\hspace*{-5mm}	
\mathcal{O}_{\tilde{W}W}^{\prime}&\rightarrow
-\frac{e\hs v^2}{\,8s_W^{}c_W^{}\,}Z^\mu\! \(s_W^2\tilde{A}_{\mu\nu}\partial_\rho A^{\nu\rho}\!+\!
s_Wc_W\tilde{A}_{\mu\nu}\partial_\rho Z^{\nu\rho}\!+\!
s_Wc_W\tilde{Z}_{\mu\nu}\partial_\rho A^{\nu\rho}\!+\!
c_W^2\tilde{Z}_{\mu\nu}\partial_\rho Z^{\nu\rho}\)\!,
\\
\hspace*{-5mm}	
\mathcal{O}_{\tilde{B}B} & \rightarrow \frac{e\hs v^2}{2s_W^{}c_W^{}}\!\!
\(\!-c_W^2\tilde{A}_{\mu\nu}A^{\nu\rho}\!+\!s_Wc_W\tilde{A}_{\mu\nu}Z^{\nu\rho}
\!+\!s_Wc_W\tilde{Z}_{\mu\nu}A^{\nu\rho}\!-\!s_W^2\tilde{Z}_{\mu\nu}Z^{\nu\rho}\hsm\)\!\hsm
(\partial^\mu Z_\rho\!+\!\partial_\rho Z^\mu),
\\
\hspace*{-5mm}	
\mathcal{O}_{\tilde{B}B}^{\prime} & \rightarrow
-\frac{e\hs v^2}{2s_Wc_W}Z^\mu \!\(c_W^2\tilde{A}_{\mu\nu}\partial_\rho A^{\nu\rho}\!-\!
s_Wc_W\tilde{A}_{\mu\nu}\partial_\rho Z^{\nu\rho} \!-\!s_Wc_W\tilde{Z}_{\mu\nu}\partial_\rho A^{\nu\rho}
\!+\!s_W^2\tilde{Z}_{\mu\nu}\partial_\rho Z^{\nu\rho}\)\!,
\\
\hspace*{-5mm}	
\mathcal{O}_{\tilde{B}W} & \rightarrow \frac{e\hs v^2}{4s_Wc_W}
\(s_Wc_W\tilde{A}_{\mu\nu}A^{\nu\rho} \!+\! c_W^2\tilde{A}_{\mu\nu}Z^{\nu\rho}\!-\!
s_W^2\tilde{Z}_{\mu\nu}A^{\nu\rho}\!-\!s_Wc_W\tilde{Z}_{\mu\nu}Z^{\nu\rho}\)\!\hsm
(\partial^\mu Z_\rho\!+\!\partial_\rho Z^\mu),
\\
\hspace*{-5mm}	
\mathcal{O}_{\tilde{B}W}^{\prime}&\rightarrow
\frac{e\hs v^2}{4s_W^{}c_W^{}}Z^\mu \!\(s_W^{}c_W^{}\tilde{A}_{\mu\nu}\partial_\rho A^{\nu\rho}\!+\!c_W^2\tilde{A}_{\mu\nu}\partial_\rho Z^{\nu\rho}
\!-\!s_W^2\tilde{Z}_{\mu\nu}\partial_\rho A^{\nu\rho}\!-\!s_Wc_W\tilde{Z}_{\mu\nu}\partial_\rho Z^{\nu\rho}\hsm\)\!,
\\
\hspace*{-5mm}	
\mathcal{O}_{\tilde{W}B}^{} & \rightarrow -\frac{e\hs v^2}{4s_W^{}c_W^{}}\!\!
\(\hsm -s_Wc_W\tilde{A}_{\mu\nu}A^{\nu\rho}\!+\!
s_W^2\tilde{A}_{\mu\nu}Z^{\nu\rho} \!-\! c_W^2\tilde{Z}_{\mu\nu}A^{\nu\rho}
\!+\!s_Wc_W\tilde{Z}_{\mu\nu}Z^{\nu\rho}\)\!\!
\(\partial^\mu Z_\rho\!+\!\partial_\rho Z^\mu\)\!.
\end{align}
\end{subequations}
}
Combining the above contributions to the nTGC vertices, we can derive the general Lorentz structures 
for the $ZZV^*$ vertices with the form factors $(f_5^Z,\hs f_5^{\gamma})$
in coordinate space that respect the residual gauge symmetry U(1)$_{\rm{em}}^{}$ of QED, 
\begin{align}
\mathcal{L}&=\frac{e}{\,2M_Z^2\,}f_5^\gamma\!\hsm 
\(c_W^2\tilde{A}_{\mu\nu}Z^{\nu\rho}\!-\!
s_W^2\tilde{Z}_{\mu\nu}A^{\nu\rho}\)\!\hsm 
(\partial^\mu Z_\rho\!+\hsm\partial_\rho Z^\mu)
-\frac{e}{\,2M_Z^2\,}
f_5^ZZ^\mu\tilde{Z}_{\mu\nu}^{}\partial_\rho^{} Z^{\nu\rho}\,.
\end{align}

\vspace*{1mm}
\subsection{\hspace*{-2.5mm}Unitarity Constraints on the CPC nTGCs}
\label{sec:2.2}
\vspace*{1.5mm}

In this Section we analyze the perturbative unitarity constraints on the CPC nTGCs
from the process $e^-e^+\!\ito\! ZZ$,
finding that these constraints are much weaker than the sensitivity reaches
of the collider analysis presented in the following Section.

We make a partial-wave expansion of the nTGC contributions to the scattering amplitude
of $\,e^-e^+\ito ZZ\,$ as follows\,\cite{UCond}:
\begin{align}\label{partialwave}
a_{J}^{} = \frac{~e^{\text{i}(\nu^{\prime}-\nu) \phi}\,}{\,32\hs\pi\,} \!\!
\int_{-1}^{1} \!\!\mathrm{d}(\cos \theta)\hs d_{\nu^{\prime} \nu}^{J}(\cos \theta)\hs \mathcal{T}_{\mathrm{nTGC}}^{s_{f} s_{\bar{f}}, \lambda_{Z_1} \lambda_{Z_2}} \, ,
\end{align}
where the differences between initial-state and final-state helicities are given by
$\nu \!=\! s_{f}^{}\hsm -\hsm s_{\bar{f}}^{}$ and
$\nu^{\prime} \!=\!\lambda_{Z_1}^{}\!-\!\lambda_{Z_2}^{}$, respectively.\
We note that it is sufficient to treat the initial-state electrons/positrons as massless,\
so we have $s_{f}^{}\!=\!-s_{\bar{f}}^{}\,$,
leading to $\nu\!=\!\pm 1\hs$.\
This implies that the $J\!=\!1$ partial wave makes the leading contribution.\
The relevant Wigner $d$ functions\,\cite{d-function} are given by
\begin{align}
d^1_{1,0}=-\frac{1}{\,\sqrt{2\,}\,}, \qquad
d^1_{1,\pm1}=\frac{1}{\,2\,}(1\!\pm\hsm\cos\theta) \, ,
\end{align}
and the relation $\,d^J_{m,m^\prime}\!=d^J_{-m,-m^\prime}\,$ holds in general.

\vs

For the process $e^-_Le^+_R\ito ZZ$,  we compute
the leading contribution to the scattering amplitude
$\mathcal{T}_{\mathrm{nTGC}}^{s_{f} s_{\bar{f}}, \lambda_{Z_1} \lambda_{Z_2}}$ as follows:
\begin{align}
\mathcal{T}_{(8)}^{LR} \,=\, \pm
\frac{~c_L^Vf_5^Ve^2s^{\frac{3}{2}}}{\sqrt{2\,}c_W^{}s_W^{}M_Z^3~}
\(\!\sin^2\!\frac{\,\theta\,}{2},\,\cos^2\!\frac{\,\theta\,}{2}\!\)\!,
\end{align}
and the corresponding leading $p$-wave amplitude $a_1^{}$ is given by
\begin{align}
\big|\Re {\tt e}(a_1^{})\big|
=\left|\frac{c_L^Vf_5^Ve^2s^{\frac{3}{2}}}{\,48\sqrt{2\,}\hs\pi\hs c_W^{}s_W^{}M_Z^3\,}\right| \!.
\end{align}
The partial-wave unitarity condition is
$|\Re{\tt e}(a_1^{})|\hsm\!<\hsm\!1$,
where we have included a factor of $2$
as the final-state $ZZ$ are identical particles\,\cite{UCond}.\
Thus, we derive the following unitarity bound on the leading contributions
to the nTGC form factor:
\begin{align}
\label{leftbound}
|f_5^V|<\left|
\frac{\,48\sqrt{2\,}\hs\pi\hs c_W^{}s_W^{}M_Z^3\,}
{c_L^Ve^2s^{\frac{3}{2}}}\right| .
\end{align}
Analogously, for the process $e^-_Re^+_L\ito ZZ$,
we derive the following unitarity bound:
\begin{align}
\label{rightbound}
|f_5^V|<\left|
\frac{\,48\sqrt{2\,}\hs\pi\hs c_W^{}s_W^{}M_Z^3\,}
{c_R^Ve^2s^{\frac{3}{2}}}\right| ,
\end{align}
which can be obtained from the unitarity bound \eqref{leftbound} by the replacement
$c_L^V\!\leftrightarrow\! c_R^V\,$.\
using these results we derive the following unitarity bounds on
the nTGC form factors $f_5^\gamma$ and $f_5^Z$:
%
\begin{align}
\label{unitarityonf5}
\big|f_5^\gamma\big| \!<\! \frac{\,48\sqrt{2\,}\hs\pi M_Z^3\,}{e^2s^{\frac{3}{2}}}
\,,
\hspace*{10mm}
\big|f_5^Z\big| \!<\!
\frac{\,96\sqrt{2\,}\hs\pi\hs c_W^{}s_W^{}M_Z^3\,}
{\,e^2(1\!-\!2s_W^2)s^{\frac{3}{2}}\,} \, .
\end{align}
%
Since the nTGC form factors $f_5^\gamma$ and $f_5^Z$ are connected to the new physics
cutoff scales $\Lambda_{\tilde{B}W}^{}$ and $\Lambda_{3Z}^{}$
of the corresponding dimension-8 nTGC operators as shown in
Eqs.(\ref{f5a}) and (\ref{f5z}),
we derive the following unitarity bounds on
$\Lambda_{\tilde{B}W}^{}$ and $\Lambda_{3Z}^{}$:
%
\begin{align}
\label{unitarityonscale}
\Lambda_{\tilde{B}W}^{} \!>
\sqrt[4]{\!\frac{e^2v^2s^{3/2}}{\,96\sqrt{2}\hs\pi\hs s_W^{}c_W^{}M_Z^{}\,}\,} \,,
\hspace*{10mm}
\Lambda_{3Z}^{} \!>\!
\sqrt[4]{\!\frac{\,e^2v^2s^{3/2}(1\!-\!2s_W^2)\,}{\,384\sqrt{2\,}\hs\pi s_W^2c_W^2M_Z^{}\,}} \,.
\end{align}

\begin{figure}[]
\begin{center}
\includegraphics[width=8cm,height=6cm]{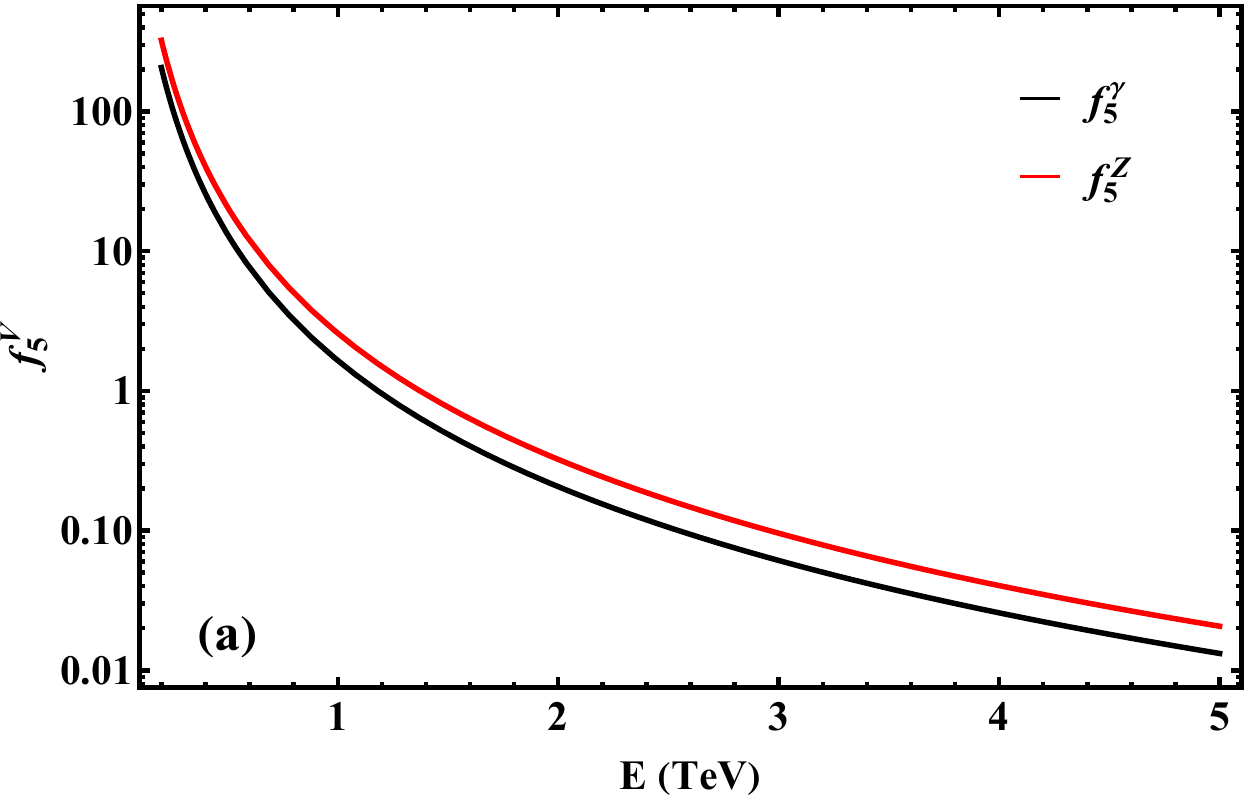}
\includegraphics[width=8cm,height=6cm]{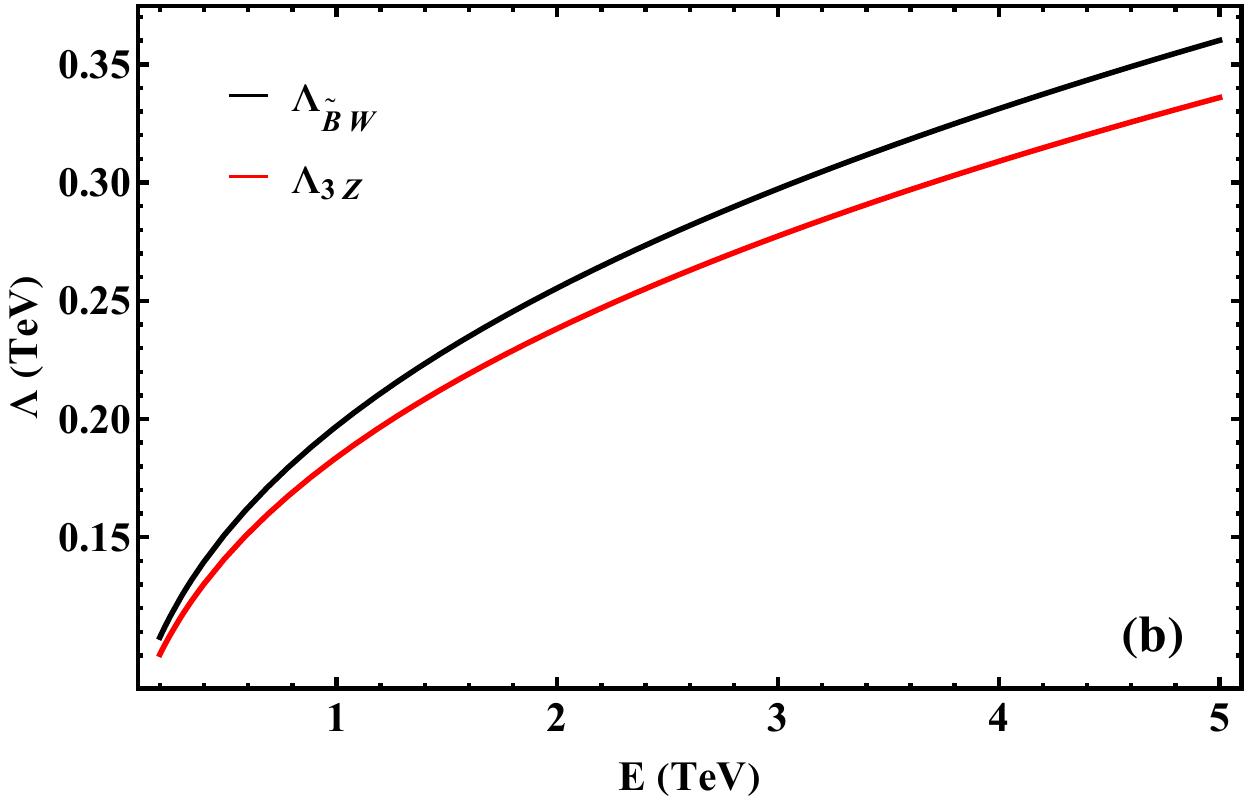}
\vspace*{-3mm}
\caption{\hspace*{-1mm}\small 
Unitarity bounds on the nTGC form factors $f_5^\gamma$ and $f_5^Z$ in plot\,(a)
and on the new physics scales $\Lambda_{\tilde{B}W}^{}$ and $\Lambda_{3Z}^{}$ 
in plot\,(b).\ These bounds are derived from the $p$-wave amplitudes for the reaction $e^-e^+\ito ZZ\hs$.\ The region below (above) each curve is allowed 
in plot\,(a) [in plot\,(b)].}
\label{unitarity}
\label{fig:1}
\end{center}
\end{figure}
\begin{table}[b]
\begin{center}
\begin{tabular}{c||c|c|c|c|c}
	\hline\hline
	& & & & &
	\\[-3.4mm]
	$\sqrt{s\,}$\,(TeV) & 0.25 & 0.5 & 1 & 3 & 5
	\\
	\hline\hline
	& & & & &
	\\[-4mm]
	$\Lambda_{\tilde{B}W}$(GeV) & 48.8 & 82 & 138 & 314 & 461
	\\
	& & & & &
	\\[-4mm]
	\hline
	& & & & &
	\\[-4mm]
	$\Lambda_{3Z}$(GeV) & 36.6 & 61.6 & 104 & 236 & 347 \\
	& & & & &
	\\[-4.5mm]
	\hline
	& & & & &
	\\[-4.3mm]
	$|f_5^\gamma|$ & 106 & 13.2 & 1.65 & 0.061 & 0.013 \\
	\hline
	& & & & &
	\\[-4mm]
	$|f_5^Z|$ & 166 & 20.7 & 2.59 & 0.096 & 0.021 \\
	\hline\hline
\end{tabular}
\end{center}
\vspace*{-4mm}
\caption{\hspace*{-1mm}\small 
{Unitarity bounds on the new physics scales $\Lambda_{\tilde{B}W}^{}$ and $\Lambda_{3Z}^{}$} (in GeV)
{of the dimension-8 nTGC operators and on the nTGC form factors $f_5^\gamma$ and $f_5^Z$.}}
\label{unitaritytable}
\label{tab:2}
\end{table}

We present the unitarity bounds (\ref{unitarityonf5}) and (\ref{unitarityonscale}) for the reaction $e^-e^+\!\ito ZZ\hs$ in Table\,\ref{unitaritytable},
for the representative $e^+e^-$ collider energies  $\sqrt{s\,}\hsm\!=\hsm\!(0.25,0.5,1,3,5)\, \text{TeV}$.\
We further present in Fig.\,\ref{fig:1} the unitarity bounds on the nTGC form factors and on the new physics scales of the nTGC operators
as functions of the $e^+e^-$ collision energy $\sqrt{s\,}\!=\!(0.25,\,0.5,\,1,3,\,5)$\,TeV for the reaction $e^-e^+\ito ZZ$.\
The unitarity bounds on the nTGC form factors $f_5^\gamma$ and $f_5^Z$ are shown in plot\,(a),
and in plot\,(b) we display the unitarity bounds on the new physics scales $\Lambda_{\!\tilde{B}W}$ and $\Lambda_{3Z}$.\
The region below (above) each curve is allowed 
in plot\,(a) [in plot\,(b)].\ 
We find that these unitarity bounds on the nTGC form factor
$(f_5^\gamma,\,f_5^Z)$ and on the new physics scales $(\Lambda_{\!\tilde{B}W}^{},\,\Lambda_{3Z}^{})$
are much weaker than our collider constraints, and hence do not affect our collider analyses in the next Section.

\section{\hspace*{-2.5mm}Scattering Amplitudes, Cross Sections and Angular Distributions}
\label{sec:3}

In this Section we study the reaction $e^-e^+ \!\ito ZZ$ including 
contributions from both the SM and the nTGCs.\
For this purpose, we compute helicity amplitudes and cross sections,
and analyze the angular distributions of its final states.\ 

\begin{figure}[h]
\centering
\includegraphics[width=9cm]{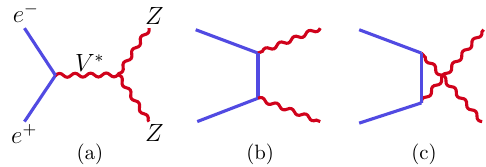}
\vspace*{-3mm}
\caption{\hspace*{-1.5mm}\small {Feynman diagrams that contribute to the reaction $e^-e^+\!\hsm\ito\! ZZ$.\
	Plot\,(a) shows the signal process containing the nTGC vertices ${ZZV^*}$,
	and plots\,(b) and (c) show the leading SM background contributions.}}
\label{nTGCpheno}
\label{fig:2}
\end{figure}

As shown in Fig.\,\ref{fig:2}, the SM contributions to the reaction
$e^-e^+ \!\ito ZZ$ are given by the $t$-channel and $u$-channel fermion-exchange diagrams, 
whereas the nTGC vertex ${ZZV^*}$ contributes via $s$-channel diagrams mediated by a virtual $Z$ or $\gamma\hs$.\
We first derive the SM contributions to the scattering amplitudes of the reaction 
$e^-(p_1)e^+(p_2)\ito Z(p_3)Z(p_4)$,
\begin{align}
\mathcal{T}_{\rm{sm}}^{} = &~ \bar{v}(p_2)\Gamma^{\mu}_Z\epsilon^*_{\mu}(p_4)
\frac{\,(\gammaslash{p}_1^{}\!-\!\gammaslash{p}_3^{})\,}{t}
\Gamma^{\nu}_Z\epsilon^*_{\nu}(p_3^{})u(p_1^{}) 
+\bar{v}(p_2^{})\Gamma^{\mu}_Z\epsilon^*_{\mu}(p_3^{})
\frac{\,(\gammaslash{p}_1^{}\!-\!\gammaslash{p}_4^{})\,}{u}
\Gamma^{\nu}_Z\epsilon^*_{\nu}(p_4^{})u(p_1^{}) \, ,
\end{align}
where we have defined the Mandelstam variables, $s\!=\!(p_1^{}\!+\!p_2^{})^2$, $t\!=\!(p_1^{}\!-\!p_3^{})^2$ and
$u\!=\!(p_1^{}\!-\!p_4^{})^2\,$.\ 
In the above the vertex function $\Gamma_Z^{\mu}$ is defined as follows: 
%
\begin{align}
\label{eq:GammaZ}
\Gamma_Z^{\mu} = 
\frac{\ii\hs e}{\,s_W^{}c_W^{}\,}\gamma^{\mu}(c_L^Z P_L^{}\!+\!c_R^Z P_R^{})\hs ,
\end{align}
%
where $P_{L,R}^{}\!=\!\frac{\,1\mp \gamma_5\,}{2}$ are projection operators, and
$(c_L^Z,c_R^Z)\!=\!(-\fr{1}{\,2\,} \!+\!s_W^2,\hs s_W^2)$ denote 
the (left,\hs right)-handed gauge couplings of the electron to the $Z$ boson, respectively.\

Next we compute the new physics amplitudes contributed by the nTGCs for the reaction 
$e^-(p_1)e^+(p_2)\\\ito Z(p_3)Z(p_4)$, 
%
\begin{align}
\mathcal{T}_{(8)}&=\, 
\bar{v}(p_2)\Gamma_V^{\mu}u(p_1)\frac{-g_{\mu\nu}}{~s-M_V^2~}
\rm{i}e\Gamma_{ZZV^*}^{\alpha \beta \nu}\epsilon_{\alpha}^*(p_3)\epsilon_{\beta}^*(p_4)
\nn\\
&=\,
\bar{v}(p_2)\Gamma_V^{\mu}u(p_1)\frac{~-eg_{\mu\nu~}}{M_Z^2}f_5^V\epsilon^{\nu\alpha\beta\rho}
(p_3^{}\!-\!p_4^{})_{\rho}\epsilon_{\alpha}^*(p_3)\epsilon_{\beta}^*(p_4) \hs ,
\end{align}
%
where we define the vertex functions $\Gamma_V^{\mu}$ ($V\!=\hsm\ga,Z$) as follows: 
\begin{subequations}
\label{eq:Gamma-V-cLR}
\begin{align}
\label{eq:Gamma-V}
\Gamma_V^{\mu} &= 
\frac{\ii\hs e}{\,s_W^{}c_W^{}\,}\gamma^{\mu}(c_L^V P_L^{}\!+\!c_R^V P_R^{})\hs ,
\\[0.5mm]
\label{eq:ee-cLR}
c_L^\gamma &=c_R^\gamma=-s_Wc_W \hs,
\hspace*{4mm}
c_L^Z =-\fr{1}{\,2\,} \!+\!s_W^2 \hs,
\hspace*{4mm}
c_R^Z = s_W^2 \hs,
\end{align}
\end{subequations}
where $c_{L}^V$ and $c_{R}^V$ arise from the left-handed and right-handed gauge couplings of electrons 
to the gauge bosons $V\!=\!Z,\gamma$, respectively.\

\vs

For the reaction $e^-(p_1^{})\,e^+(p_2^{})\ito Z(p_3^{})Z(p_4^{})$,
we parametrize the external momenta as follows:
\begin{align}
& p_1^{} =\frac{\sqrt{s\,}\,}{2}(1,0,0,1), \hspace*{5mm}
p_2^{} =\frac{\sqrt{s\,}\,}{2}(1,0,0,-1),\,
\nn\\
& p_3^{} = \!\bigg(\!\hsm\frac{\sqrt{s\,}\,}{2},q\sin\theta,0,q\cos\theta\!\bigg),
\hspace*{5mm}
p_4^{} =\!\bigg(\!\hsm\frac{\sqrt{s\,}\,}{2},-q\sin\theta,0,-q\cos\theta\!\bigg) \, ,
\end{align}
where $q\!=\!\sqrt{\frac{\,s\,}{4}\!-\!M_Z^2\,}$ is the magnitude of each $Z$ boson's momentum and $\theta$ is the scattering angle between the directions
of the incident $e^-(p_1^{})$ and the outgoing $Z(p_3^{})$.

\vs 

In order to characterize the final state $Z$ boson pair
$Z(p_3^{},\lambda)Z(p_4^{},\lambda')$ with its 9 helicity combinations,
we define a $3\!\times\! 3$ matrix amplitude $\mathcal{T}$ whose $i$-th row and $j$-th column
element is defined as
$\mathcal{T}(i,j)\equiv\mathcal{T}(\lambda\!+\!2,\lambda'\!+\!2)$.\
We derive the following SM contributions to the scattering amplitudes:
{\small 
\begin{align}
& \hspace*{-2mm}
\mathcal{T}_{\rm{sm}}^{\rm{LR}}=
\frac{2\hs e^2{c_L^Z}^2}{~s_W^2 c_W^2(4M_Z^4\!-\!4\sin^2\!\theta M_Z^2s
	\!+\! s^2\sin^2\!\theta )~}\times
\nn\\[1mm]
& \hspace*{-2mm}
\begin{pmatrix}\!\!
	-s\hs M_Z^2\sin\hsm 2\theta
	& \sqrt{2s\,}M_Z^{}(s\hs\sin^2\!\theta \!-\! 4M_Z^2\sin^2\!\hsm\frac{\,\theta\,}{2})
	& -2s(s\!-\!2M_Z^2) \sin\hsm\theta\sin^2\!\hsm\frac{\,\theta\,}{2}
	\\[1.5mm]
	\!\!\sqrt{2s\,}M_Z\big(s\sin^2\!\theta \!-\! 4M_Z^2\cos^2\!\hsm\frac{\theta}{2}\big)
	& 2\hs s\hs M_Z^2\sin 2\theta
	&-\sqrt{2s\,} M_Z^{}\(s\hs\sin^2\hsm \theta \!-\!4M_Z^2\sin^2\!\hsm\frac{\,\theta\,}{2}\)\!\!
	\\[1mm]
	\!\! 2\hs s\hs (s \!-\! 2M_Z^2)\sin\hsm\theta\cos^2\hsmx\!\frac{\,\theta\,}{2}
	& -\sqrt{2\hs s\,} M_Z^{} \!\(s\hs\sin^2\!\theta\!-\! 4M_Z^2\cos^2\!\hsmx\frac{\,\theta\,}{2}\)
	&-s\hs M_Z^2\sin\hsm 2\theta
\end{pmatrix} \hsm\!,
\end{align}
}
for the process $e^-_Le^+_R\ito ZZ\hs$, and
{\small 
\begin{align}
& \hspace*{-2mm}
\mathcal{T}_{\rm{sm}}^{\rm{RL}}=
\frac{2\hs e^2{c_R^Z}^2}{~s_W^2 c_W^2
	(4M_Z^4 \!-\! 4\hs s \hs\sin^2\hsm\theta M_Z^2 \!+\!s^2\sin^2\hsm\theta)~}
\times 
\nn\\[1mm]
& \hspace*{-2mm}
\begin{pmatrix}
	-s\hs M_Z^2\sin \hsm 2\theta
	& \sqrt{2s\,}M_Z (s\sin^2\!\theta \!-\! 4M_Z^2\cos^2\!\hsm\frac{\,\theta\,}{2}) &
	2\hs s\hs (s\!-\!2M_Z^2)\sin\hsm\theta\cos^2\!\hsm\frac{\,\theta\,}{2}\!\!
	\\[1mm]
	\!\!\sqrt{2\hs s\,}\hs M_Z^{} (s\sin^2\!\theta \!-\! 4M_Z^2\sin^2\!\frac{\theta}{2})
	& 2\hs s\hs M_Z^2\sin\hsmx 2\theta
	& -\sqrt{2s\,}M_Z^{} (s\sin^2\hsm\!\theta \!-\! 4M_Z^2\cos^2\!\hsm\frac{\,\theta\,}{2}) \!\!
	\\[1.5mm]
	-2\hs s\hs (s\!-\!2M_Z^2)\sin\hsm\theta\hs \sin^2\!\!\frac{\,\theta\,}{2}
	& -\sqrt{2s\,}M_Z^{}(s\sin^2\!\theta \!-\! 4M_Z^2\sin^2\!\hsmx\frac{\,\theta\,}{2})
	& -s\hs M_Z^2\sin\hsm 2\theta
\end{pmatrix} \!\!.
\end{align}
}
for the process $e^-_Re^+_L\ito ZZ\hs$.\

As mentioned above, the $ZZV^*$ coupling only contributes to the helicity amplitude
for $e^-e^+\!\ito ZZ\hs$ with one longitudinally-polarized $Z$ boson and one
transversely-polarized $Z$ boson:
\beqs
\label{eq:TLR+TRL8}
\begin{align}
\mathcal{T}_{(8)}^{\rm{LR}} &= 
\frac{~c_L^Vf_5^Ve^2\hsm\sqrt{s\,}(s\!-\!4M_Z^2)~}{\sqrt{2\,}c_W^{}s_W^{}M_Z^3}\!\!
\begin{pmatrix}
\! 0 &\sin^2\!\frac{\theta}{2} & 0 \!\\
\! \cos^2\!\frac{\theta}{2} & 0 & -\sin^2\!\frac{\theta}{2}\!
\\
\! 0 & -\cos^2\!\frac{\theta}{2} & 0 \!
\end{pmatrix} \!\hsm ,
\label{TLR8}
\\[1mm]
\mathcal{T}_{(8)}^{\rm{RL}} & 
= \frac{~c_R^Vf_5^Ve^2\hsm\sqrt{s\,}(s\!-\!4M_Z^2)~}{\sqrt{2\,}c_W^{}s_W^{}M_Z^3}\!\!
\begin{pmatrix}
\! 0 & -\cos^2\!\frac{\theta}{2} & 0\\
-\sin^2\frac{\theta}{2}&0&\cos^2\frac{\theta}{2} \!
\\
\! 0 & \sin^2\!\frac{\theta}{2} & 0 \!
\end{pmatrix} \!\hsm .
\label{TRL8}
\end{align}
\eeqs
Making a high-energy expansion, we deduce the 
high-energy behaviors of the helicity amplitudes from the constributions of the SM and the nTGCs, respectively:
\beqs
\label{eq:Tsm-T8-Edep}
\begin{align}
& \mathcal{T}_{\rm{sm}}^{} =
\begin{pmatrix}
\!O\hsm\Big(\!\frac{M_Z^2}{E^2}\!\Big) & O\hsm\Big(\!\frac{M_Z}{E}\!\Big) & O(E^0)\!
\\[2mm]
\!\!O\hsm\Big(\!\hsm\frac{M_Z}{E}\!\hsm\Big) & O\Big(\hsm\!\frac{M_Z^2}{E^2}\!\Big) & O\hsm\Big(\hsm\!\frac{M_Z}{E}\hsm\!\Big)\!
\\[2mm]
\!O(E^0) & O\hsm\Big(\hsm\!\frac{M_Z}{E}\!\hsm\Big) & O\hsm\Big(\hsm\!\frac{M_Z^2}{E^2}\hsm\!\Big) \!
\end{pmatrix} \hsm\!,
\label{smhigh}
\\[1mm]
& \mathcal{T}_{(8)}^{} =
\begin{pmatrix}
\! 0 & O\Big(\!\hsm\frac{E^3}{M_Z^3\Lambda^4}\!\hsm\Big) & 0 \!
\\[1mm]
\!\! O\hsm\Big(\!\hsm\frac{E^3}{M_Z^3\Lambda^4}\hsm\!\hsm\Big) 
& 0 & O\hsm \Big(\!\hsm\frac{E^3}{M_Z^3\Lambda^4}\!\hsm\Big) \!\!
\\[1mm]
\! 0 & O\hsm\Big(\!\hsm\frac{E^3}{M_Z^3\Lambda^4}\hsm\!\Big) & 0 \!
\end{pmatrix} \hsm\!,
\label{d8high}
\end{align}
\eeqs
where we denote the energy scale $E\!=\!\sqrt{s\,}$.\ 

We see from Eq.(\ref{smhigh}) that the SM amplitude with the final states $Z_L^{}Z_T^{}$ would behave like 
$\mathcal{T}_{\rm{sm}}(Z_L^{}Z_T^{})\!=\! O\hsm\Big(\!\frac{M_Z}{\sqrt{s\,}\,}\!\Big)$ at high energies.\  
This behavior can be understood via the electroweak equivalence theorem (ET) that connects the scattering amplitude of longitudinal gauge bosons to the scattering amplitude of the corresponding Goldstone bosons\,\cite{ET}.\
For the current scattering process we have the following ET relation: 
\beq
\label{eq:ET-ee-ZZLT}
\TT [Z_L^{},\hsm Z_T^{}] \,=\, 
\TT [-\ii\pi^0\!,\hsm Z_T^{}] + B[v,\hsm Z_T^{}]\,,
\eeq
where $\pi^0$ denotes the neutral Goldstone boson that is converted to the longitudinal mode $Z_L^{}$ 
via the Higgs mechanim\,\cite{Higgs}.\ 
In Eq.\eqref{eq:ET-ee-ZZLT}, the $B$ term is defined as
$B\!=\!\TT [v, Z_T]$, 
with $v\!\equiv\!v_\mu^{}Z^\mu$ and 
$v^\mu \!=\!\ep_L^{\mu}\!-\!\ep_S^\mu\!=\!O(M_Z^{}/E)\hs$.\ 
Since the initial-state fermion masses are negligible, the SM contributions contain only the
$t$- and $u$-channel exchanges.\ Thus  we can deduce 
\begin{align}
\mathcal{T}_{\rm{sm}}[\pi^0 \!,\hsm Z_T]\simeq 0\hs,  \qquad 
\mathcal{T}_{\rm{sm}}[Z_L^{},\hsm Z_T^{}]\simeq B 
= O\hsm\bigg(\hsm\!\frac{M_Z}{\sqrt{s\,}\,}\hsm\!\bigg)  ,
\end{align}
which explains the high-energy behavior of $\mathcal{T}_{\rm{sm}}^{}[Z_LZ_T]$.\
Moreover,  Eq.(\ref{smhigh}) shows that
$\TT_{\rm{sm}}^{}[Z_L^{}Z_L^{}]\!=\!O(M_Z^2/E^2)$.\
This is because the ET gives
\beq 
\label{eq:ET-ee-ZZLL}
\TT [Z_L^{},\hsm Z_L^{}] \,=\, 
-\TT [\pi^0\!,\pi^0] + B[LL]\,,
\eeq 
where $\TT [\pi^0\!,\pi^0]\!=\!0$ holds because the initial-state fermions $e^\mp$ have very small masses, so that the Yukawa couplings between the Higgs boson (Goldstone boson) and fermions can be ignored.\ 
Hence, for the ET identity \eqref{eq:ET-ee-ZZLL}, we deduce the following: 
%
\begin{align}
\TT [Z_L^{},\hsm Z_L^{}] =  B[LL]
= \TT[-\ii\pi^0\!,v]\!+\!\TT[v,-\ii\pi^0]\!+\!\TT[v,v] = \TT[v,v] = O\hsm\bigg(\hsm\!\frac{M_Z^2}{E^2}\!\bigg) .
\end{align}
%
We can then compute the SM contributions to amplitude-squared as follows:
\begin{align}
\label{smampsquare}
\overline{|\mathcal{T}_{\rm{sm}}|^2}[ZZ] = &
\frac{~e^4s\Big(\!{c_L^Z}^4\!\!+\!{c_R^Z}^4\Big)\!\!
\left[16M_Z^6\big(1\!+\!\cos^2\!\theta\big)\!-\!4\big(3\!-\!7\cos^2\!\theta\!+\!4\hs s \cos^4\!\theta\big)M_Z^4\right]~}
{c_W^4 s_W^4 \hsm\big(4M_Z^4\!-\!4\hs s \sin^2\!\theta M_Z^2 \!+\! s^2\sin^2\!\theta \big)^{\!2}}
\notag\\
& -\frac{~e^4s\Big(\!{c_L^Z}^4\!\!+\!{c_R^Z}^4\hsm\Big)\!\!
\left[8\hs s^2\sin^2\!\theta\cos^2\!\theta M_Z^2 \!+\!s^3\hsm\big(\!\cos^4\!\theta\!-\!1\big)\hsm\right]~}
{c_W^4 s_W^4 \hsm\big(4M_Z^4\!-\!4\hs s \sin^2\!\theta M_Z^2 \!+\hsm s^2\sin^2\!\theta \big)^{\!2}} \, ,
\end{align}
where we average over the spins of the initial-state fermions and sum over the helicity combinations of the final states.\ 
The squared scattering amplitudes for the SM contributions to the final states $Z_L^{}Z_T^{}$ and  $Z_T^{}Z_L^{}$
take the following form:
\begin{align}
\label{smLT}
\overline{|\mathcal{T}_{\rm{sm}}|^2}[Z_LZ_T]
=\frac{~2\hs e^4sM_Z^2\hsm\Big(\!{c_L^Z}^4\!\!+\!{c_R^Z}^4\hsm\Big)\!\!
\left[ 4\big(1\!+\!\cos^2\!\theta\big)\hsm M_Z^4\!-\!4\hs s \sin^2\!\theta M_Z^2\!+\!s^2\hsm \sin^4\!\theta \right]~}
{c_W^4 s_W^4 \big(4M_Z^4\!-\!4\hs s \sin^2\!\theta\hs M_Z^2\!+\! s^2\sin^2\!\theta \big)^{\!2}} \, .
\end{align}
We have verified that our above formulas \eqref{smampsquare} and \eqref{smLT} agree with the literature\,\cite{Degrande:2013kka}.\ 
We can further derive the SM contributions to the squared amplitudes for the final states $Z_L^{}Z_L^{}$ and $Z_T^{}Z_T^{}$:  
\begin{subequations}
\begin{align}
\label{eq:Amp2-SM-ZLZL}
\hspace*{-3mm}
\overline{|\mathcal{T}_{\rm{sm}}|^2}[Z_L^{}Z_L^{}] &=
\frac{e^4s^2M_Z^4\Big(\!{c_L^Z}^4\!\!+\hsm {c_R^Z}^4\hsm\Big)\!\sin^2\!2\theta}
{~c_W^4 s_W^4\hsm \big(4M_Z^4\!-\!4\hs s \sin^2\!\theta M_Z^2\!+\!s^2\sin^2\!\theta \big)^{\!2}~} \,,
\\[1mm]
\label{eq:Amp2-SM-ZTZT}
\hspace*{-3mm}
\overline{|\mathcal{T}_{\rm{sm}}|^2}[Z_T^{}Z_T^{}] &=
\frac{~e^4s^2\hsm\Big(\!{c_L^Z}^4\!\!+\!{c_R^Z}^4\hsm\Big)\!\!
	\left[4\big(1\!+\!2\cos^2\!\theta\big)M_Z^4\!-\!4\hs s \big(1\!+\!\cos^2\!\theta\big)\hsm M_Z^2
	\!+\! s^2\big(1\!+\!\cos^2\!\theta\big)\hsm\right]\!\sin^2\!\theta~}
{8c_W^4 s_W^4 \big(4M_Z^4\!-\!4\hs s \sin^2\!\theta\hs M_Z^2\!+\hsm s^2\sin^2\!\theta \big)^{\!2}} \,.
\end{align}
\end{subequations}

As we have discussed earlier, the new physics of $ZZV^*$ nTGCs contribute only to the reaction
$e^-(p_1)e^+(p_2)\!\ito\! Z(p_3)Z(p_4)$ with the final states $Z_L^{}Z_T^{}$ and $Z_T^{}Z_L^{}$.\  
Computing the interference between the contributions from the dimension-8 nTGC operator and from the SM, 
we find the following form:
\begin{align}
2\Re\frak{e}\Big(\hsm\over{\mathcal{T}_{\rm{sm}}\mathcal{T}_{(8)}^*}\Big)\! [Z_LZ_T] 
= 
\frac{~2\hs s \hs e^4\hsm f_5^V \!\big(c_L^V{c_L^Z}^2\!\!-\hsm c_R^V{c_R^Z}^2\big)\!
\big(4M_Z^2\hsm -\hsm s\big)\!\big[2M_Z^2\big(1\!+\!\cos^2\!\theta\big)\!-\!s\sin^2\!\theta\big]~}
{c_W^3 s_W^3M_Z^2 \big(4M_Z^4\!-\! 4\hs s \sin^2\!\theta M_Z^2 \!+\!s^2\sin^2\!\theta \big)} \,.
\end{align}
%
This interference term is of ${O}(\Lambda^{-4})$.\ We can further compute the squared amplitude
of ${O}(\Lambda^{-8})$  from the dimension-8 nTGC contributions as follows:
\begin{align}
\overline{|\mathcal{T}_{(8)}|^2}[Z_L^{}Z_T^{}] \hs =\hs 
\frac{~e^4s{f_5^V}^2\hsm\big({c_L^V}^2\!\!+\!{c_R^V}^2\big)\!\big(s\!-\!4M_Z^2\big)^{\!2}\hsm\big(3\!+\!\cos\hsm 2\theta\big)~}
{4\hs s_W^2c_W^2M_Z^6}. 
\end{align}
According to the energy power counting results \eqref{eq:Tsm-T8-Edep}, 
we can infer the leading high-energy behavior of each contribution to the squared amplitude $|\TT|^2$:
\beq 
|\TT_{\rm{sm}}|^2 \!\sim\! E^0, \hspace*{4mm} 
\sum_{\lambda,\lambda'}\!\TT_{\rm{sm}}^{}(\lambda\lambda')\TT_{(8)}^*(\lambda\lambda') \!\sim\! E^2\hs,
\hspace*{4mm} 
|\TT_{(8)}|^2 \!\sim\! E^6 .
\eeq 
However, we recall that the $Z$ boson is unstable and decays inside the detector.\ 
The intermediate-state $Z$ bosons with different polarizations interfere, 
and the leading high-energy dependence of $\TT_{\rm{sm}}^{}\TT_{(8)}^*$ is $E^3$. We need to analyze the angular distributions of final-state particles 
from $Z$ boson decays to obtain information about the interference of 
different $Z$ polarizations, which is discussed in the following section.\ 

\begin{figure}[t]
\begin{center}
\includegraphics[height=8cm,width=11cm]{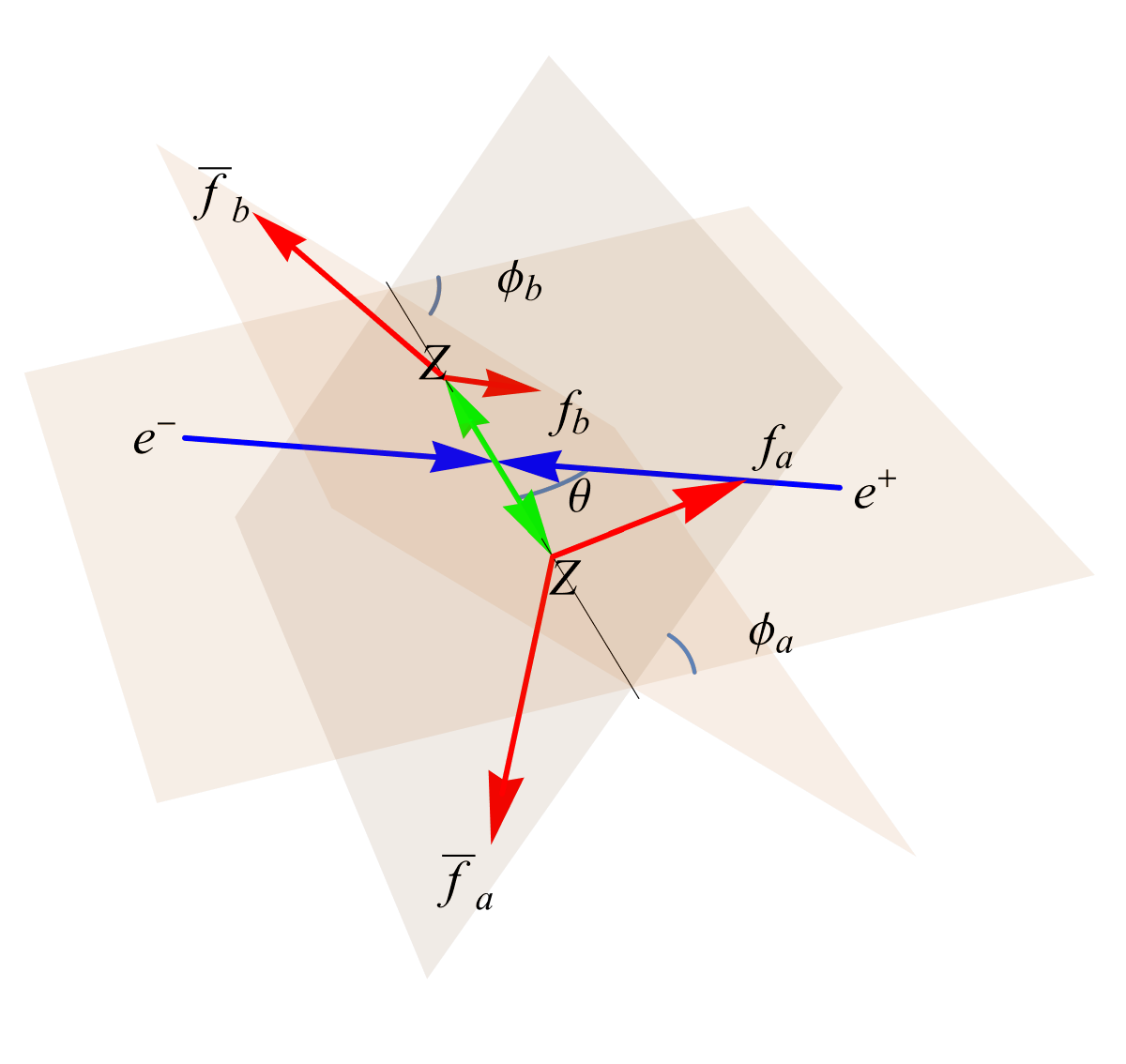}
\vspace*{-10mm}
\caption{\hspace*{-0.5mm}\small {Kinematic structure of the reaction $\,e^-e^+\!\!\to ZZ\,$ followed by the fermionic decays 
		$\,Z\hsm\ito\hsm f\bar{f}\hs$,\, in the $\,e^-e^+$ collision frame.}}
\vspace*{-1mm}
\end{center}
\label{kinematics}
\label{fig:3}
\end{figure}

Using these results, we derive the total cross section of the reaction $e^-e^+\ito ZZ$ as follows:
\begin{align}
\sigma \,=~&
\frac{~e^4 \big({c_L^Z}^4\!\!+\!{c_R^Z}^4\big)\! \!
\left[\hsm -2\hs s \beta (s\!-\!2 M_Z^2)\!+\!\big(4 M_Z^4\!+\!s^2\big)\!
\ln\!\frac{\,s\hs (1+\beta)-2 M_Z^2\,}{\,s\hs (1-\beta) -2 M_Z^2\,}\right]~}
{16\hs\pi\hs  c_W^4 s_W^4 \big(s\!-\!2 M_Z^2\big)s^2}
\notag\\
& +\frac{~e^4f_5^V\!\big(c_L^V{c_L^Z}^2\!-\!c_R^V{c_R^Z}^2\big) \!\!
\left[ s\hs (s\!+\!2 M_Z^2)\beta \!+\!4 M_Z^2 \big(s\!-\!M_Z^2\big)
\rm{arcoth}\hsm\Big(\hsm\!\frac{\,2 M_Z^2-s\,}{\beta\hs s}\hsm\!\Big)\!\right]~}
{16\hs\pi\hs  c_W^3s_W^3M_Z^2\, s^2}
\notag\\
&+\frac{~e^4{f_5^V}^2\hsm\big({c_L^V}^2\!\!+\!{c_R^V}^2\big)\hsm\beta^{5}s^2~}{192\hs\pi\hs c_W^2 s_W^2M_Z^6} \, ,
\label{eq:sigma-eeZZ} 
\end{align}
where $\beta\!=\! p_Z^{}/E_Z^{}\!=\hsm\!\sqrt{1\!-\!4M_Z^2/s\,}$
and the symbol ``$\rm{arcoth}$'' denotes the inverse function of the hyperbolic tangent function $\coth\hs$.\ 
In the high-energy limit, we can deduce the following behaviors for each contribution to the total cross section:
\beq 
\label{eq:sigma-expand}
\sigma =O(s^{-1})+O\hsm\bigg(\!\!\frac{\,s^0M_Z^2\,}{\,\Lambda^4\,}\!\hsm\bigg)
+ O\hsm\bigg(\!\!\frac{\,s^2M_Z^2\,}{\,\Lambda^8\,}\!\hsm\bigg) .
\eeq
We see that the SM contribution scales as $s^{-1}$ and decreases with energy, 
whereas the interference term scales as $s^0\!/\hsm\Lambda^4$ and thus 
is insensitive to the energy.\ Finally, the last contribution (arising from the squared amplitude)
scales as $s^2\!/\hsm\Lambda^8$ and increases with energy rapidly.

\begin{figure}[t]
\centering
\includegraphics[width=7.5cm,height=5.5cm]{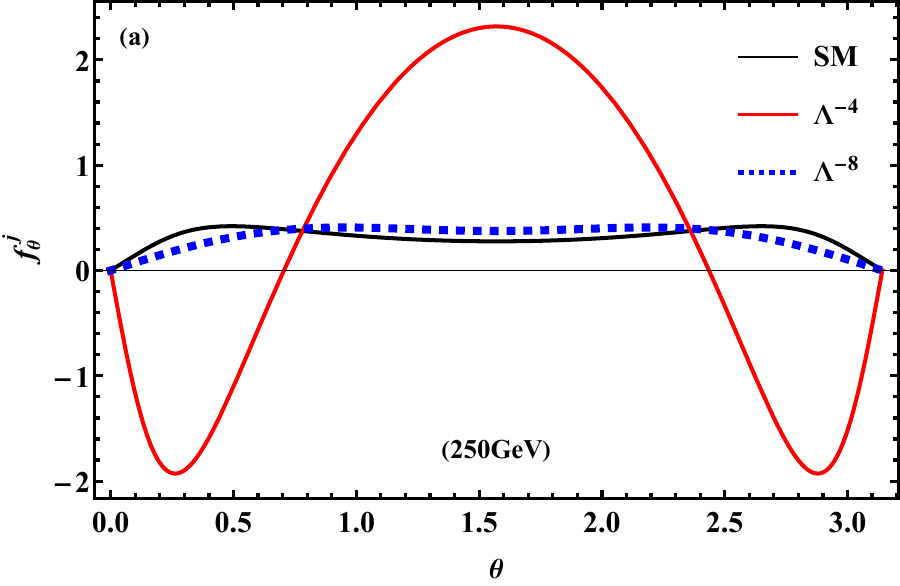}
\includegraphics[width=7.5cm,height=5.5cm]{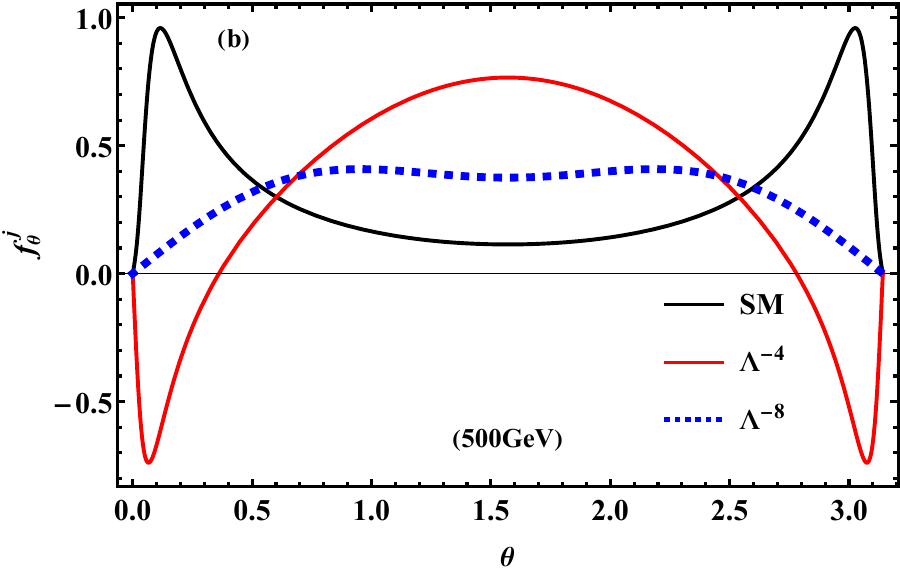}
\\[1.5mm]
\includegraphics[width=7.5cm,height=5.5cm]{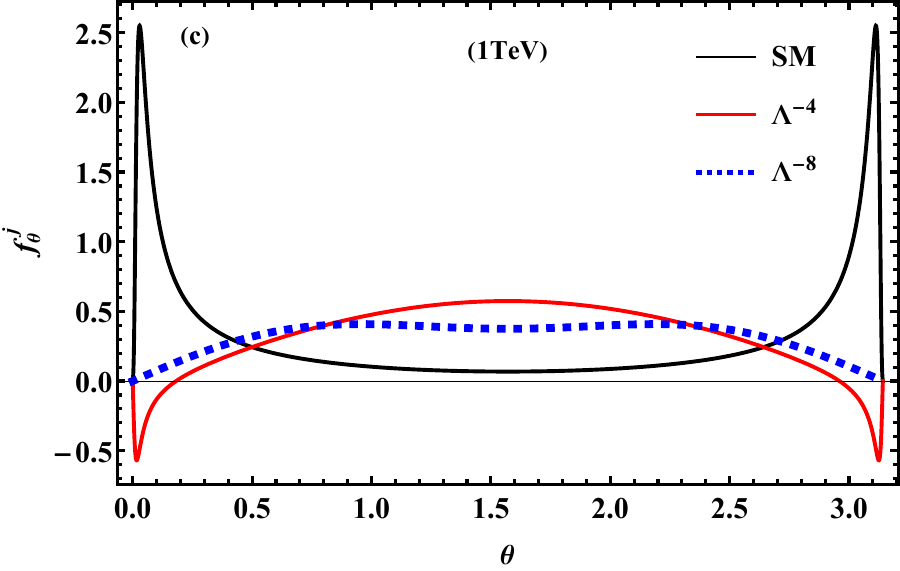}
\includegraphics[width=7.5cm,height=5.5cm]{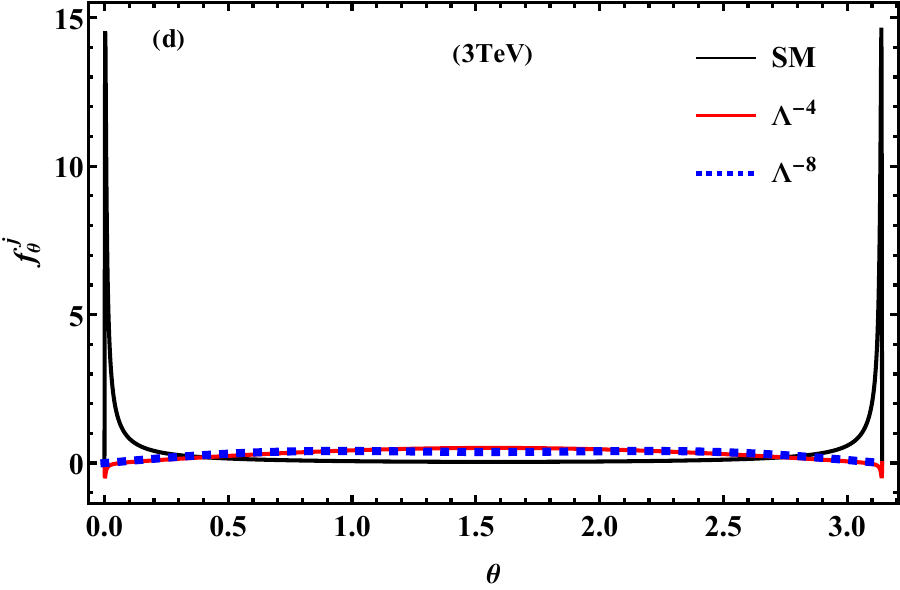}
\vspace*{-3mm}
\caption{\hspace*{-1.5mm}\small 
Normalized angular distributions in the polar scattering angle $\theta$ 
for different collider energies,
$\sqrt{s\,}\!=\!(0.25, 0.5, 1, 3)$\,TeV, are shown in plots\,(a)-(d) respectively.\ In each plot, the black, red and blue curves denote the contributions from the SM, the ${O}(\cut^{-4})$ term, and the ${O}(\cut^{-8})$ term, respectively.\ The $e^+e^-$ collision energy chosen for each plot
is shown in the parentheses, such as 
(250{\hs}GeV), (500{\hs}GeV), and so on.\   
}
\vspace*{2mm}
\label{fig:ftheta}
\label{fig:4}
\end{figure}

For the current analysis, we define the following normalized angular distribution functions:
\begin{equation}
f^j_{\theta} = \frac{1}{\,\sigma_j^{}\,}\frac{\dif\sigma_j}{\dif\theta} \, ,
\end{equation}
where $\sigma_j^{}$ (with $j\!=\!0,1,2$) represents the SM contribution ($\sigma_0^{}$), 
the ${O}(\Lambda^{-4})$ contribution ($\sigma_1$), and the ${O}(\Lambda^{-8})$ contribution ($\sigma_2$).\ 
Then, we compute the angular distribution functions as follows:
{\small 
\beqs      
\label{eq:f-theta-012}
\begin{align}
f_{\theta}^0 &=
\frac{~s^2\beta\big(2M_Z^2\!-\!s\big)\hsm\!
	\left\{\! 192M_Z^6\!-\!32\hs s M_Z^4\!-\!8\hs s^2M_Z^2 \!+\!5\hs s^3\!-\!s\hs\beta^2\hsm  
	\big[4\cos\hsm 2\theta\big(s\!+\!2M_Z^2\big)^{\!2}\hsm\!+\!s^2\beta^2\hsmx \cos\hsm 4\theta\big]\!\right\}\hsm\sin\hsm\theta~}
{16 \!\left[2\hs s \beta \big(s\!-\!2M_Z^2\big)\!+\!(s^2\!+\!4M_Z^4)\hsm
	\ln\!\frac{\,s(1-\beta) -2 M_Z^2\,}{\,s(1+\beta)-2 M_Z^2\,}\hsmx\right]\!\!
	\big(4M_Z^4\!+\!s^2\beta^2\sin^2\!\theta\big)^2} \,, 
\\[0.7mm]
f_{\theta}^1 &=
-\frac{s^3\beta ^3\!\left[ 6M_Z^2\!-\!s\!+\!\big(s\!+\!2M_Z^2\big)\cos\hsm 2\theta\right]\hsm\sin\hsm \theta}
{~4\!\left[\hsm s\hs\beta \big(s\!+\!2M_Z^2\big)\!+\!4M_Z^2\big(s\!-\!M_Z^2\big) \rm{arcoth}\hsm\frac{\,2M_Z^2-s\,}{\beta\hs s}\!\right] 
	\!\!\big(4M_Z^4\!+\!s^2\hs\beta^2\sin^2\!\theta\big)~} \, ,
\\
f_{\theta}^2 &=\frac{3}{\,16\,} \big(3+\cos2\theta\big)\hsmx\sin\hsm\theta \, .
\end{align}
\eeqs
}
\hspace*{-3mm}
We see in Fig.\,\ref{fig:ftheta} that the SM distribution is concentrated around the forward and backward scattering regions 
with $\theta$ close to $0^\deg$ or $180^\deg$.\ This feature is particularly clear for large collision energies
$\sqrt{s\,}\!\gg\!\! M_Z^{}\hs$.\ 
It also shows that the interference distributions of 
${O}(\Lambda^{-4})$ have negative sign in the small $\sin\hsm\theta$ region, namely, when the scattering angle $\theta$ obeys the condition
$|\hsm\sin\hsm\theta|\!<\!2M_Z^{}/(s\!+\!2M_Z^2)^{1/2}\hs$.

Next, we further incorporate the fermionic decays of the final states $ZZ\hs$.\ 
We first consider $Z$ decay in its rest frame, $Z(q_1^{})\ito f(k_1)\bar{f}(k_2)$, 
and express the momenta of the final-state fermions as follows:
\\[-7mm]
\beqs
\begin{align}
k_{1}^{} &= k\big(1, \sin \hsm\theta_{a}\hsm\cos\hsm\phi_{a}, \sin\hsm\theta_{a}\hsm\sin\hsm\phi_{a}, \cos\hsm\theta_{a}\big) \hs ,
\\
k_{2}^{} &= k\big(1,-\hsm\sin\hsm\theta_{a} \hsm\cos\hsm \phi_{a},-\hsm\sin \hsm\theta_{a} \hsm\sin \hsm\phi_{a},-\hsm\cos\hsm\theta_{a}\big) 
\hs,
\end{align}
\eeqs
where $k\!=\!\frac{1}{2} M_Z^{}$ holds since we treat the final-state fermions as effectively massless.\ 
Here the positive direction $\hat{z}_a^{}$ in the $Z$ rest frame is chosen to be
the $Z$ moving direction in the laboratory frame,
and $\theta_a^{}$ denotes the angle between the positive direction $\hat{z}_a^{}$
and the fermion $f$ direction of motion in the $Z$ rest frame.\ 
We define $\phi_a^{}$ as the angle between the $e^-e^+$ scattering plane 
and the $Z$-decay plane in the $e^-e^+$ center-of-mass frame.\
Similarly, we can express the kinematics of the other $Z$ decay 
in terms of the angular variables $\theta_{b}^{}$ and $\phi_{b}^{}\hs$.\ This kinematic structure is illustrated in Fig.\,\ref{fig:3}.

Then, we express the helicity amplitude for the reaction $e^-e^+\!\ito ZZ\ito f_a \bar f_a f_b^{}\bar f_b^{}$ as follows:
\begin{align}
{\TT} = &
~\frac{~e^2M_Z^2\hs\mathcal{D}_{\!Z}^{}(q_1^{})\mathcal{D}_{\!Z}^{}(q_2^{})~}{2\hs s_W^2c_W^2}\hspace*{-2mm}
\sum_{\lambda,\lambda'=-1}^1 \hspace*{-3mm}
{e^{-\text{i}(\lambda\phi_{a}+\lambda'\phi_{b})}}
\hs\mathcal{T}_{ss'}[\lambda\lambda'] \nn \\ 
& \times\!\bigg\{\!(1\!-\!\delta_{\lambda}^0)\hsmx\big[f_R^a(1\!-\!\lambda\cos\hsm\theta_{a})\!-\!
f_L^a(1\!+\!\lambda\cos\hsm\theta_{a})\big]\!+\!\sqrt{2\,} \delta_{\lambda}^0
\sin\hsm\theta_a(f_R^a\!+\!f_L^a)\!\bigg\}
\nn 
\\ 
& \times\!\bigg\{\! (1\!-\!\delta_{\lambda'}^0)\hsmx\big[f_R^b(1\!-\!\lambda'\hsm\cos\hsm\theta_b^{})
\!-\!f_L^b(1\!+\!\lambda'\hsm\cos\hsm\theta_b^{})\big]\!+\!\sqrt{2\,} \delta_{\lambda'}^0
\sin\hsm\theta_b^{}(f_R^b\!+\!f_L^b)\!\bigg\} ,
\label{eq:T-ZZ-4f}
\end{align}
where $D_{\!Z}^{}$ is the propagator of the $Z$ boson, and 
$\,(f_L^{a,b}\hsm ,\hs f_R^{a,b})\!=\! (f_L^{}\delta_{\!\si,-\frac{1}{2}}^{},\hs f_R^{}\delta_{\!\si,\frac{1}{2}}^{})$ 
denote the gauge couplings of the fermions $a$ and $b$ with helicity $\sigma\hs$.\  
In the above, the gauge couplings $(f_L^{},f_R^{})\!=\! (T_3^{} \!-\! Q\hs s_W^2,-Qs_W^2)$ 
denote the gauge couplings of the (left,\,right)-handed fermions.\ 
In Eq.\eqref{eq:T-ZZ-4f}, ${\TT}_{ss'}[\lambda\lambda']$ 
represents the on-shell helicity amplitude of the scattering process $e^{-} e^{+} \ito ZZ\,$,
\begin{equation}
{\TT}_{ss'}^{}[\lambda\lambda'] \,=\,  
\mathcal{T}_{\rm{sm}}^{ss'}[\lambda\lambda'] + \mathcal{T}_{(8)}^{ss'}[\lambda\lambda'] \,,
\end{equation}
which include the contributions from both the SM and the dimension-8 nTGC operators.

Next, we derive the following normalized angular distributions 
$f^j_{\theta_{a}}\!\!=\hsm\!\frac{1}{\,\sigma_j}\frac{\dif\sigma_j}{\dif\theta_{a}}$ 
as functions of the angle $\theta_{a}^{}$:
\begin{subequations}
\label{eq:f-theta-a012}  
\begin{align}
f^0_{\theta_{a}} =~ &
\frac{~3\sin\hsm\theta_a^{}\big(s\!-\!2M_Z^2\big)\!\!\left[3s\!-\!10M_Z^2\hsm +\!\big(s\!+\!2M_Z^2\big)\!\cos\hsm 2\theta_a\hsm\right]~}
{~16\hs\beta\!\hsm 
	\left[\hsm s\big(s\!-\!2M_Z^2\big)\beta \!+\!\big(s^2\!+\!4M_Z^4\big)\rm{arcoth}\hsm\frac{\,2M_Z^2-s~}{\beta\hs s}\hsm\right]~}
\notag\\
& +\frac{~6\sin\hsm\theta_a^{}\!\left[\hsm\big(1\!+\!\cos^2\!\theta_a^{}\big)\!\big(s^3\!-\!4\hs s^2M_Z^2\!+\!4\hs sM_Z^4\!+\!8M_Z^6\big)
	\!-\!32M_Z^6\right]\!
	\rm{arcoth}\hsm\frac{\,2M_Z^2-s~}{\beta\hs s}~}
{16\hs\beta^2s\!\hsm\left[s\big(s\!-\!2M_Z^2\big)\beta\!+\!\big(s^2\!+\!4M_Z^4\big)\rm{arcoth}\frac{2M_Z^2-s}{\beta s}\right]~} \, ,
\label{eq:fthetaa0}
\\
f^1_{\theta_{a}} =~ & f^2_{\theta_{a}}
=\frac{3}{\,32\,}\sin\hsm\theta_a^{}\big(5 \hsm -\hsm  \cos\hsm 2\theta_a^{}\big)  ,
\label{eq:fthetaa1}
\end{align}
\end{subequations}
and the following normalized angular distributions  $f^j_{\phi_{a}}\!\!=\!\frac{1}{\,\sigma_j}\frac{\dif\sigma_j}{\dif\phi_{a}}$ 
with respect to the angle $\phi_{a}^{}$:
\beqs
\label{eq:fphi-a012}
\begin{align}
f^0_{\phi_{a}} =\,& \frac{1}{\,2\hs\pi\,}
\!+\!\frac{3\hs\pi\sqrt{s\,} \beta M_Z^3 f_-^2\big({c_L^Z}^4\!\!-\!{c_R^Z}^4\big)\hsm\cos\hsm\phi_a}
{~8f_+^2 \big({c_L^Z}^4\!\!+\hsm {c_R^Z}^4\big)\!\!
\left[s\big(s\!-\!2M_Z^2\big)\beta\!+\!\big(s^2\!+\!4M_Z^4\big) \rm{arcoth}\hsm\frac{\,2M_Z^2-s\,}{\beta\hs s}\right]~}
\notag\\
&-\!\frac{~M_Z^2\big(s\!-\!2M_Z^2\big)\!\!\left[s\hs\beta \!+\!\big(s\!-\!2M_Z^2\big)
\rm{arcoth}\hsm\frac{\,2M_Z^2-s\,}{\beta\hs s}\hsm\right]\!\cos\hsm 2\phi_a^{}~}
{~2\hs\pi \beta^2 s\!\hsm 
\left[\hsm s\big(s\!-\!2M_Z^2\big)\beta\!+\!\big(s^2\!+\!4M_Z^4\big) \rm{arcoth}\hsm\frac{\,2M_Z^2-s\,}{\beta\hs s}\right]~} \,,
\label{eq:fphi0}
\\
f^1_{\phi_{a}} =\,& 
\frac{1}{\,2\hs\pi\,} \!-\! 
\frac{3\hs\pi\beta^3 f_-^2\hs s^{\frac{3}{2}}\hsm\big(3s\!-\!8M_Z^2\big)\!\big(c_L^V{c_L^Z}^2\!\!+\!c_R^V{c_R^Z}^2\big) 
\!\cos\hsm\phi_a^{}}
{~256M_Z^{}f_+^2\hsm\big(c_L^V{c_L^Z}^2\!\!-\!c_R^V{c_R^Z}^2\big)\!\!
\left[\hsm s\big(s\!+\!2M_Z^2\big)\beta \!+\! 4M_Z^2\big(s\!-\!M_Z^2\big)\rm{arcoth}\hsm\frac{\,2M_Z^2-s\,}{\beta\hs s}\right]~
}\notag\\
&-\frac{\left[ s\big(s\!-\!2M_Z^2\big)\beta\hsm +\hsm 4M_Z^4\hs\rm{arcoth}\frac{2M_Z^2-s}{\beta s}\right]\!\cos\hsm 2\phi_a^{}}
{~8\hs\pi\!\left[\hsm s\big(s\!+\!2M_Z^2\big)\beta\hsm +\hsm 4M_Z^2\big(s\!-\!M_Z^2\big)\rm{arcoth}\hsm\frac{\,2M_Z^2-s\,}{\beta\hs s}\right]~} \, ,\label{eq:fphi1}
\\
f^2_{\phi_{a}} =\,&
\frac{1}{\,2\hs\pi\,} - \frac{~\cos\hsm 2\phi_a^{}\,}{16 \pi } \, ,
\label{eq:fphi2}
\end{align}	
\eeqs
where we have defined notations 
$f_\pm^2\hsm\!=\hsm\! f_L^2\!\pm\! f_R^2\hs$, 
with $(f_L^{},f_R^{})\hsm\!=\hsm\! (T_3^{} \!-\! Q\hs s_W^2,-Qs_W^2)$ 
denoting gauge couplings of the (left,\,right)-handed fermions 
in the final states.\
The other notations ($c_L^{V},\hs c_R^{V}$) denote the gauge couplings of the (left,\,right)-handed $e^\mp$ 
with the SM gauge boson $V\,(=\!Z,\gamma )$ 
as defined in Eq.\eqref{eq:ee-cLR}.\  
We note that 
$f^1_{\theta_{a}} \!\!=\! f^2_{\theta_{a}}$ in Eq.\eqref{eq:fthetaa1}.\ 
In general,  
the angular distributions $f_{\theta_a}^i$ depend on the $ZZ$ helicity combinations (TT, TL, LL).\ However, we see in Eq.\eqref{eq:TLR+TRL8} that 
$\mathcal T_{\rm{sm}}\mathcal T_{(8)}^*$ and $|\mathcal T_{(8)}|^2$ 
both contain only the (TL) combination, hence the distributions 
$f^1_{\theta_a}$ and $f^2_{\theta_a}$ are the same.

\begin{figure}[t]
\centering
\includegraphics[width=7.5cm,height=5.5cm]{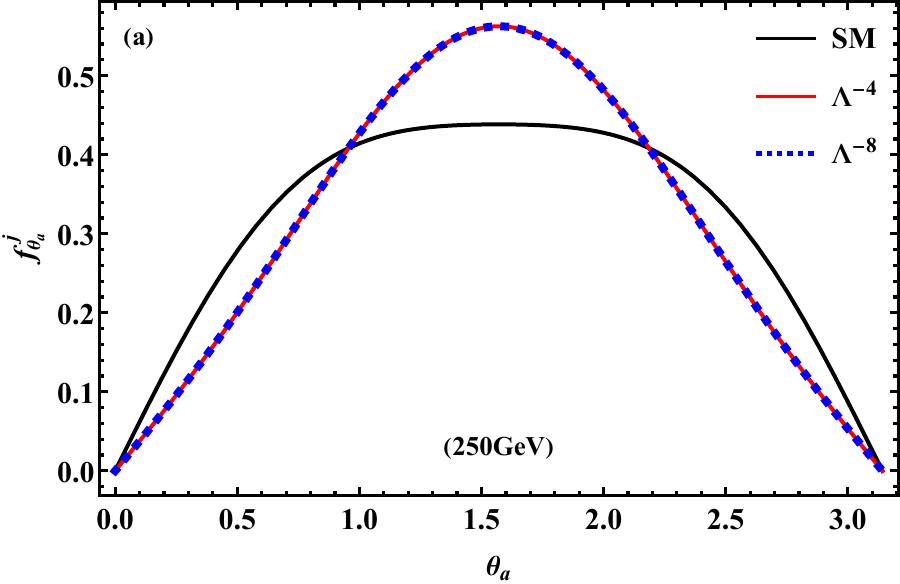}
\includegraphics[width=7.5cm,height=5.5cm]{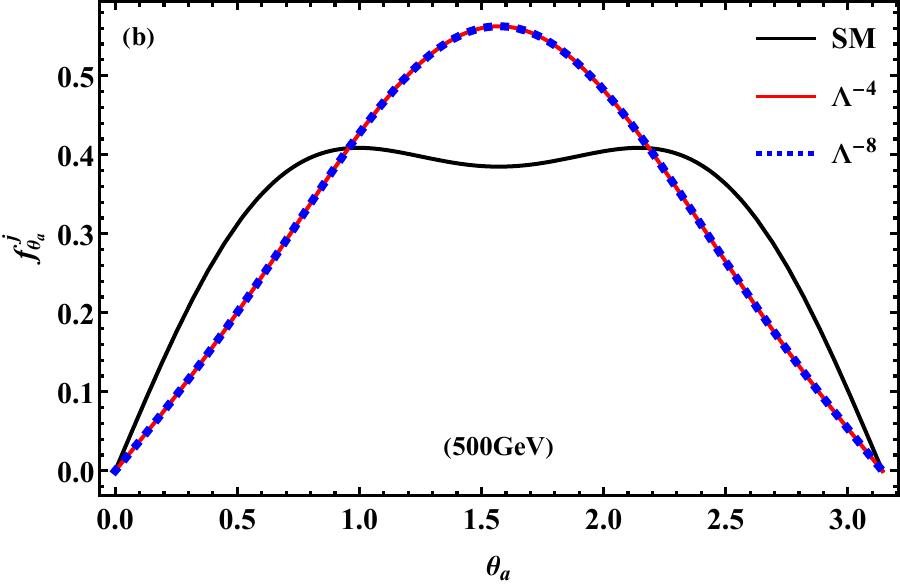}
\\[1mm]
\includegraphics[width=7.5cm,height=5.5cm]{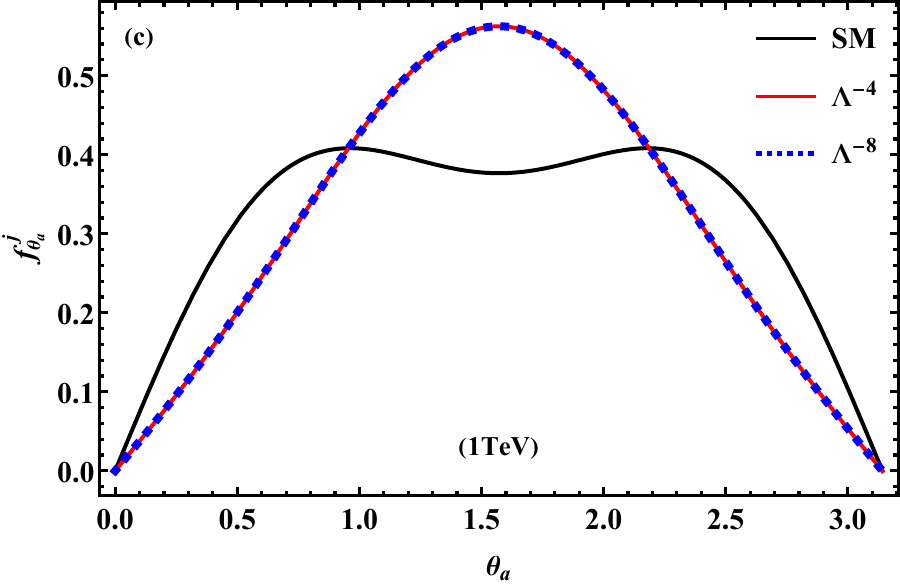}
\includegraphics[width=7.5cm,height=5.5cm]{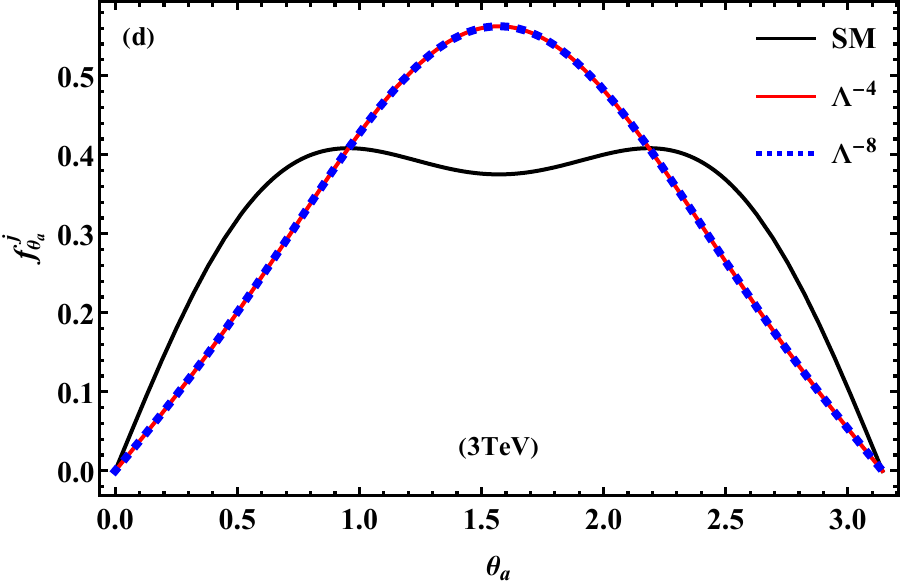}
\vspace*{-3mm}
\caption{\hspace*{-1.5mm}\small 
Normalized angular distribution in the polar angle $\theta_a$ 
in the $Z$ decay frame for different collider energies,
$\sqrt{s\,}\!=\!(0.25, 0.5, 1, 3)$\,TeV, as shown 
in plots\,(a)-(d) respectively.\  
In each plot, the black, red and blue curves denote the 
contributions from the SM, the ${O}(\cut^{-4})$ interference, and the ${O}(\cut^{-8})$ term respectively, 
where the red and blue curves exactly overlap.\
The $e^+e^-$ collision energy chosen for each plot
is shown in the parentheses, such as 
(250{\hs}GeV), (500{\hs}GeV), and so on.\  
}
\label{fig:thetaa}
\label{fig:5}
\vspace*{5mm}
\end{figure}

\begin{figure}[t]
\centering
\includegraphics[width=7.5cm,height=5.5cm]{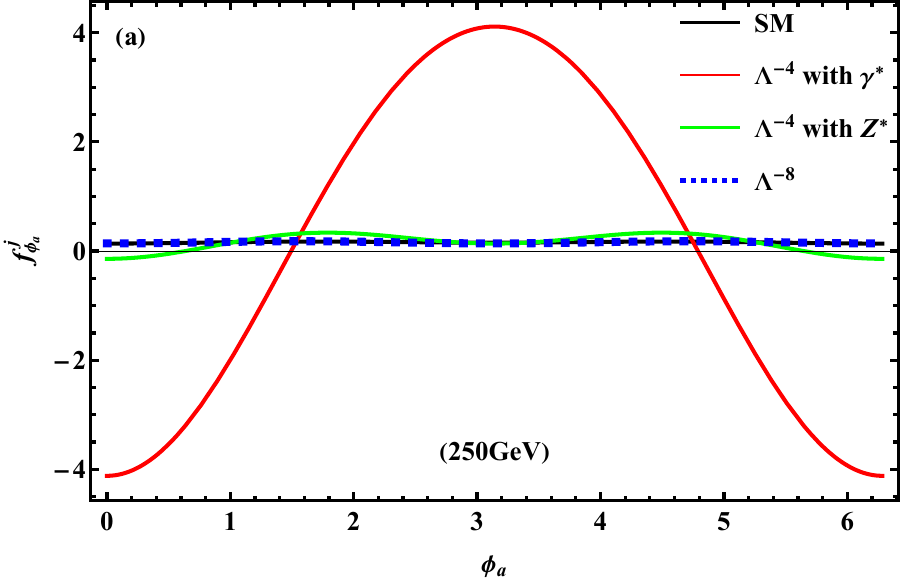}
\includegraphics[width=7.5cm,height=5.5cm]{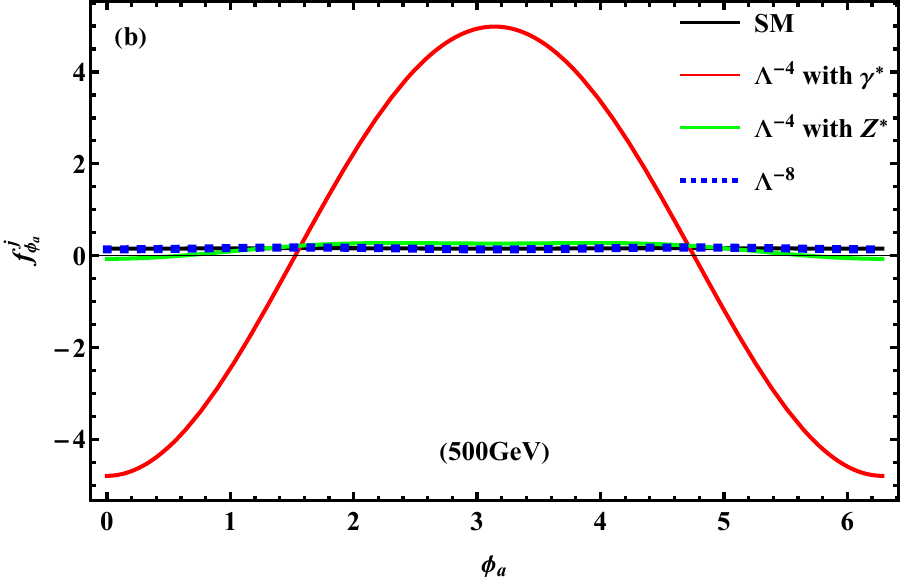}
\\[1.5mm]
\includegraphics[width=7.5cm,height=5.5cm]{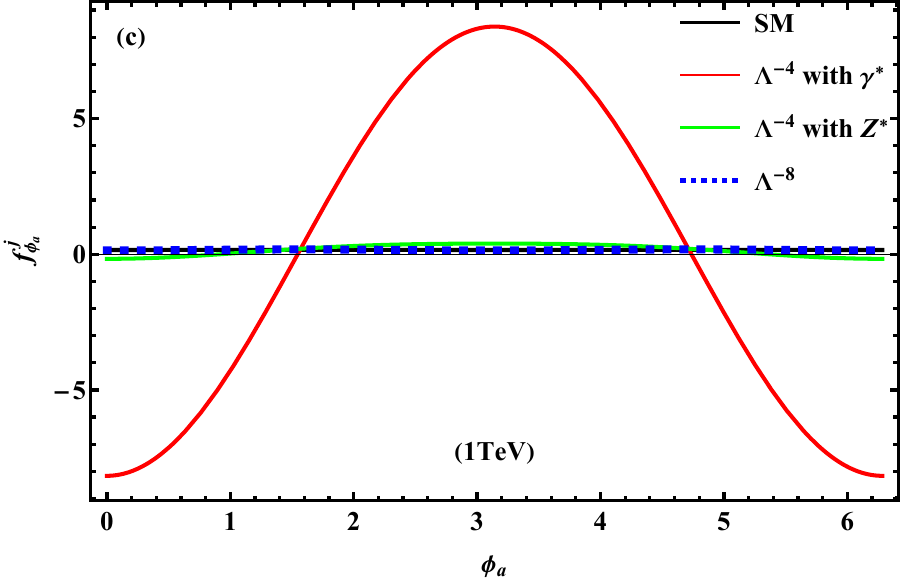}
\includegraphics[width=7.5cm,height=5.5cm]{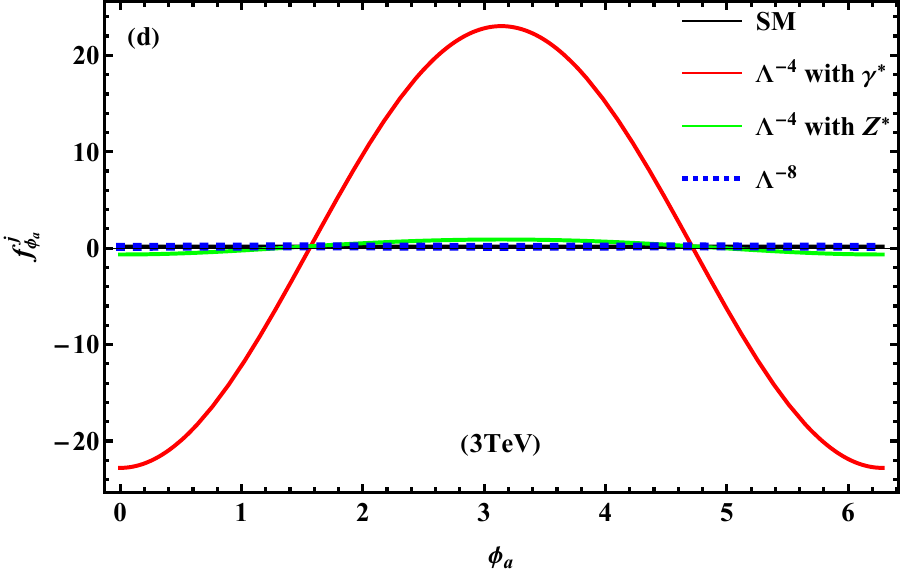}
\vspace*{-3mm}
\caption{\hspace*{-1mm}\small 
Normalized angular distributions in the azimuthal angle $\phi_a^{}$ 
of down-type quarks for different collider energies,
$\sqrt{s\,}\!=\!(0.25, 0.5, 1, 3)$\,TeV, are shown in plots\,(a)-(d) respectively.\  
In each plot, the black, red, green and blue curves denote the contributions 
from the SM, the ${O}(\Lambda^{-4})$ interference with $\gamma^*$, the ${O}(\Lambda^{-4})$ interference with $Z^*$, 
and the ${O}(\Lambda^{-8})$ term, respectively, 
where the blue and black curves nearly overlap.\
The $e^+e^-$ collision energy chosen for each plot is shown 
in the parentheses, such as (250{\hs}GeV), (500{\hs}GeV), and so on.\  
}
\label{fig:phia}
\label{fig:6}
\end{figure}

Using the above analytical formulas, we present numerical results for the normalized angular distribution functions of 
$\theta$, $\theta_a^{}$, and $\phi_a^{}$ in Figs.\,\ref{fig:ftheta}, \ref{fig:thetaa}, and \ref{fig:phia}, respectively.\ 
In each figure, the four plots have input different collision energies, 
$\sqrt{s\,}\!=\!(0.25, 0.5, 1, 3)$\,TeV, 
corresponding to the expected collision energies of the CEPC, FCC-ee, ILC and its possible energy upgrades, and the design energy of CLIC.\ 
In Figs.\,\ref{fig:ftheta} and \ref{fig:thetaa} we use black, red and blue curves to denote the contributions from the SM, 
the ${O}(\Lambda^{-4})$ interference, and the ${O}(\Lambda^{-8})$ term, respectively.\ 
In Fig.\,\ref{fig:phia}, we use black, red, green and blue curves to denote the contributions 
from the SM, the ${O}(\Lambda^{-4})$ interference with $\gamma^*$, the ${O}(\Lambda^{-4})$ interference with $Z^*$, 
and the ${O}(\Lambda^{-8})$ term, respectively.\ 
We find that the distributions with respect to $\theta_a^{}$ and $\phi_a^{}$ are similar to 
the corresponding distributions for the $Z\ga$ final states given in \cite{Ellis:2020ljj}\cite{Ellis:2019zex}.\ 
The SM and pure dimension-8 distributions in $\phi_a^{}$ are rather flat, 
whereas the ${O}(\Lambda^{-4})$ interference terms are dominated by cosine functions.

It is of interest to examine the high-energy behaviors of the angular distribution functions.\ 
In the case of the angular distribution respect to the scattering angle $\theta$, the distribution from the SM, 
namely $f^0_\theta$, increases rapidly with collision energy in the small $\sin\theta$ region, whereas the distribution 
from the ${O}(\Lambda^{-4})$ interference term, $f^1_\theta$, decreases with energy and the distribution 
from the pure ${O}(\Lambda^{-8})$ dimension-8 term, $f^2_\theta$, is rather flat.\ 
For all the functions $f^j_{\theta_a}$, the coefficients of all trigonometric functions approach constants 
in the high-energy limit $s\!\gg\! M_Z^2$.\ 
This is why Fig.\,\ref{fig:thetaa} shows that the distributions in $\theta_a^{}$ are not sensitive 
to the collision energy $\sqrt{s\,}$, as we vary the collision energy over the range 
$\sqrt{s\,}\!=\!(0.25, 0.5, 1, 3)$\,TeV in the four plots.\ 
In the high-energy limit $s\!\!\gg\!\! M_Z^2\hs$ we can derive the following angular distributions with respect to
the angle $\phi_a$: 
\beqs
\begin{align}
f^0_{\phi_{a}} &= \frac{1}{\,2\hs\pi\,} \,,
\\
f^1_{\phi_{a}} &= \frac{1}{\,2\hs\pi\,} 
\!-\!\frac{~9\hs\pi\big(c_L^V{c_L^Z}^2\!\!+\!c_R^V{c_R^Z}^2\big)\hsm\big(f_L^2\!-\!f_R^2\big)
\sqrt{s\,}\cos\hsm\phi_{a}^{}~}
{256\big(c_L^V{c_L^Z}^2\!\!-\!c_R^V{c_R^Z}^2\big)\hsm\big(f_L^2\!+\!f_R^2\big)M_Z}
\!-\!\frac{~\cos\hsm 2\phi_a^{}\,}{8\pi } \,,
\\
f^2_{\phi_{a}} &=  \frac{1}{\,2\hs\pi\,} \!-\!\frac{~\cos\hsm 2\phi_a^{}\,}{16\hs\pi} \, .
\label{eq:f2-phia}
\end{align}
\eeqs
From the above formulas, we see that 
the angular function $f^0_{\phi_a}$ approaches the constant term 
$\frac{1}{2\pi}$ for $s\!\gg\! M_Z^2\hs$.\ 
We see in Eq.\eqref{eq:fphi2} or Eq.\,\eqref{eq:f2-phia}
that the value of the angular function $f^2_{\phi_a}$ fluctuates around $\frac{1}{2\pi}$ 
with a magnitude of only $\frac{1}{\,16\hs\pi\,}$, 
which is why in Fig.\,\ref{fig:phia} the angular functions $f^0_{\phi_a}$ and $f^2_{\phi_a}$ 
(shown as the black and blue curves) appear fairly flat and largely overlap with each other.\ 
In contrast, for the case of the angular function $f^1_{\phi_a}$, the coefficient of $\cos\hsm\phi_a$ 
is enhanced by an energy factor $\sqrt{s\,}/M_Z$, and can be much larger than the constant term.

\section{\hspace*{-2.5mm}Probing \boldmath{${ZZV^*}$} Couplings at \boldmath{$e^+e^-$} Colliders}
\label{sec:4}

In this Section we analyze systematically the capabilities of $ZZ$ production in $e^+e^-$ collisions followed by fermionic decays.\ 
We derive the nTGC signal sensitivity using the differential angular distributions of the cross section with respect to 
all angles $\left(\theta,\,\theta_a^{},\,\phi_a^{},\,\theta_b^{},\,\phi_b^{}\right)$ for the case of $Z$ bosons 
decaying to either leptons or quarks in Section\,\ref{sec:4.1}.\ Then, in Section\,\ref{sec:4.2}, we integrate 
over the angles $\theta_b^{}$ and $\phi_b^{}$, which may be regarded as decay angles of $Z$ bosons to neutrinos, 
thereby incorporating the neutrino decay channel of $Z$ bosons.\ After this, we demonstrate in Sections\,\ref{sec:4.3} 
and \ref{sec:4.4} that the signal sensitivity can be significantly enhanced through the implementation of 
machine learning and the use of polarized $e^\mp$ beams.\ 
Finally, in Section \ref{sec:4.5} we compare our results with those obtained in previous studies that considered the $Z\gamma$ final state.

\subsection{\hspace*{-2.5mm}Analyzing \boldmath{$ZZ$} Production via Leptonic/Hadronic Channels}
\label{sec:4.1}

For the present analysis, we define the observables:
\begin{align}\label{OMC}
\mathbb{O}_{j} \,\equiv\, \sigma_{j}^{}\!\int_\Omega^{} \!\!\dif\Omega\,  |f_{\Omega}^{j}| \,,
\end{align}
where the angular distribution function is
%
\begin{align}
f_{\Omega}^{j} &\,=\, \frac{\d^5\hsm\sigma_j}{~\sigma_j^{}\dif\theta\dif\theta_a^{}\!\dif\theta_b^{}\!\dif\phi_a^{}\!\dif\phi_b^{}~} \, ,
\label{fj}
\end{align}
%
and $\dif\Omega\!=\!\dif\theta\dif\theta_a^{}\!\dif\phi_a^{}\!\dif\theta_b^{}\!\dif\phi_b^{}\hs$.\ 
We note that Eq.(\ref{OMC}) integrates the absolute value of the angular distribution function $f_{\Omega}^{j}$ 
over all the phase space, where 
$|f_{\Omega}^{j}|\!=\!\pm f_{\Omega}^{j}$ 
for $f_{\Omega}^{j}\hsm\gtrless\hsm 0\hs$.\ 
In this way, the angular integration over $|f_{\Omega}^{j}|$ is enhanced
without any cancellations between the $f_{\Omega}^{j}\hsm>\hsm 0$ regions
and $f_{\Omega}^{j}\hsm<\hsm 0$ regions.\

Then, we compute the signal significance, 
\begin{align}
\label{eq:significance-Z}
\mathcal{Z}(f_a\bar{f}_af_b^{}\bar{f}_b^{}) =
\frac{|S_I^{}|}{~\Delta_{B_I}^{}\,}
=\frac{~|\mathbb{O}_1(f_a\bar{f}_af_b^{}\bar{f}_b^{})|~}
{\,\sqrt{\hsm\sigma_0^{}(f_a\bar{f}_af_b^{}\bar{f}_b^{})\,}\,}\!\times\!\sqrt{\mathcal{L}\!\times\!\epsilon\,} \, ,
\end{align}
where $\mathcal{L}$ is the integrated luminosity and $\epsilon$ is the detection efficiency.\ 
When evaluating the signal significance by using the above formula \eqref{eq:significance-Z}, 
we include only the final-state leptons and quarks, since neutrinos cannot be detected directly.\ 
We note that reducible backgrounds contribute to the production process $e^+e^-\hsm\ito 4f\hs$.\ 
Most of these can be effectively eliminated by applying the $Z$ on-shell condition 
$|M_{ff}^{}\!-\!M_Z^{}|\!<\!10$\,GeV.\ 
Since the final-state fermions are nearly massless, there are certain soft and collinear divergences 
arising from these reducible backgrounds.\ The implementation of kinematic cuts on the separation angle 
$\Delta R(f\bar{f})$ and transverse momentum $P_T^{}(f)$ can significantly suppress the contributions
of these reducible backgrounds.\  Our MadGraph\,\cite{Alwall:2014hca} simulations demonstrate that 
the reducible backgrounds can be reduced to below 5\% of the total backgrounds for processes excluding 
$e^-e^+$ final states; and to $O(10\%)$ for processes containing $e^-e^+$ final states.\ 
The optimal cuts are process-dependent, varying with both the final-state configurations 
and collision energies\,\cite{XeeZZ}.\ Notably, the relative contribution of the reducible backgrounds 
become smaller as the $e^+e^-$ collision energy rises.\  
We further note that the SM backgrounds $\sigma_0^{}(f_a\bar{f}_af_b^{}\bar{f}_b^{})$ contribute 
to the significance \eqref{eq:significance-Z} through the square-root of ${\sigma_0^{}\,}$.\ 
Hence, for the present analysis, the effects of reducible backgrounds on the significance \eqref{eq:significance-Z}
are rather minor and mainly negligible.\  
For this analysis we can focus on the signal extraction from irreducible backgrounds.\ 
A dedicated backgrounds simulation including both the irreducible and reducible backgrounds  
will be beneficial for future detector-level analyses.

The differential angular distribution $f^j_\theta$  with respect to the angle $\theta$ is presented 
in Fig.\,\ref{fig:ftheta}, and we observe that the SM background is predominantly concentrated 
within the forward and backward regions.\ 
Furthermore, as the scattering energy increases, the relative contributions of these SM backgrounds within the forward and backward regions increase.\ Therefore, imposing a cut on the scattering angle $\theta$ (such that $\sin\theta\!>\!\sin\hsm\delta\hs$) 
serves to suppress the SM backgrounds, and thus improves the significance of signals.

\begin{table}[t]
\begin{center}
\begin{tabular}{c|c||c|c|c|c}
	\hline\hline
	&&&&&
	\\[-4mm]
	$\sqrt{s\,}$\,(TeV)  &  $\mathcal{L}$\,({ab}$^{-1})$  &  
	$|f_5^\gamma|_{2\sigma}^{}$ & $|f_5^\gamma|_{5\sigma}^{}$ & $|f_5^Z|_{2\sigma}^{}$ & $|f_5^Z|_{5\sigma}^{}$
	\\[-4mm]
	&&&&&	
\\
\hline\hline 
&&&&&\\[-4.2mm]
	0.25 & 20 & $2.8\!\times\!10^{-4}$ & $7.0\!\times\!10^{-4}$ & $3.5\!\times\!10^{-4}$ & $8.9\!\times\!10^{-4}$
	\\
	\hline
	&&&&&\\[-4.2mm]
	0.25 & 5 & $5.6\!\times\!10^{-4}$ & $1.4\!\times\!10^{-3}$ & $7.1\!\times\!10^{-4}$ & $1.8\!\times\!10^{-3}$
	\\
	\hline
	&&&&&\\[-4.2mm]
	0.5 & 5 & $5.5\!\times\!10^{-5}$ & $1.4\!\times\!10^{-4}$ & $1.2\!\times\!10^{-4}$ & $2.9\!\times\!10^{-4}$
	\\
	\hline
	&&&&&\\[-4.2mm]
	1 & 5 & $1.1\!\times\!10^{-5}$ & $2.9\!\times\!10^{-5}$ & $2.9\!\times\!10^{-5}$ & $7.3\!\times\!10^{-5}$
	\\
	\hline
	&&&&&\\[-4.2mm]
	3 & 5 & $1.2\!\times\!10^{-6}$ & $3.0\!\times\!10^{-6}$ & $3.4\!\times\!10^{-6}$ & $8.4\!\times\!10^{-6}$
	\\
	\hline
	&&&&&\\[-4.2mm]
	5 & 5 & $4.3\!\times\!10^{-7}$ & $1.1\!\times\!10^{-6}$ & $1.2\!\times\!10^{-6}$ & $3.1\!\times\!10^{-6}$
	\\
	\hline\hline
\end{tabular}
\end{center}
\vspace*{-5mm} 
\caption{\hspace*{-1mm}\small 
{Sensitivity reaches for the nTGC form factors $f_5^\gamma$ and $f_5^Z$ at the $2\sigma$ and $5\sigma$ levels, 
	as derived by analyzing the reaction $e^-e^+\!\ito f_a\bar{f}_af_b^{}\bar{f}_b^{}$ 
	(with $f\!=\!\ell,q$), for a representative integrated luminosity} $\mathcal{L}\!=\!5\,\text{ab}^{-1}$ at selected collision energies
{and an ideal detection efficiency $\hs\epsilon\!=\!100\%\hs$.}}
\label{tab:3}
\end{table}

With the angular cut $\delta\!=\!(0,0.36,0.33,0.34,0.34)$ for $\sqrt{s\,}\!=\!(0.25,0.5,1,3,5)$TeV, we derive
the results in Table\,\ref{tab:3} for the sensitivities to the nTGC form factors $(f_5^\gamma,f_5^Z)$ and
the results in Table\,\ref{tab:4} for the sensitivities to the new physics cutoff scales $\cut_j^{}$ of the dimension-8 nTGC operators.\ 
In the present analysis, for ease of comparison we choose 
an integrated luminosity $\mathcal{L}\!=\!5\,\text{ab}^{-1}\!$ 
as the universal benchmark\,\cite{CEPC-CDR}-\cite{ILC-CLIC}
for all collider energies.\
In addition, for the $e^-e^+$ collision energy of 250\,GeV, we choose
another benchmark of integrated luminosity 
$\mathcal{L}\!=\!20\,\text{ab}^{-1}\!$ 
based on the CEPC TDR\,\cite{CEPC-TDR}.\ 
For any other choice of integrated luminosity, one can obtain the corresponding
sensitivities by rescaling.\ 

\vs 

For reference, the FCC-ee Feasibility Study Report\,\cite{FCC} quotes a
baseline integrated luminosity 10.8\,ab$^{-1}$ 
for the 240\,GeV collision energy and another 3.12\,ab$^{-1}$ 
for the $(340\!-\!350){\hs}$GeV \& 365{\hs}GeV collision energies.\ 
The proposed project of LCF (Linear Collider Facility)\,\cite{LCF-CERN} 
at CERN sets an envisioned integrated luminosity of 
$3\,\rm{ab}^{-1}\!$ for the 250\,GeV collision energy and 
$8\,\rm{ab}^{-1}\!$ for the 550\,GeV collision energy.\ 
The proposed CLIC project\,\cite{CLIC}\cite{CLIC-Phys} at CERN plans 
to collect integrated luminosities of 4.3\,ab$^{-1}$ 
for the 380\,GeV (and 350\,GeV) collision energies  
and of 4.0\,ab$^{-1}$ for the 1.5\,TeV collision energy.\  
The proposed ILC project\,\cite{ILC-CLIC} is expected to deliver 
an integrated luminosities of 2\,ab$^{-1}$ at the 250\,GeV collision energy
and 4\,ab$^{-1}$ at the 500\,GeV collision energy, with additional
200\,fb$^{-1}$ at the 350\,GeV collision energy.

\begin{table}[t]
\begin{center}
\begin{tabular}{c|c||c|c|c|c}
\hline\hline
& & & & &
\\[-4mm]
	$\sqrt{s\,}$\,(TeV) & $\mathcal{L}$\,(ab$^{-1})$ & $\Lambda_{\tilde{B}W}^{2\sigma}$ & $\Lambda_{\tilde{B}W}^{5\sigma}$ & $\Lambda_{3Z}^{2\sigma}$ & $\Lambda_{3Z}^{5\sigma}$
\\[-4mm]
& & & & &
\\
\hline\hline 
&&&&&\\[-4.2mm]
	0.25 & 20 & 1.2 & .96 & .96 & .76\\
	\hline
	0.25 & 5 & 1.0 & .81 & .81 & .64\\
	\hline
	0.5 & 5 & 1.8 & 1.4 & 1.3 & 1.0\\
	\hline
	1 & 5 & 2.7 & 2.1 & 1.8 & 1.4\\
	\hline
	3 & 5 & 4.7 & 3.8 & 3.1 & 2.4\\
	\hline
	5 & 5 & 6.1 & 4.9 & 4.0 & 3.1\\
	\hline\hline
\end{tabular}
\end{center}
\vspace*{-5mm}
\caption{\hspace*{-0.5mm}\small 
{Sensitivity reaches on the new physics scales 
$\Lambda_{\tilde{B}W}$ and $\Lambda_{3Z}$} (in TeV) 
{via the reaction $e^-e^+\ito f_a\bar{f}_af_b^{}\bar{f}_b^{}$\,, 
	$(\text{with}\,f\!=\!\ell,q)$.\ The sensitivity reaches are shown at the $2\sigma$ and $5\sigma$ levels and 
	for different collider energies.\ For illustration, we assume a representative integrated luminosity}  
$\mathcal{L}\!=\!5\,\text{ab}^{-1}\!$ {and an ideal detection efficiency $\epsilon\!=\!100\%\,$.}}
\label{tab:4} 
\end{table}

The sensitivity bounds on the new physics scale $\Lambda_{\!\tilde{B}W}^{}$ are consistently stronger than those on $\Lambda_{3Z}^{}$, 
and this advantage grows with increasing scattering energy.\
Limits on the new physics scale $\Lambda$ generally strengthen as the scattering energy increases, primarily because the differences between the differential angular distributions of the interference signal $f^1_\Omega$ and the SM background $f^0_\Omega$ become more pronounced 
at higher scattering energies.\

In passing, we note that for the present analysis of $ZZ$ production 
at the $e^+e^-$ colliders, the dominant contribution comes from 
the interference term between the SM and the nTGC contributions, 
whereas the contribution of the quadratic nTGC term is negligibly small.\ 
For instance, we find that when the contribution 
of the interference term reaches a significance  
$\mathcal Z_{\rm{int}}^{}\!\!=\!2$, the significance contributed 
by the quadratic term is only  
$\mathcal Z_{\rm{qua}}^{}(f_5^\gamma)\!=\!(0.006, 0.009, 0.016, 0.047)$ 
and 
$\mathcal Z_{\rm{qua}}^{}(f_5^Z)\!=\!(0.003, 0.014, 0.036, 0.13)$ 
for $\sqrt{s} \!=\! (0.25, 0.5, 1, 3)$\,TeV,
showing that the contribution of the interference term always dominates 
the sensitivity.

\subsection{\hspace*{-2.5mm}Analyzing the Invisible Decay Channel \boldmath{$Z\hsm\ito\nu\bar{\nu}$}}
\label{sec:4.2}

\begin{table}[t]
	\begin{center}
		\begin{tabular}{c|c||r|r|r}
			\hline\hline
			&&&&
			\\[-3.7mm]
			$\sqrt{s\,}$\,(TeV)  &  $\mathcal{L}\,(\text{ab}^{-1})$  
			&  $|f_5^\gamma|_{\!f\bar{f}}^{}$\hspace*{9mm}  & $|f_5^\gamma|_{\nu\bar{\nu}}^{}$\hspace*{9mm}  
			&  $|f_5^\gamma|_{\text{tot}}^{}$\hspace*{9mm}  
\\[1mm]
\hline\hline
			&&&&\\[-4mm]
			0.25 & 20 & $(2.8,\,7.0)\!\times\!10^{-4}$ & $(.63,\,1.6)\!\times\!10^{-3}$ & $(2.6,\,6.4)\!\times\!10^{-4}$
			\\
			\hline
			&&&&\\[-4.2mm]
			0.25 & 5 & $(.56,\,1.4)\!\times\!10^{-3}$ & $(1.3,\,3.2)\!\times\!10^{-3}$ & $(.51,\,1.3)\!\times\!10^{-3}$\\
			\hline
			&&&&\\[-4.2mm]
			0.5 & 5 & $(.55,\,1.4)\!\times\!10^{-4}$ & $(1.2,\,2.9)\!\times\!10^{-4}$ & $(.50,\,1.2)\!\times\!10^{-4}$\\
			\hline
			&&&&\\[-4.2mm]
			1 & 5 & $(1.1,\,2.9)\!\times\!10^{-5}$ & $(2.4,\,6.0)\!\times\!10^{-5}$ & $(1.0,\,2.6)\!\times\!10^{-5}$\\
			\hline
			&&&&\\[-4.2mm]
			3 & 5 & $(1.2,\,3.0)\!\times\!10^{-6}$ & $(2.5,\,6.3)\!\times\!10^{-6}$ & $(1.1,\,2.7)\!\times\!10^{-6}$\\
			\hline
			&&&&\\[-4.2mm]
			5 & 5 & $(.43,\,1.1)\!\times\!10^{-6}$ & $(.90,\,2.3)\!\times\!10^{-6}$ & $(3.9,\,9.7)\!\times\!10^{-7}$\\
\hline\hline
\end{tabular}
\end{center}
\vspace*{-5mm}
\caption{\hspace*{-1mm}\small {Sensitivity reaches for the nTGC form factor $f_5^\gamma$ at the} ($2\sigma,5\sigma$) 
{level, as derived by analyzing the reactions 
$e^-e^+\!\ito\! f_a\bar{f}_af_b^{}\bar{f}_b^{}$, $e^-e^+\!\ito\! f_a\bar{f}_a\nu\bar{\nu}$ (with $f\!\!=\!\ell,q)$ 
and their combination at selected collision energies, for a representative integrated luminosity} $\hs\mathcal{L}\!=\!5\,\text{ab}^{-1}$ 
{and an ideal detection efficiency $\epsilon\!=\!100\%\hs$.}}
\label{tab:5}
\end{table}

To compute the nTGC signal significance for the reaction $e^-e^+\ito ZZ\ito f\bar{f}\nu\bar{\nu}$ (with $f\!=\!\ell,q$), 
we consider $\theta_b^{}$ and $\phi_b^{}$ to be the angles of the decay of one $Z$ boson into neutrinos and integrate over them.\ 
Then, we compute the signal significance, with $\theta_a^{}$ and $\phi_a^{}$ 
being the angles of the other $Z$ boson decays into leptons or quarks.

For this analysis, we define the following observable:
\begin{align}
{\mO}_{j}^{\rm{n}}\,\equiv\, \sigma_{j}^{}\!
\int_{\Omega^\prime}^{}\!\!\dif\Omega^\prime\hs |f_{\Omega^\prime}^{j}| \, ,
\end{align}
where the superscript ``{\hs}n'' of ${\mO}_{j}^{\rm{n}}$ denotes one of the final state $Z$ bosons decaying into neutrinos, 
and $\dif\Omega^\prime\!=\!\dif\theta\dif\theta_a^{}\!\dif\phi_a^{}$.\ 
In the above, the partially integrated angular distribution function $f_{\Omega^\prime}^j$ is given by
%
\begin{align}
f_{\Omega^\prime}^j &= \int f_{\Omega}^{j}\dif\theta_{b}^{}\dif\phi_{b}^{} \,.
\label{fjn}
\end{align}
%
With these, we compute the nTGC signal significance:
\begin{align}
\mathcal{Z}_{\rm n}^{}(f_a^{}\bar{f}_a^{}) = 
\frac{|S_I|}{~\Delta_{B_I}\,}
=\frac{~|\mathbb{O}_1^{\rm n}(f_a\bar{f}_a)|~}{\sqrt{\sigma_0^{}(f_a\bar{f}_a)\,}\,}\!\times\!\sqrt{\mathcal{L}\!\times\!\epsilon\,} \,.
\end{align}
We then derive the total signal significance for probing the nTGCs:
\begin{align}
\mathcal{Z}_{\text{tot}}^{} \,=\,
\left[\sum_{(a,b)\in(\ell,q)}\hspace*{-5mm}\mathcal{Z}^2(f_a\bar{f}_af_b\bar{f}_b)
\,+\!\!\sum_{a\in(\ell,q)}\hspace*{-2.5mm}\mathcal{Z}_{\rm n}^2(f_a\bar{f}_a)\!\right]^{\!\!\frac{1}{2}} \!,
\end{align}
by using the angular cut $\delta_{\rm n}^{}\!=\!(0.02,0.30,0.31,0.31,0.33)$ for $\mathcal{Z}_{\rm n}^{}$ 
at each $e^+e^-$ collider energy $\sqrt{s\,}\!=\!(0.25,0.5,1,3,5)\hs${TeV}.\ 
We present our findings for the sensitivity reaches on the nTGC form factors $(f_5^\gamma,f_5^Z)$ in Tables\,\ref{tab:5} and \ref{tab:6},
and on the nTGC new physics scales in Table\,\ref{tab:7}.

\begin{table}[t]
\vspace*{3mm}	
\begin{center}
\begin{tabular}{c|c||r|r|r}
	\hline\hline
	&&&&\\[-4mm]
	$\sqrt{s\,}$\,(TeV)  &  $\mathcal{L}$\,({ab}$^{-1})$  
	&  $|f_5^Z|_{\!f\bar{f}}^{}$\hspace*{9mm} & $|f_5^Z|_{\nu\bar{\nu}}^{}$\hspace*{9mm} 
	& $|f_5^Z|_{\text{tot}}^{}$\hspace*{9mm} \\
&&&&\\[-4mm]
\hline\hline
	&&&&\\[-4.2mm]
	0.25 & 20 & $(3.5,\,8.9)\!\times\!10^{-4}$ & $(.57,\,1.4)\!\times\!10^{-3}$ & $(3.0,\,7.5)\!\times\!10^{-4}$
	\\
	\hline
	0.25 & 5 & $(.71,\,1.8)\!\times\!10^{-3}$ & $(1.1,\,2.9)\!\times\!10^{-3}$ & $(.60,\,1.5)\!\times\!10^{-3}$\\
	\hline
	0.5 & 5 & $(1.2,\,2.9)\!\times\!10^{-4}$ & $(2.3,\,5.9)\!\times\!10^{-4}$ & $(1.0,\,2.6)\!\times\!10^{-4}$\\
	\hline
	1 & 5 & $(2.9,\,7.3)\!\times\!10^{-5}$ & $(.67,\,1.7)\!\times\!10^{-4}$ & $(2.7,\,6.7)\!\times\!10^{-5}$\\
	\hline
	3 & 5 & $(3.4,\,8.4)\!\times\!10^{-6}$ & $(.86,\,2.2)\!\times\!10^{-5}$ & $(3.1,\,7.8)\!\times\!10^{-6}$\\
	\hline
	5 & 5 & $(1.2,\,3.1)\!\times\!10^{-6}$ & $(3.2,\,8.0)\!\times\!10^{-6}$ & $(1.1,\,2.9)\!\times\!10^{-6}$\\
	\hline\hline
\end{tabular}
\end{center}
\vspace*{-5mm}
\caption{\hspace*{-1.5mm}\small {Sensitivity reaches for the nTGC form factor $f_5^Z$ at the} 
($2\sigma,5\sigma$) {level, as derived by analyzing the reactions 
	$e^-e^+\!\ito f_a\bar{f}_af_b\bar{f}_b$, $e^-e^+\!\ito f_a\bar{f}_a\nu\bar{\nu}$ (with $f\!=\!\ell,q)$ and their combination at selected collision energies, 
	for a representative integrated luminosity} $\hs\mathcal{L}\!=\!5\,\text{ab}^{-1}$ 
{and an ideal detection efficiency $\hs\epsilon\!=\!100\%\hs$.}}
\label{tab:6}
\end{table}

It can be seen that the limits on $\Lambda^{\nu\bar{\nu}}$ are less stringent than those on $\Lambda^{f\bar{f}}$. This is due to the lower branching ratio of the $Z$ boson decays to neutrinos as compared to its decays to charged fermions (including leptons and quarks):
\begin{align}
\text{Br}[Z\!\ito\nu\bar{\nu}]< \text{Br}[Z\!\ito \ell\bar{\ell},q\bar{q}]\hs,
\end{align}
and the absence of decay angle information.\ 
Consequently, incorporating the contribution from the neutrino decay channel of the $Z$ boson yields only a modest improvement 
in the limit on the new physics scale $\Lambda$, as compared to the results obtained in Section \ref{sec:4.1} 
that include only $Z$ boson decays to charged fermions.

\begin{table}[t]
\vspace*{3mm}
\begin{center}
\tabcolsep 1pt
\begin{tabular}{c|c||c|c|c|c|c|c}
	\hline\hline
	&&&&&&&\\[-4mm]
	~$\sqrt{s\,}$\,(TeV)~ & ~$\mathcal{L}$\,({ab}$^{-1})$~ & $\Lambda_{\tilde{B}W}^{f\bar{f}}$ & $\Lambda_{\tilde{B}W}^{\nu\bar{\nu}}$ & $\Lambda_{\tilde{B}W}^{\text{tot}}$ & $\Lambda_{3Z}^{f\bar{f}}$ & $\Lambda_{3Z}^{\nu\bar{\nu}}$ &$\Lambda_{3Z}^{\text{tot}}$
\\[1.2mm]
\hline\hline
&&&&&&&\\[-4mm]
	0.25 & 20 & (1.2,\,.96) & ~(.99,\,.78)~ & (1.2,\,.98) & ~(.96,\,.76)~ & ~(.85,\,.68)~ & ~(1.0,\,.79)~ \\
	\hline
	&&&&&&&\\[-4.2mm] 
	0.25 & 5 & \,(1.0,\,.81)\, & \,(.83,\,.66)\, & \,(1.0,\,.83)\, & \,(.81,\,.64)\, & \,(.72,\,.57)\, & \,(.84,\,.67)\,\\
	\hline
	&&&&&&&\\[-4.2mm] 
	0.5 & 5 & (1.8,\,1.4) & (1.5,\,1.2) & (1.9,\,1.5) & (1.3,\,1.0) & ~(1.1,\,.85)~ &(1.3,\,1.0)\\
	\hline
	&&&&&&&\\[-4.2mm]
	1 & 5 & (2.7,\,2.1) & (2.2,\,1.8) & (2.8,\,2.2) & (1.8,\,1.4) & (1.5,\,1.2) &(1.8,\,1.5)\\
	\hline
	&&&&&&&\\[-4.2mm]
	3 & 5 & (4.7,\,3.8) & (3.9,\,3.1) & (4.8,\,3.9) & (3.1,\,2.4) & (2.4,\,1.9) &(3.1,\,2.5)\\
	\hline
	&&&&&&&\\[-4.2mm]
	5 & 5 & (6.1,\,4.9) & (5.1,\,4.0) & (6.3,\,5.0) & (4.0,\,3.1) & (3.1,\,2.5) &(4.0,\,3.2)\\
	\hline\hline
\end{tabular}
\end{center}
\vspace*{-4mm}
\caption{\hspace*{-1mm}\small {Sensitivity reaches on the new physics scales $\Lambda_{\tilde{B}W}^{}$ and $\Lambda_{3Z}^{}$} (in TeV) 
{through the reactions $e^-e^+\!\ito f_a\bar{f}_af_b^{}\bar{f}_b^{}$ and  $e^-e^+\!\ito f_a\bar{f}_a\nu\bar{\nu}$ 
(with $f\!=\!\ell,q)$ plus their combination.\ 
The bounds are shown at the} $(2\sigma,5\sigma)$ {level and 
for different $e^+e^-$ collider energies.\ 
For illustration, we choose a representative integrated luminosity} 
$\hs\mathcal{L}\!=\!5\,\text{ab}^{-1}$ {and an ideal detection efficiency $\hs\epsilon\!=\!100\%\hs$.}}
\label{tab:7}
\end{table}


\vspace*{1.5mm}
\subsection{\hspace*{-2.5mm}Probability Cuts from Machine Learning}
\label{sec:4.3}
\vspace*{0.5mm}

In this Section we explore the use of machine learning (ML) 
to improve the signal significance.\ 
ML has become a transformative tool in collider physics, addressing challenges in data analysis, pattern recognition, 
and theoretical modeling\,\cite{Plehn:2022ftl}.\ 
ML algorithms (such as boosted decision trees, deep neural networks,
nearest neighbor analysis, and support vector machines)
are widely used to distinguish between rare particle collision signals 
and large background noise sources.\ 
These methods optimize event classification by learning complex correlations 
in high-dimensional datasets.\ 

\vs 

For this analysis, we employ the Mathematica's built-in function 
\texttt{Classify} to perform machine learning for distinguishing between 
the nTGC signals and the SM background events.\ 
The \texttt{Classify} function incorporates many machine learning algorithms
and automatically identifies the optimal algorithm 
for the specific classification task 
to deliver the optimized performance.\ 
This implementation operates autonomously without requiring manual 
hyperparameter tuning.\

\vs 

The core methodology is to use large samples of (labeled) simulated scattering events (explicitly tagged as belonging to the 
signal or background categories), 
which we use to train a classification algorithm 
to discover automatically optimal criteria for
distinguishing between the signal and backgrounds 
on the basis of measurements of physical quantities such as energies, momenta, angles, 
and their combinations within the events.\ After training, the classification algorithm functions as a classifier, 
which computes the probability that any experimental event belongs to the signal or background category.

\begin{figure}[t]
\centering
\includegraphics[width=7.7cm,height=5.8cm]{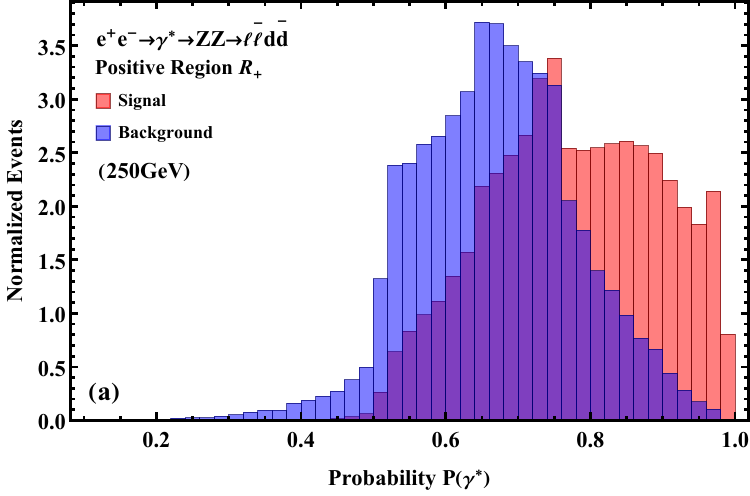}
\includegraphics[width=7.7cm,height=5.8cm]{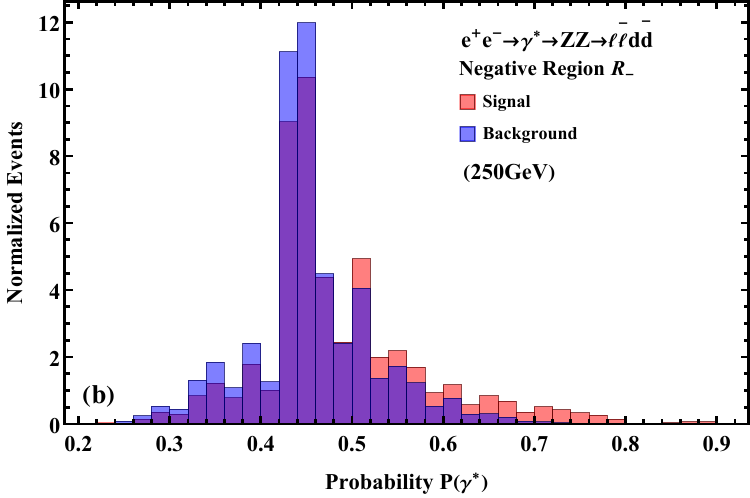}
\\[2mm]
\hspace*{2mm}
\includegraphics[width=7.7cm,height=5.8cm]{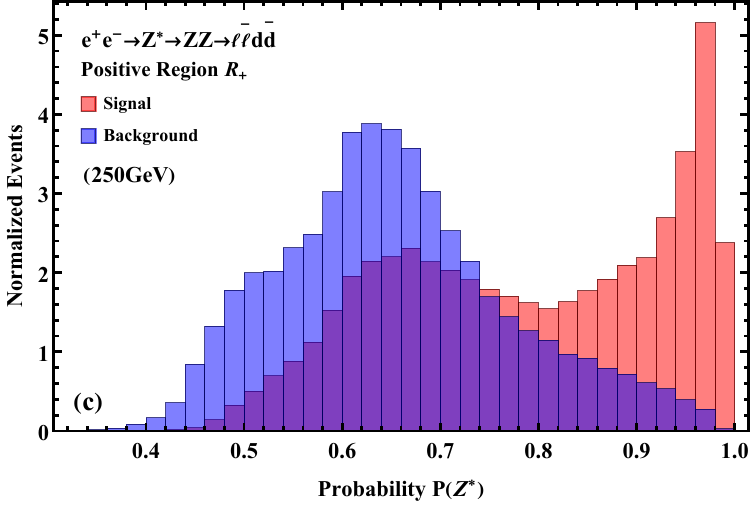}
\includegraphics[width=7.7cm,height=5.8cm]{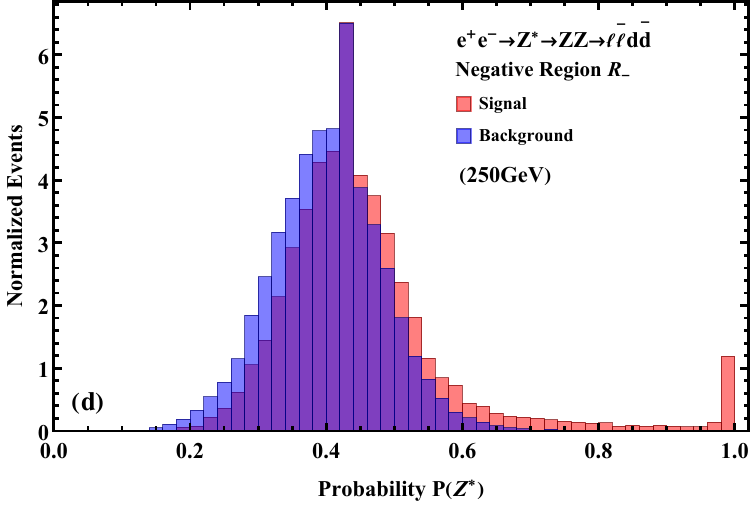}
\vspace*{-2.5mm}  
\caption{\hspace*{-1mm}\small 
Probability distributions of events being signals and backgrounds respectively,
for the case of $e^+e^-\!\ito V^*\!\ito ZZ\!\ito (\ell\bar{\ell})(d\bar{d})\,$ 
at a given collider energy  $\sqrt{s\,}\!=\!250${\hs}GeV
[marked as (250{\hs}GeV) in each plot].\ 
Here we choose the particular hadronic decay channel $Z\ito d\bar{d}$ [with $d$ denoting the down-type quarks ($d,s,b$)] for illustration.\
Plots\,(a) and (b) present the probability distributions of events
for the angular regions $R_+$ and $R_-$ with the $\gamma^*$ contributions, respectively;
plots\,(c) and (d) show the probability distributions of events
for the angular regions $R_+$ and $R_-$ with the $Z^*$ contributions, respectively.  
}
\label{probabilitydistribution}
\label{fig:7}
\vspace*{4mm}
\end{figure}
\begin{figure}[h]
\centering
\includegraphics[width=7.7cm,height=5.8cm]{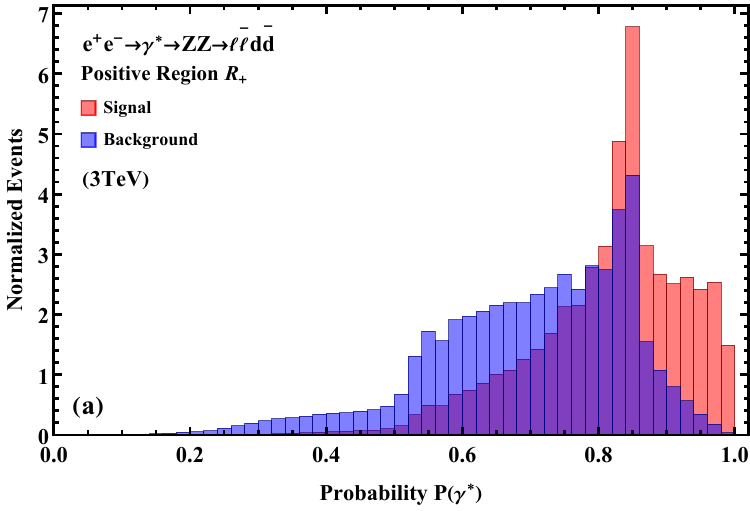}
\includegraphics[width=7.7cm,height=5.8cm]{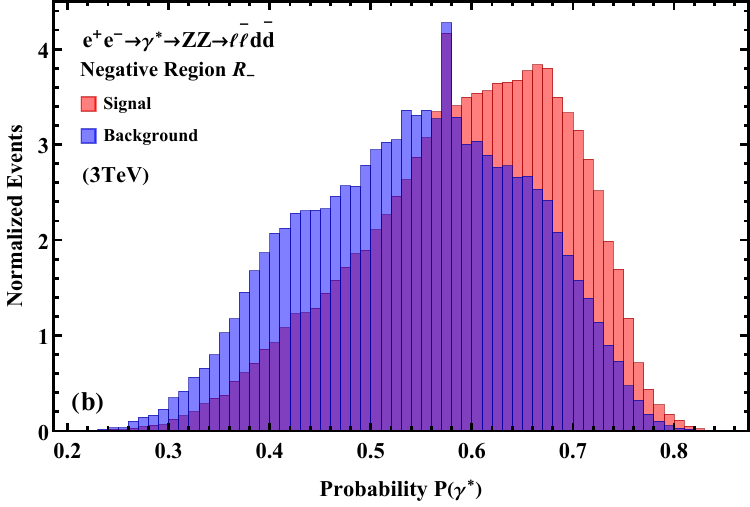}
\\[2mm]
\hspace*{2mm}
\includegraphics[width=7.7cm,height=5.8cm]{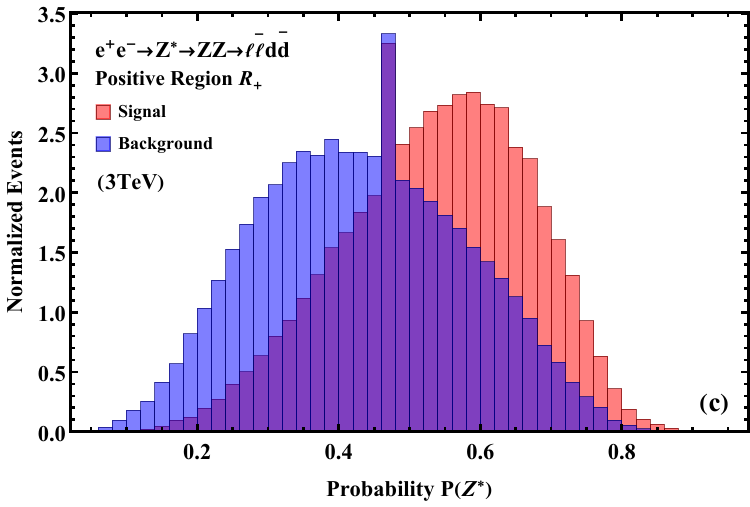}
\includegraphics[width=7.7cm,height=5.8cm]{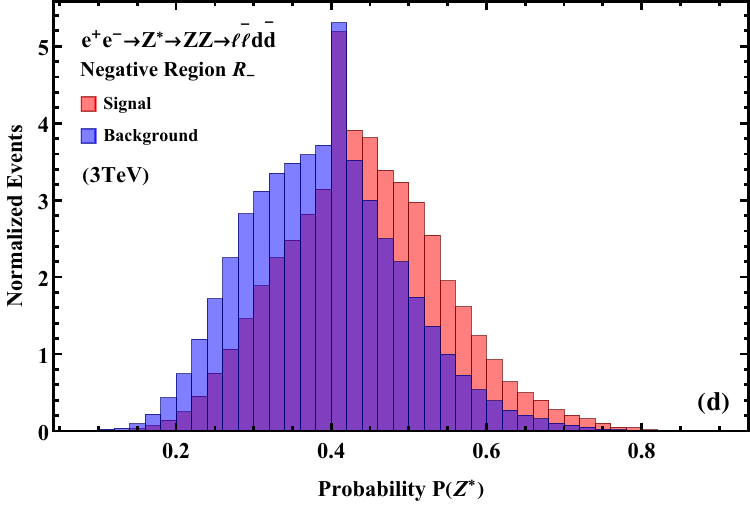}
\vspace*{-2.5mm}  
\caption{\hspace*{-1mm}\small 
Probability distributions of events being signals and backgrounds respectively,
for the case of $e^+e^-\!\hsm\ito V^*\!\hsm\ito\! ZZ\!\ito (\ell\bar{\ell})(d\bar{d})\,$ at a given collider energy  
$\sqrt{s\,}\!=\!3$\,TeV [marked as (3{\hs}TeV) in each plot].\ 
Here we choose the particular hadronic decay channel 
$Z\ito d\bar{d}$ [with $d$ denoting the down-type quarks ($d,s,b$)] for illustration.\  
Plots\,(a) and (b) present the probability distributions of events
for the angular regions $R_+$ and $R_-$ with the $\gamma^*$ contributions, respectively;
plots\,(c) and (d) show the probability distributions of events
for the angular regions $R_+$ and $R_-$ with the $Z^*$ contributions, respectively.
}
\label{fig:8new}
\vspace*{1mm}
\end{figure}

We proceed by dividing the phase space of the final states into a positive region $R_+$ 
and a negative region $R_-$ based on the signs of the angular distribution $f^1_\Omega$ in Eq.\eqref{fj} 
or $f^1_{\Omega^\prime}$ in Eq.\eqref{fjn}:
\begin{subequations}
\begin{align}
R_+^{}\!:&\quad f^1_\Omega\!>\!0\hs, \,~\text{or,}~\, f^1_{\Omega^\prime}\!>\!0 \,,
\\
R_-^{}\!:&\quad f^1_\Omega\!<\!0\hs, ~\,\text{or,}\,~ f^1_{\Omega^\prime}\!<\!0 \,.
\end{align}
\end{subequations}
Subsequently, we categorize the signal $S$ and background $B$ into positive dataset ($S_+,B_+$) 
and negative dataset ($S_-,B_-$) according to $R_+$ and $R_-$.

In this theoretical study, we generate events in {\tt Mathematica} and then leverage its built-in ML function {\tt Classify} 
with variables ($\theta$, $\theta_a$, $\theta_b^{}$, $\phi_a$, $\phi_b^{}$) to distinguish the nTGC signals from SM backgrounds.\ 
For the positive portion of the signal and background dataset ($S_+$ and $B_+$), we train a classifier $C^+_{\text{sig-bkg}}$ 
capable of distinguishing between signals and backgrounds, and providing the probability that each event belongs 
to either category.\ We apply the same approach to the negative dataset using a classifier $C^-_{\text{sig-bkg}}$.\ 
For instance, we illustrate in Fig.\,\ref{probabilitydistribution} and Fig.\,\ref{fig:8new} 
the probability distributions 
for the collider energies $\sqrt{s\,}\!=\!250$\,GeV and $\sqrt{s\,}\!=\!3$\,TeV, respectively.\
Comparing Fig.\,\ref{fig:7} with Fig.\,\ref{fig:8new}, we see that probability distributions significantly change 
as the collider energy increases from $\sqrt{s\,}\!=\!250$\,GeV to $\sqrt{s\,}\!=\!3$\,TeV.\ 
Furthermore, it is shown that the higher collider energy can help to raise the probability of discriminating 
the signals from backgrounds.\

%

In order to achieve higher sensitivity, we discriminate between signals and backgrounds 
by using the event probability distributions that are computed via ML.\ 
We divide events into bins of the probability distribution, whose widths are chosen as 
$\Delta P\!=\!0.1\hs$.\ 
We compute the significance $\mathcal{Z}_{\rm{bin}}^{}$ for each bin, 
and construct the following significance measure:
\begin{align}
\mathcal{Z}=\sqrt{\,\sum\mathcal{Z}_{\text{bin}}^2~} ~ .
\label{MLbin}
\end{align}
Finally, we compute the total signal significance $\mathcal{Z}_{\text{tot}}^{}$  
by combining the significances from both the positive region $\mathcal{Z}_+^{}$ and negative region $\mathcal{Z}_-^{}$:
\begin{align}
\mathcal{Z}_{\text{tot}}^{}=\sqrt{\mathcal{Z}_+^2+\mathcal{Z}_-^2~} ~.
\end{align}
Employing this approach, we have derived the sensitivity reaches on the nTGC form factors in Table\,\ref{ff} 
and on the nTGC new physics cutoff scales in Table\,\ref{scale}.

\begin{table}[t]
\begin{center}
\begin{tabular}{c|c||c|c|c|c}
	\hline\hline
	&&&&&\\[-4mm]
	$\sqrt{s\,}$\,(TeV)  &  $\mathcal{L}$\,({ab}$^{-1})$  &  $|f_5^\gamma|_{2\sigma}$ & $|f_5^\gamma|_{5\sigma}$ & $|f_5^Z|_{2\sigma}$ & $|f_5^Z|_{5\sigma}$
	\\[0.5mm]
	\hline\hline
	&&&&&\\[-4mm]
	0.25 & 20 & $2.0\!\times\!10^{-4}$ & $4.9\!\times\!10^{-4}$ & $2.0\!\times\!10^{-4}$ & $4.9\!\times\!10^{-4}$
	\\
	\hline
	&&&&&\\[-4mm]
	0.25 & 5 & $3.9\!\times\!10^{-4}$ & $9.8\!\times\!10^{-4}$ & $3.9\!\times\!10^{-4}$ & $9.8\!\times\!10^{-4}$\\
	\hline
	&&&&&\\[-4mm]
	0.5 & 5 & $4.0\!\times\!10^{-5}$ & $1.0\!\times\!10^{-4}$ & $8.2\!\times\!10^{-5}$ & $2.1\!\times\!10^{-4}$\\
	\hline
	&&&&&\\[-4mm]
	1 & 5 & $8.3\!\times\!10^{-6}$ & $2.1\!\times\!10^{-5}$ & $2.1\!\times\!10^{-5}$ & $5.3\!\times\!10^{-5}$\\
	\hline
	&&&&&\\[-4mm]
	3 & 5 & $8.4\!\times\!10^{-7}$ & $2.1\!\times\!10^{-6}$ & $2.5\!\times\!10^{-6}$ & $6.2\!\times\!10^{-6}$\\
	\hline
	&&&&&\\[-4mm]
	5 & 5 & $2.9\!\times\!10^{-7}$ & $7.3\!\times\!10^{-7}$ & $8.6\!\times\!10^{-7}$ & $2.2\!\times\!10^{-6}$\\
	\hline\hline
\end{tabular}
\end{center}
\vspace*{-4.5mm}
\caption{\hspace*{-0.5mm}\small 
{Sensitivities to the nTGC form factors $f_5^\gamma$ and $f_5^Z$ at the $2\sigma$ and $5\sigma$ level after using ML, as derived by analyzing the combination of $e^-e^+\!\ito f_a\bar{f}_af_b\bar{f}_b$ and $e^-e^+\!\ito f_a\bar{f}_a\nu\bar{\nu}$
	(with $f\!=\hsm\ell,q)$, with an ideal detection efficiency $\hs\epsilon\hsm =\!100\%\hs$.}}
\label{ff}
\label{tab:8}
\vspace*{3mm}
\end{table}
\begin{table}[]
\begin{center}
\begin{tabular}{c|c||c|c|c|c}
	\hline\hline
	&&&&&\\[-4mm]
	$\sqrt{s\,}$\,(TeV) & $\mathcal{L}$\,({ab}$^{-1})$ & $\Lambda_{\tilde{B}W}^{\text{tot},2\sigma}$ & $\Lambda_{\tilde{B}W}^{\text{tot},5\sigma}$ & $\Lambda_{3Z}^{\text{tot},2\sigma}$ & $\Lambda_{3Z}^{\text{tot},5\sigma}$
	\\[1mm]
	\hline\hline
	&&&&&\\[-4mm]
	0.25 & 20 & 1.3 & 1.1 & 1.1 & .88\\
	\hline
	&&&&&\\[-4mm]
	0.25 & 5 & 1.1 & .88 & .93 & .74\\
	\hline
	&&&&&\\[-4mm]
	0.5 & 5 & 2.0 & 1.6 & 1.4 & 1.1\\
	\hline
	&&&&&\\[-4mm]
	1 & 5 & 2.9 & 2.3 & 1.9 & 1.5\\
	\hline
	&&&&&\\[-4mm]
	3 & 5 & 5.2 & 4.1 & 3.3 & 2.6\\
	\hline
	&&&&&\\[-4mm]
	5 & 5 & 6.7 & 5.4 & 4.3 & 3.4\\
	\hline\hline
\end{tabular}
\end{center}
\vspace*{-4.5mm}
\caption{\hspace*{-1mm}\small 
{Sensitivity reaches for the new physics scales $\Lambda_{\tilde{B}W}$ and $\Lambda_{3Z}$} (in TeV) 
{via the combination of $e^-e^+\!\ito f_a\bar{f}_af_b\bar{f}_b$ and $e^-e^+\!\ito f_a\bar{f}_a\nu\bar{\nu}$
	(with $f\!=\!\ell,q)$ after using ML.\ The ranges are shown at the $2\sigma$ and $5\sigma$ level 
	and for different energies.\ 
	For illustration, we choose a representative integrated luminosity} $\hs\mathcal{L}\!=\hsm 5\,\text{ab}^{-1}$ 
{and an ideal detection efficiency $\hs\epsilon\hsm =\!100\%\hs$.}}
\label{scale}
\label{tab:9}
\vspace*{3mm}
\end{table}

We find that after applying machine learning to distinguish nTGC signals and SM backgrounds, 
the phenomenological constraints on the new physics cutoff scale $\Lambda$ are significantly enhanced as compared to the results 
prior to using machine learning.\ 
Specifically, in comparison with the results presented in Table\,\ref{tab:5} and Table\,\ref{tab:6} of Section \ref{sec:4.2}, 
machine learning enhances the constraints on the form factor $f^\gamma_5$ across different collider energies 
by 19\% to 25\%.\ The smallest enhancement of $19\%$ is realized at $\sqrt{s\,}\hsm\!=\hsm\!500\, \text{GeV}$, 
whereas the largest enhancement of 25\% occurs at $\sqrt{s\,}\hsm\!=\hsm\!5\hs \, \text{TeV}$.\  
Similarly, machine learning improves the constraints on the form factor $f^Z_5$ by about $(20\hsmx -\hsmx 35)\%$.\ 
Here the smallest enhancement of 20\% is found at the collider energy $\sqrt{s\,}\hsm\!=\hsm\!1\hs \, \text{TeV}$.\ 
Conversely, the strongest enhancement of 35\% is achieved at the collider energy $\sqrt{s\,}\hsm\!=\hsm\!250\, \text{GeV}$.

Machine learning also improves the sensitivity reaches  
for the new physics scale $\Lambda_{\tilde{B}W}^{}$ across various collider energies by about $(5.6\hsm -\hsm 7.5)\%\hs$.\ 
The smallest improvement of 5.6\% is realized at $\sqrt{s\,}\hsm\!=\hsm\!500\hs$GeV, 
whereas the largest improvement of 7.5\% is achieved at $\sqrt{s\,}\hsm\!=\hsm\!5\hs\text{TeV}$.\  
Furthermore, machine learning improves the sensitivity reaches for $\Lambda_{3Z}^{}$ by about $(5.7 - 11)\%$ 
across different collider energies.\ Here the smallest improvement of 5.7\% occurs at $\sqrt{s\,}\hsm\!=\hsm\!1\hs\text{TeV}$, 
and the largest improvement of 11\% is found at $\sqrt{s\,}\hsm\!=\hsm\!250\hs\text{GeV}$.

\subsection{\hspace*{-2.5mm}Improvements from Polarized Electron and Positron Beams}
\label{sec:4.4}
\vspace*{1.5mm}

\begin{table}[t]
\begin{center}
\tabcolsep 1pt
\begin{tabular}{c|c|r|r|r|r}
	\hline\hline
	&&&&&\\[-4mm]
	$~\sqrt{s\,}$~  &  $\mathcal{L}$  &  $|f_5^\gamma|_{2\sigma}^{}$\hspace*{9mm} 
	& $|f_5^\gamma|_{5\sigma}^{}$\hspace*{9mm}  & $|f_5^Z|_{2\sigma}^{}$\hspace*{9mm}  & $|f_5^Z|_{5\sigma}^{}$\hspace*{9mm} 
	\\
	~~(\text{TeV})~~ & ~($\text{ab}^{-1}$)~ & (\text{unpol},\,\text{pol})\hspace*{5mm}  
	& (\text{unpol},\,\text{pol})\hspace*{5mm}  & (\text{unpol},\,\text{pol})\hspace*{5mm}  & (\text{unpol},\,\text{pol})\hspace*{5mm} 
\\[0.5mm]
\hline\hline
&&&&&\\[-4mm]
0.25 & 20 & $(2.0,.96)\!\times\!10^{-4}$~ & $(4.9,2.4)\!\times\!10^{-4}$~ & ~$(2.0,.90)\!\times\!10^{-4}$~ & $(4.9,2.3)\!\times\!10^{-4}$~
\\
\hline
&&&&&\\[-4.2mm]
0.25 & 5 & ~$(3.9,1.9)\!\times\!10^{-4}$~ & $(9.8,4.8)\!\times\!10^{-4}$~ & $(3.9,1.8)\!\times\!10^{-4}$~ & $(9.8,4.5)\!\times\!10^{-4}$~
	\\
	\hline
	&&&&&\\[-4.2mm]
	0.5 & 5 & $(4.0,3.2)\!\times\!10^{-5}$~ & $(1.0,.81)\!\times\!10^{-4}$~ & $(8.2,5.2)\!\times\!10^{-5}$~ & ~$(2.1,1.3)\!\times\!10^{-4}$~
	\\
	\hline
	&&&&&\\[-4.2mm]
	1 & 5 & $(8.3,7.0)\!\times\!10^{-6}$~ & $(2.1,1.7)\!\times\!10^{-5}$~ & $(2.1,1.2)\!\times\!10^{-5}$~ & $(5.3,3.0)\!\times\!10^{-5}$~
	\\
	\hline
	&&&&&\\[-4.2mm]
	3 & 5 & ~$(8.4,7.6)\!\times\!10^{-7}$~ & ~$(2.1,1.9)\!\times\!10^{-6}$~ & ~$(2.5,1.3)\!\times\!10^{-6}$~ & ~$(6.2,3.2)\!\times\!10^{-6}$~
	\\
	\hline
	&&&&&\\[-4.2mm]
	5 & 5 & ~$(2.9,2.6)\!\times\!10^{-7}$~ & ~$(7.3,6.4)\!\times\!10^{-7}$~ & ~$(8.6,4.2)\!\times\!10^{-7}$~ & ~$(2.2,1.1)\!\times\!10^{-6}$~
	\\
	\hline\hline
\end{tabular}
\end{center}
\vspace*{-5mm}
\caption{\hspace*{-1mm}\small 
{Sensitivity reaches for the nTGC form factors $f_5^\gamma$ and $f_5^Z$ at the $2\sigma$ and $5\sigma$ level, 
as derived by analyzing the combination of $e^-e^+\!\ito f_a\bar{f}_af_b\bar{f}_b$ and $e^-e^+\!\ito\hsm f_a\bar{f}_a\nu\bar{\nu}$
(with $f\!=\!\ell,q)$, for (unpolarized,\,polarized) $e^{\mp}$ 
beams in each entry, 
with the $e^{\mp}$ beam polarizations $(P_L^e, P_R^{\bar{e}})\!=\!(0.9, 0.65)$ and an ideal detection efficiency $\epsilon\!=\!100\%\hs$.}}
\label{tab:10}
\end{table}
\begin{table}[]
\begin{center}
\begin{tabular}{c|c||c|c|c|c}
\hline\hline 
&&&&&\\[-4mm]
$\sqrt{s\,}$\,(TeV) & $\mathcal{L}~$ & $\Lambda_{\tilde{B}W}^{\text{tot},2\sigma}$ & $\Lambda_{\tilde{B}W}^{\text{tot},5\sigma}$ & $\Lambda_{3Z}^{\text{tot},2\sigma}$ & $\Lambda_{3Z}^{\text{tot},5\sigma}$
\\
(\text{energy}) & $(\text{ab}^{-1})$ & (\text{unpol},\,\text{pol}) & (\text{unpol},\,\text{pol}) & (\text{unpol},\,\text{pol}) & (\text{unpol},\,\text{pol})
\\[0.5mm]
\hline\hline
&&&&&\\[-4mm]
	0.25 & 20 & (1.3,\,1.6) & (1.1,\,1.3) & (1.1,\,1.3) & (.88,\,1.1) \\
	\hline
	&&&&&\\[-4.2mm]
	0.25 & 5 & (1.1,\,1.3) & (.88,\,1.1) & (.93,\,1.1) & (.74,\,.90) \\
	\hline
	&&&&&\\[-4.2mm]
	0.5 & 5 & (2.0,\,2.1) & (1.6,\,1.6) & (1.4,\,1.5) & (1.1,\,1.2) \\
	\hline
	&&&&&\\[-4.2mm]
	1 & 5 & (2.9,\,3.0) & (2.3,\,2.4) & (1.9,\,2.2) & (1.5,\,1.8) \\
	\hline
	&&&&&\\[-4.2mm]
	3 & 5 & (5.2,\,5.3) & (4.1,\,4.2) & (3.3,\,3.9) & (2.6,\,3.1) \\
	\hline
	&&&&&\\[-4.2mm]
	5 & 5 & (6.7,\,6.9) & (5.4,\,5.5) & (4.3,\,5.2) & (3.4,\,4.1) \\
	\hline\hline
\end{tabular}
\end{center}
\vspace*{-4.5mm}
\caption{\hspace*{-0.5mm}\small {Sensitivity reaches for the new physics cutoff scales 
	$\Lambda_{\tilde{B}W}^{}$ and $\Lambda_{3Z}^{}$} (in TeV) 
{through the combination of $e^-e^+\hsm\ito f_a\bar{f}_af_b^{}\bar{f}_b^{}$ and $e^-e^+\hsm\ito f_a\bar{f}_a\nu\bar{\nu}$
	$(\text{with}\,f\!=\!\ell,q)$, with (unpolarized,\,polarized) $e^{\mp}$ beams in each entry.\ 
	The bounds are presented at the $2\sigma$ and $5\sigma$ levels 
	and for different collision energies.\ For illustration, we choose a representative integrated luminosity} 
$\mathcal{L}\!=\!5\,\text{ab}^{-1}$, {an ideal detection efficiency $\epsilon\!=\!100\%$, and the $e^{\mp}$ beam polarizations 
	$(P_L^e, P_R^{\bar{e}})\!=\!(0.9, 0.65)$.}}
\label{tab:11}
\vspace*{5mm}
\end{table}

In this subsection, we study the role of polarized $e^-$ and $e^+$ beams, assuming 
$(P_L^e,P_R^{\bar{e}})\!=\!(0.9,0.65)$ for the fractions of 
(left,\,right)-handed (electrons,\,positrons) 
in the $(e^-\hsm ,\hs e^+)$ beams\,\cite{eePol}\cite{LCF-CERN}, 
respectively.\footnote{%
The degree of longitudinal beam polarization for electron $e^-$ or positron $e^+$
is defined as $\,\widehat{P}\!=\!P_R^{}\!-\!P_L^{}$ \cite{eePol}.\ 
Because the sum of left-handed and right-handed fractions equals one ($P_L^{}\!+\!P_R^{}\!=\!1$),
the left-handed and right-handed fractions of $e^-$ and $e^+$ 
can be derived as follows:
$P_{L,R}^e\!=\!\fr{1}{2}(1\!\mp\!\widehat{P}^e)$\, and
$P_{L,R}^{\bar{e}}\!=\!\fr{1}{2}(1\!\mp\!\widehat{P}^{\bar{e}})$.\ 
For example, an unpolarized beam of $e^-$ or $e^+$ has
a vanishing degree of polarization $\,\widehat{P}\!=\!0$\,,
whereas a polarized $e^-$ beam with a fraction $P_L^e=90\%$ has $\,\widehat{P}^e\!=\!-0.8$\,
and a polarized $e^+$ beam with a fraction $P_R^{\bar{e}}=65\%$
has $\,\widehat{P}^{\bar{e}}\!=\!0.3$\,.}\
We present the sensitivity reaches on the nTGC form factors 
$(f_5^\gamma,\,f_5^Z)$ in Table\,\ref{tab:10}
and on the nTGC new physics scales $\cut_j^{}$ in Tables\,\ref{tab:11},
where we compare the sensitivity reaches in the (unpolarized, polarized)
cases in each entry.\

\vs 

Compared to the unpolarized results, polarized $e^\mp$ beams enhance the sensitivity reaches 
on the form factor $f^\gamma_5$ across various $e^-e^+$ collider energies by $(8.4\!-\!51)\%$.\ 
The smallest enhancement of 8.4\% is realized at $\sqrt{s\,}\!=\!3\hs$TeV, 
whereas the largest enhancement of 51\% is achieved at 
$\sqrt{s\,}\!=\!250\hs$GeV.\ 
Moreover, polarized $e^\mp$ beams improve the sensitivity bounds on the form factor $f^Z_5$ by about $(36\hsm -\hsm 54)\%$ 
across different collider energies.\ Here the smallest enhancement of 36\% is found at the collision energy $\sqrt{s\,}\!=\!500\,\text{GeV}$, 
and the largest enhancement of 54\% occurs at the collision energy $\sqrt{s\,}\!=\!250\,$GeV.\

\vs 

We find that in most scenarios the beam polarizations can enhance $\Lambda_{\tilde{B}W}^{}$ 
by about 0.2\,TeV as compared to the unpolarized case.\ However, the improvement of the sensitivity reaches for 
the new physics scale $\Lambda_{3Z}^{}$ facilitated by the polarized $e^\mp$ beams is significantly more pronounced.\ 
Specifically, relative to the unpolarized results, the polarized $e^\mp$ beams can improve the sensitivity reaches for 
$\Lambda_{3Z}^{}$ across different collision energies by about $(12\hsm -\hsm 21)\%\hs$.\ 
The smallest improvement of 12\% occurs at the collision energy $\sqrt{s\,}\!=\!0.5\hs$TeV,  
whereas the largest improvement of 21\% is achieved at the collision energy $\sqrt{s\,}\!=\!0.25\hs$TeV.\

\tabcolsep 1pt
\begin{table}[t]
\begin{center}
\begin{tabular}{c|c||c|c|c|c}
\hline\hline 
&&&&&\\[-4mm]
$\sqrt{s\,}$\,(TeV) & $\mathcal{L}~$ & $|f_5^\gamma|_{2\sigma}^{}$ & $|f_5^Z|_{2\sigma}^{}$ & $\Lambda_{\tilde{B}W}^{\text{tot},2\sigma}$ & $\Lambda_{3Z}^{\text{tot},2\sigma}$
\\
(\text{energy}) & $(\text{ab}^{-1})$ & (\text{unpol},\,\text{pol},\,\text{mixed}) & (\text{unpol},\,\text{pol},\,\text{mixed}) & (\text{unpol},\,\text{pol},\,\text{mixed}) & (\text{unpol},\,\text{pol},\,\text{mixed})
\\[0.5mm]
\hline\hline
			&&&&&\\[-4mm]
			0.25 & 20 & $\,(2.0,\,.96,\,1.2)\!\times\!10^{-4}\,$ & $\,(2.0,\,.90,\,1.2)\!\times\!10^{-4}\,$ & 
			(1.3,\,1.6,\,1.5) & (1.1,\,1.3,\,1.3) \\
			\hline
			&&&&&\\[-4.2mm]
			0.25 & 5 & $(3.9,\,1.9,\,2.4)\!\times\!10^{-4}$ & $(3.9,\,1.8,\,2.3)\!\times\!10^{-4}$ & (1.1,\,1.3,\,1.3) & (.93,\,1.1,\,1.1) \\
			\hline
			&&&&&\\[-4.2mm]
			0.5 & 5 & $(4.0,\,3.2,\,3.6)\!\times\!10^{-5}$ & $(8.2,\,5.2,\,6.3)\!\times\!10^{-5}$ & (2.0,\,2.1,\,2.0) & (1.4,\,1.5,\,1.5) \\
			\hline
			&&&&&\\[-4.2mm]
			1 & 5 & $(8.3,\,7.0,\,7.5)\!\times\!10^{-6}$ & $(2.1,\,1.2,\,1.5)\!\times\!10^{-5}$ & (2.9,\,3.0,\,3.0) & (1.9,\,2.2,\,2.1) \\
			\hline
			&&&&&\\[-4.2mm]
			3 & 5 & $(8.4,\,7.6,\,8.0)\!\times\!10^{-7}$ & $(2.5,\,1.3,\,1.6)\!\times\!10^{-6}$ & (5.2,\,5.3,\,5.2) & (3.3,\,3.9,\,3.7) \\
			\hline
			&&&&&\\[-4.2mm]
			5 & 5 & $(2.9,\,2.6,\,2.7)\!\times\!10^{-7}$ & $(8.6,\,4.2,\,5.3)\!\times\!10^{-7}$ & (6.7,\,6.9,\,6.8) & (4.3,\,5.2,\,4.9) \\
\hline\hline
\end{tabular}
\end{center}
\vspace*{-4.5mm}
\caption{\hspace*{-0.5mm}\small 
Same as Tables\,\ref{tab:10}-\ref{tab:11} for the $2\sigma$ sensitivity bounds,
but the sensitivity reaches for the mixed setup (including 
half of the data taken from the unpolarized operation
and the other half of the data taken from the 
polarized operation) with beam polarizations $(P_L^e,{\hs}P_R^{\bar{e}})\!=\!(0.9,{\hs}0.65)$.}
\label{tab:12new}
\vspace*{5mm}
\end{table}

Finally, we note that each $e^+e^-$ collider will first operate with
unpolarized $e^\mp$ beams and then operate with polarized $e^\mp$ beams.\
Thus, it is desired to consider an (ideally) mixed setup where
one half of the total data is taken from the first-phase operation with unpolarized beams and the other half of the total data is from 
the second-phase operation with polarized beams.\ 
This mixed setup requires to combine the sensitivity bound using 
the unpolarized data with the sensitivity bound using 
the polarized data.\  We present the sensitivity reaches 
on the nTGC form factors $(f_5^{\ga},\hs f_5^Z)$ and 
the corresponding nTGC new physics scales 
$(\cut_{3Z}^{},\cut_{}^{\tilde{B}W})$ in Table\,\ref{tab:12new}.\  
It shows that the sensitivity bounds 
in the mixed setup always lie between
the sensitivity bounds of the unpolarized and polarized cases.\
As we will demonstrate in Section\,\ref{sec:4.5}, the mixed
setup is important for optimizing the correlation bounds (contours)
for each pair of nTGC parameters by avoiding certain poorly constrained
parameter space in the case of the polarized beams.

\subsection{\hspace*{-2.5mm}Analyzing Correlations of nTGC Parameters by Machine Learning}
\label{sec:4.5}
\vspace*{1.5mm}

In this subsection, we use machine learning to analyze correlations 
between two nTGC form factors or two dimension-8 nTGC operators.\ 
We find that using the machine learning method can significantly improve 
the sensitivities of probes of correlations between the nTGC parameters.\  

\begin{figure}[]
\vspace*{-5mm}
\hspace*{4mm}
\includegraphics[width=7.7cm,height=7.4cm]{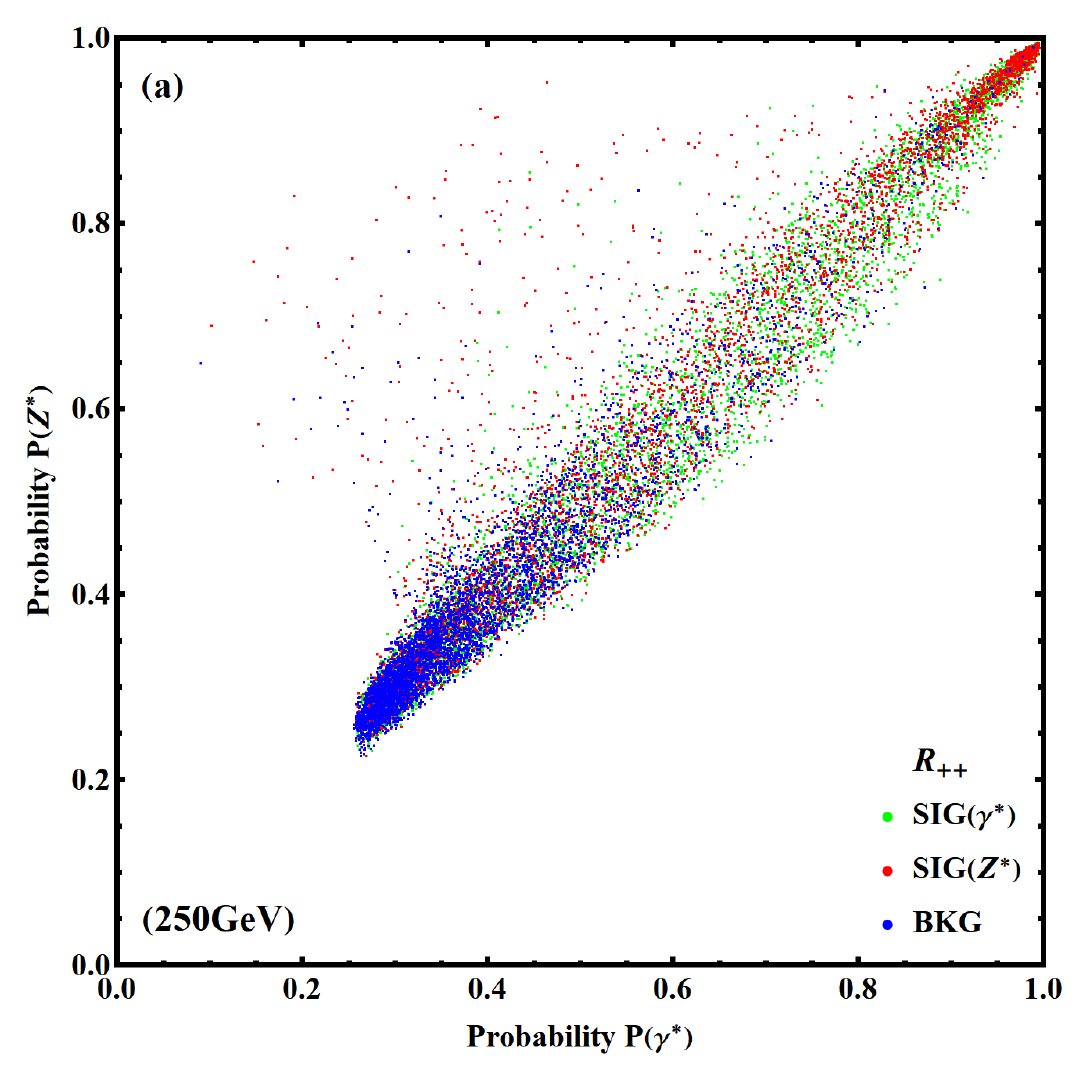}
\hspace*{-1mm}
\includegraphics[width=7.7cm,height=7.4cm]{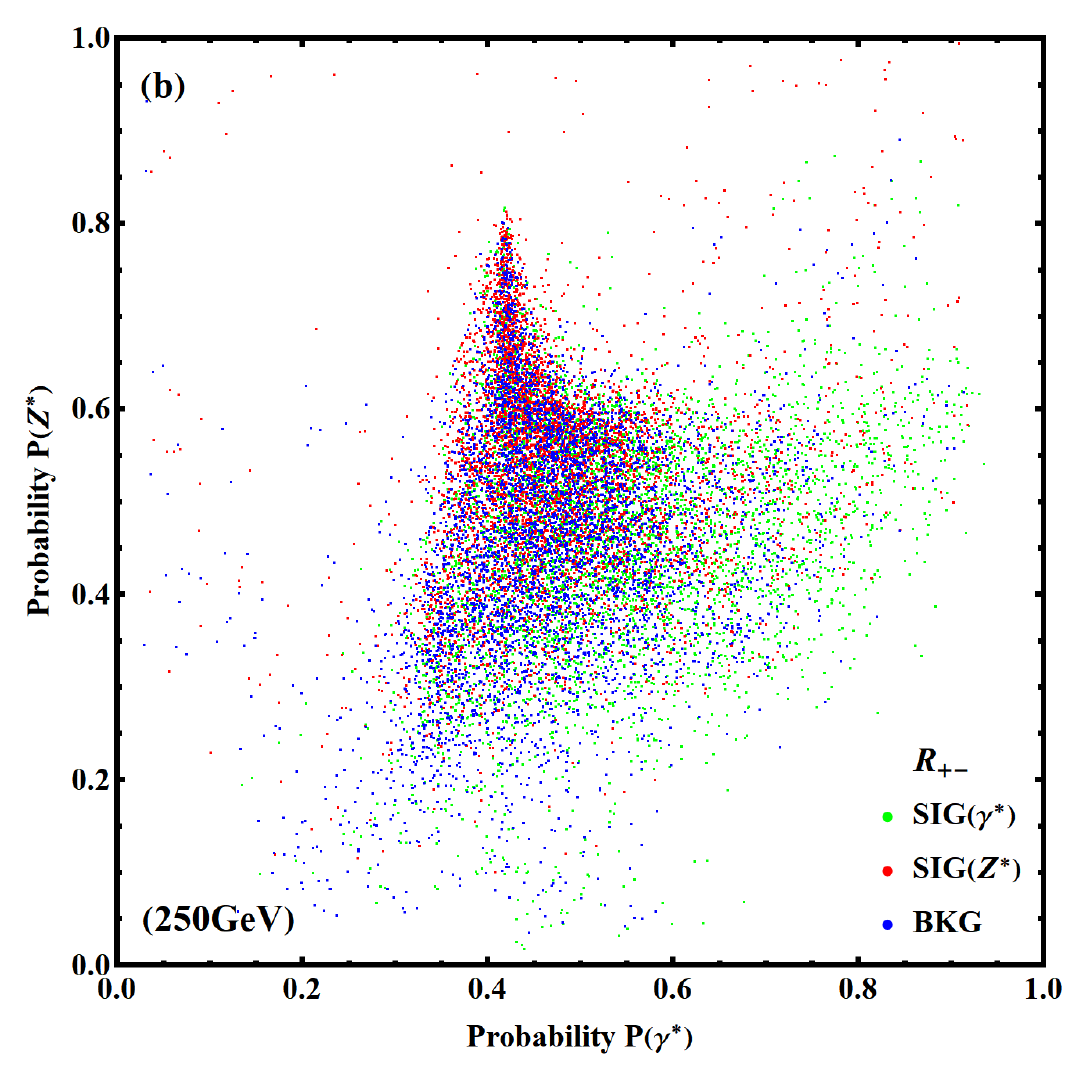}
\\[1.5mm]
\hspace*{4mm}
\includegraphics[width=7.7cm,height=7.4cm]{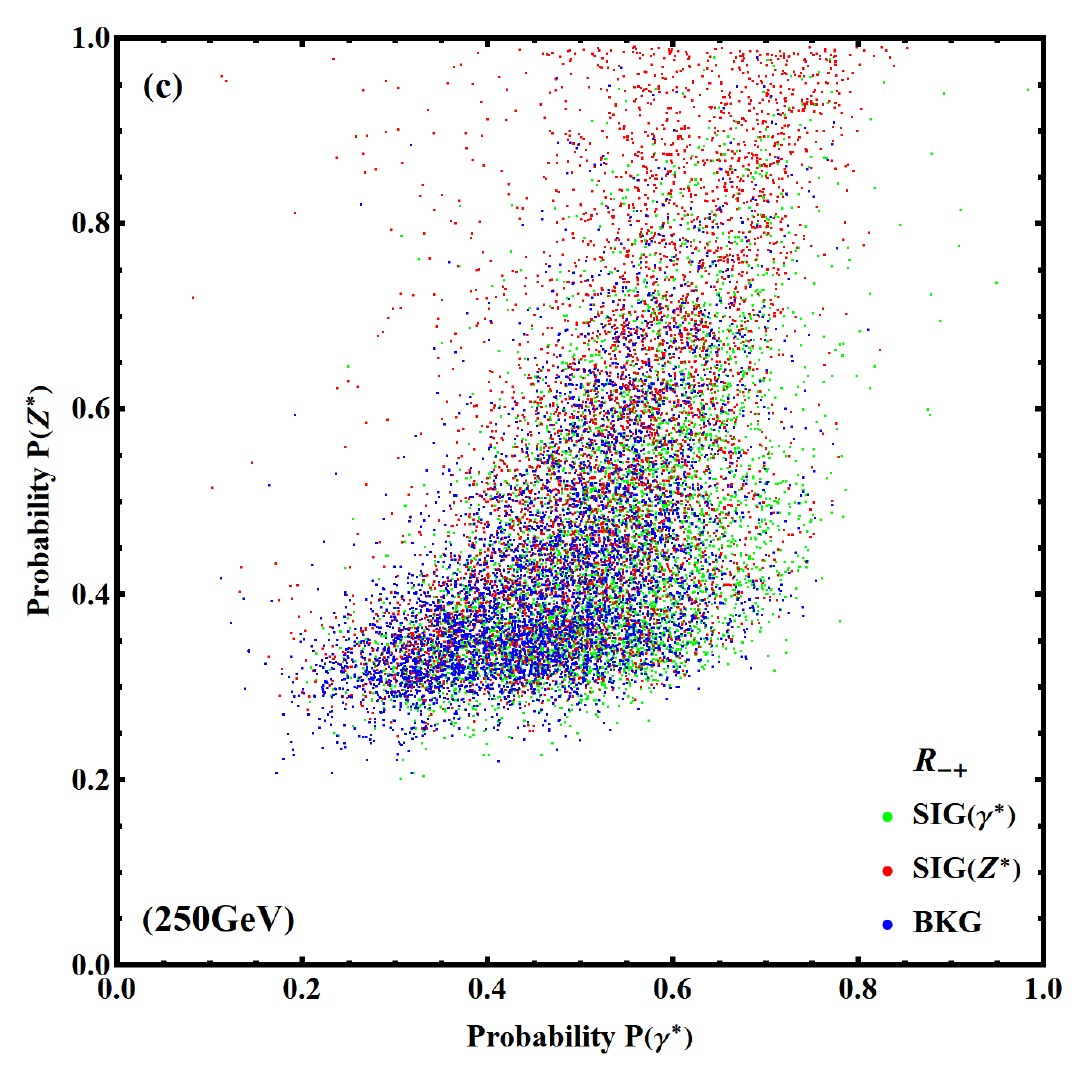}
\hspace*{-1mm}
\includegraphics[width=7.7cm,height=7.4cm]{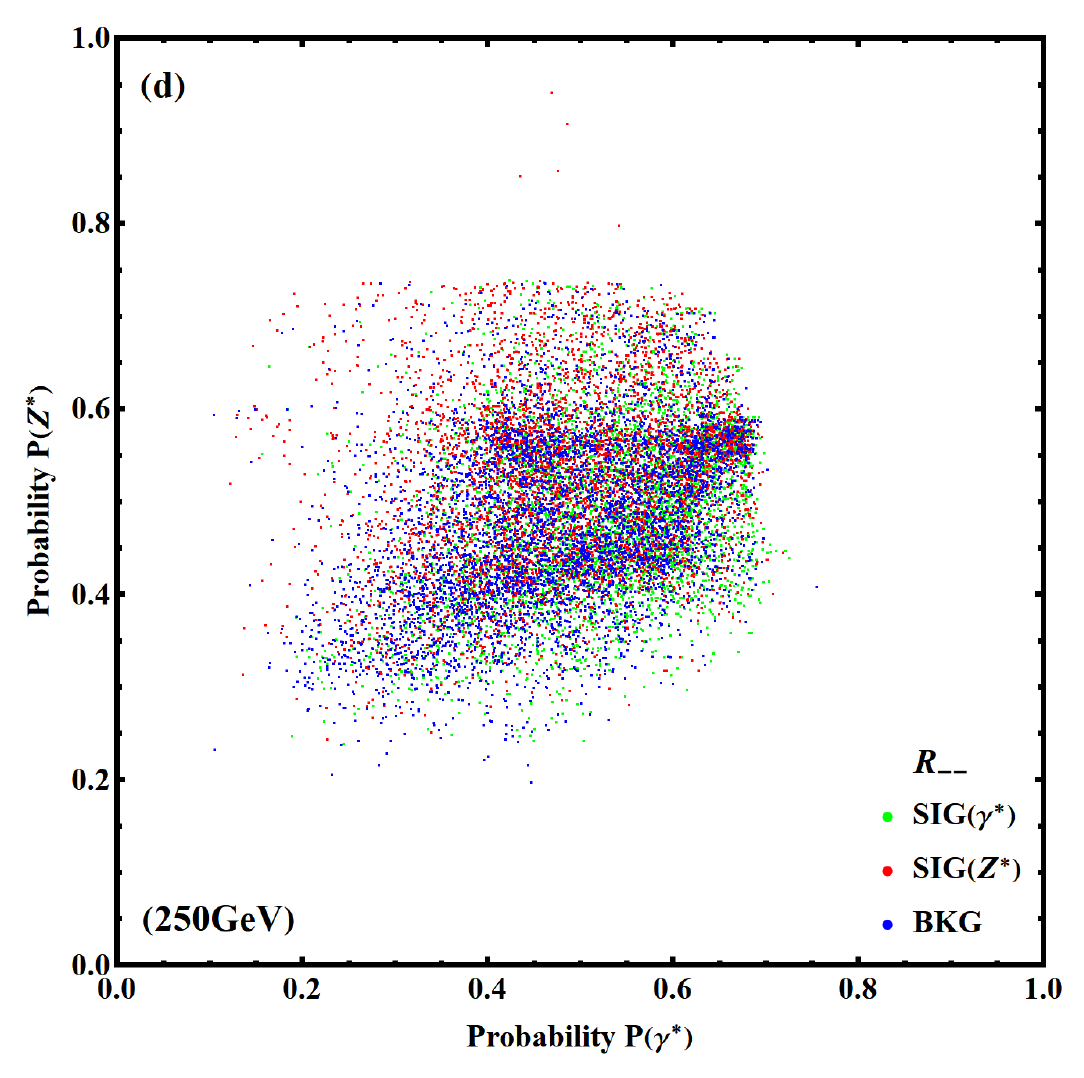}
\vspace*{-2mm}
\caption{\hspace*{-0.5mm}\small 
Distribution of probabilities $P(\gamma^*)$ and $P(Z^*)$ of simulated events belonging to $\gamma^*$ signal or $Z^*$ signal category within the 4 regions of Eq.\eqref{eq:R-ij} for the case of 
$e^+e^-\!\!\ito\! V^*\!\ito\! ZZ\!\hsm\ito\! (\ell\bar{\ell})(d\bar{d})\,$ 
at a collider energy $\sqrt{s\,}\!=\!250$\,GeV
[marked as (250{\hs}GeV) in every plot], 
where $d$ denotes the down-type quarks ($d,s,b$).\  
Plots\,(a)-(d) present the probability distributions of events
for the angular regions
$R_{++}$, $R_{+-}$, $R_{-+}$, and $R_{--}$, respectively.\
The (green,\,red,\,blue) points in each plot represent 
the ($\gamma^*$\!,\,$Z^*$\!,\,SM) contributions respectively.}
\label{fig:9}
\end{figure}
\begin{figure}[]
\vspace*{-5mm}
\hspace*{4mm}
\includegraphics[width=7.7cm,height=7.4cm]{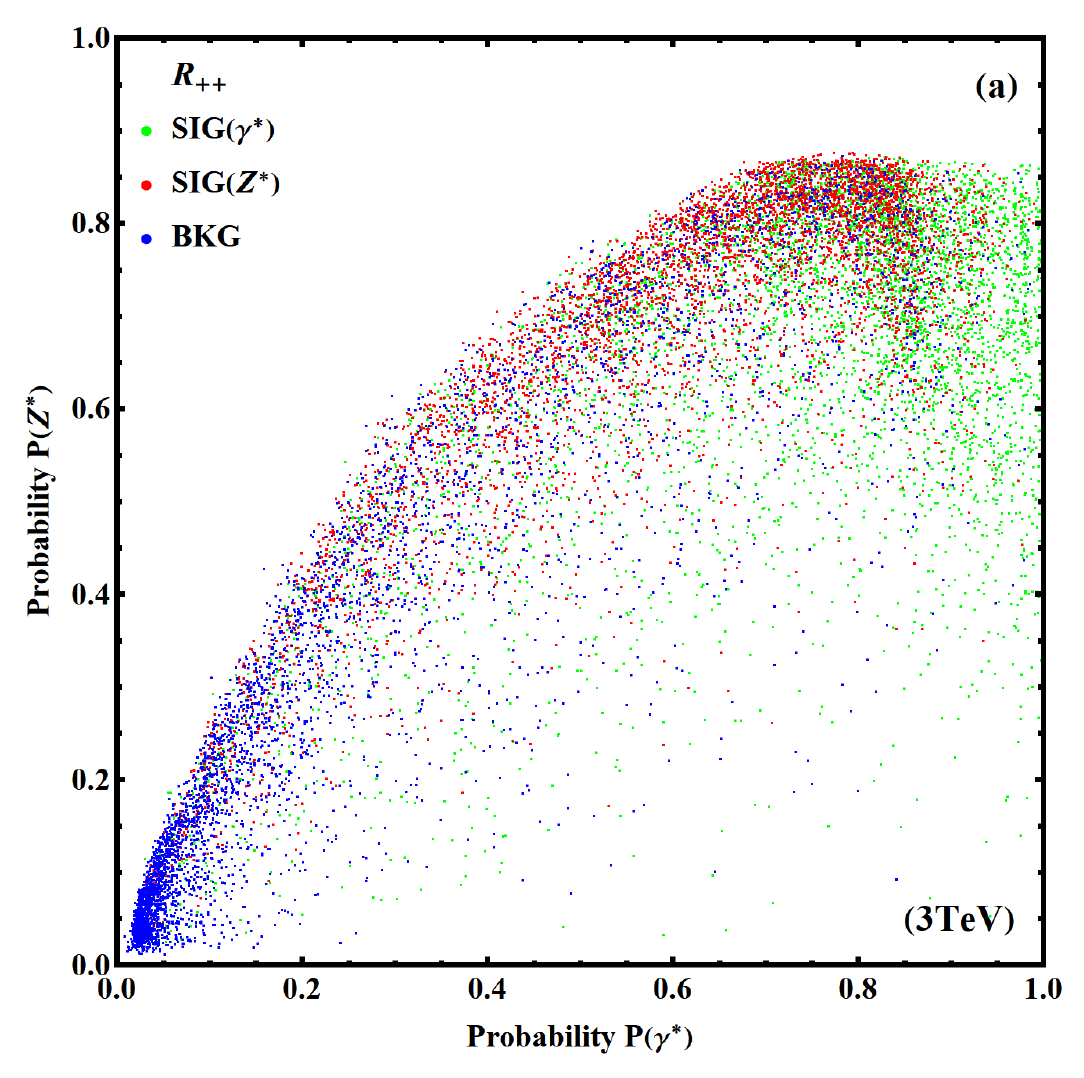}
\hspace*{-1mm}
\includegraphics[width=7.7cm,height=7.4cm]{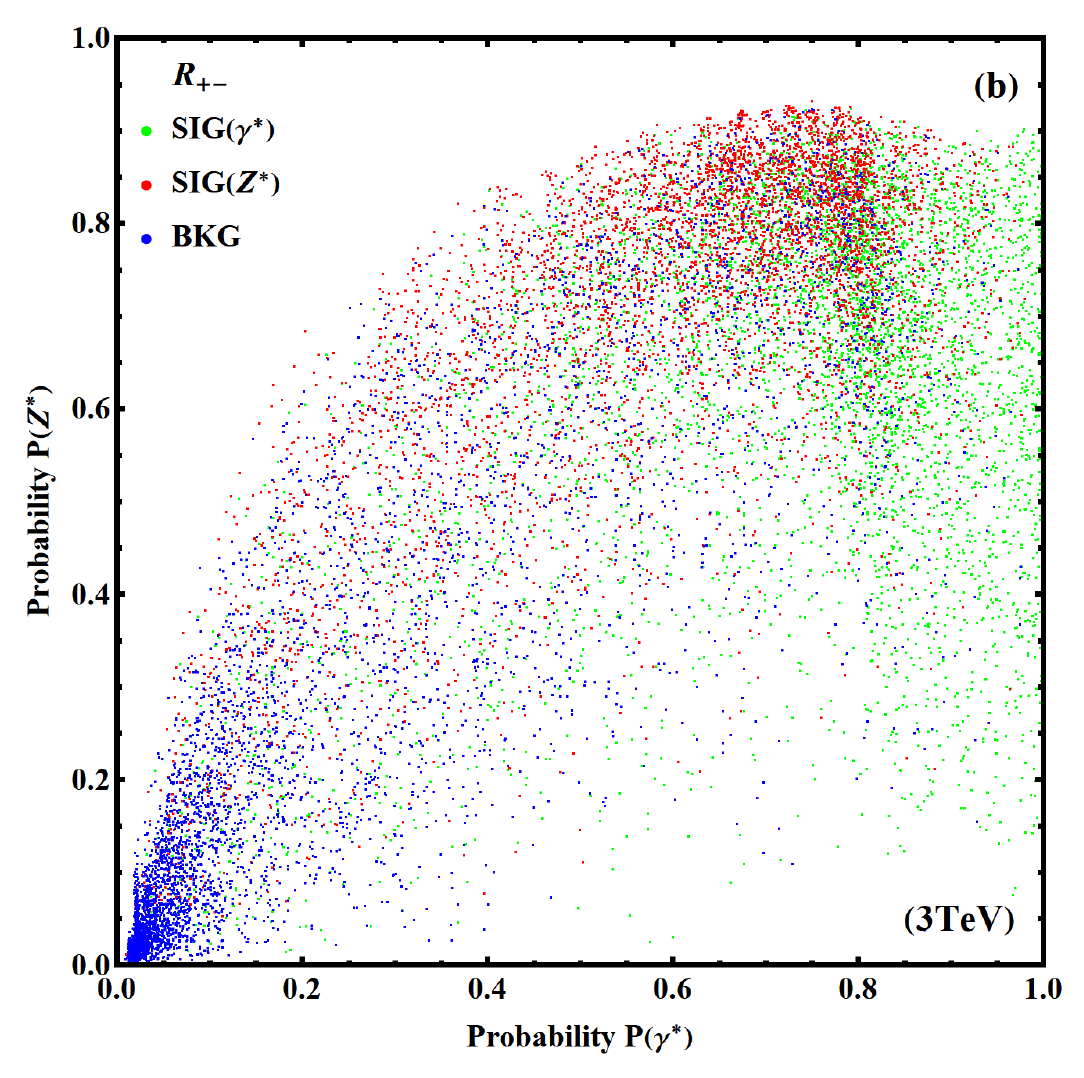}
\\[1.5mm]
\hspace*{4mm}
\includegraphics[width=7.7cm,height=7.4cm]{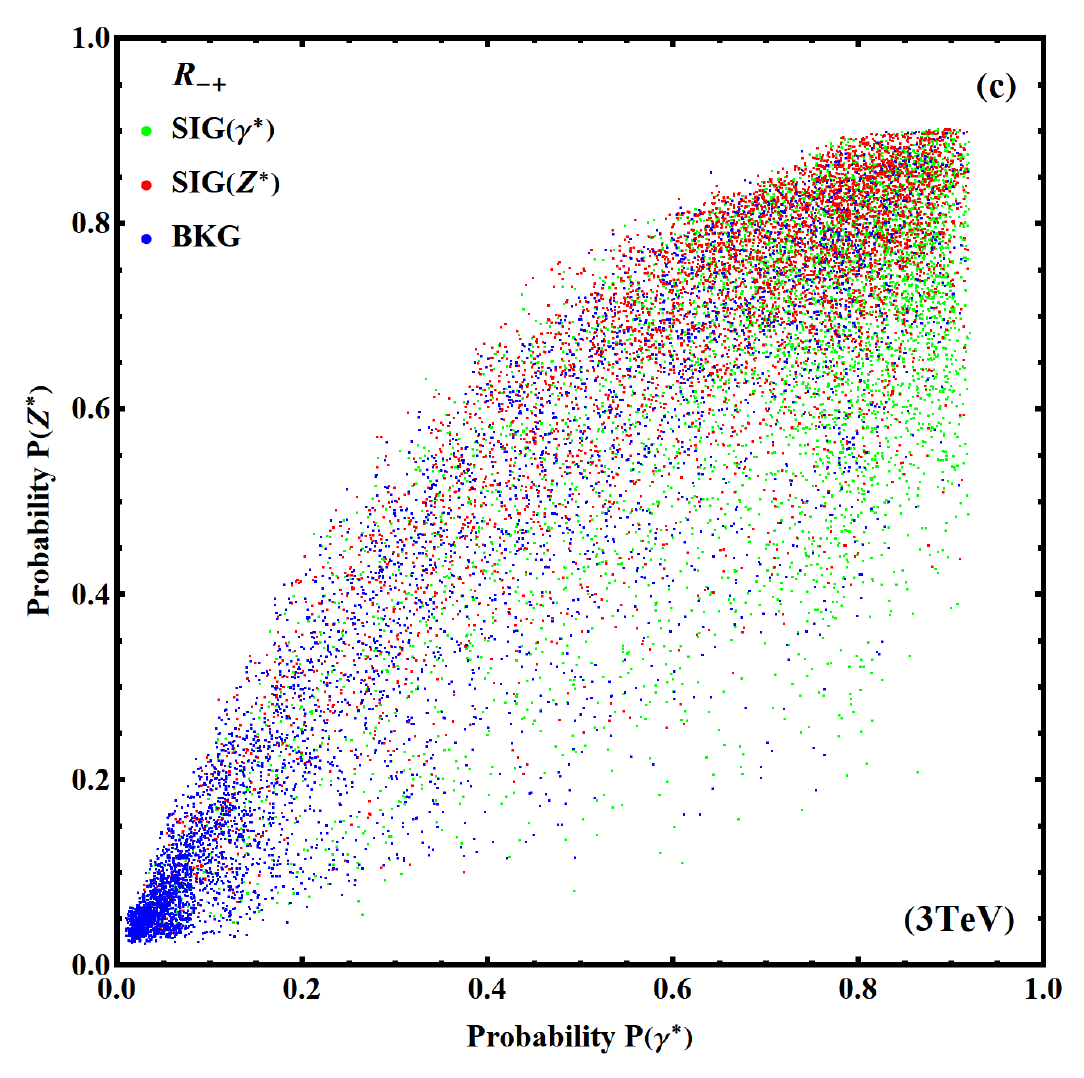}
\hspace*{-1mm}
\includegraphics[width=7.7cm,height=7.4cm]{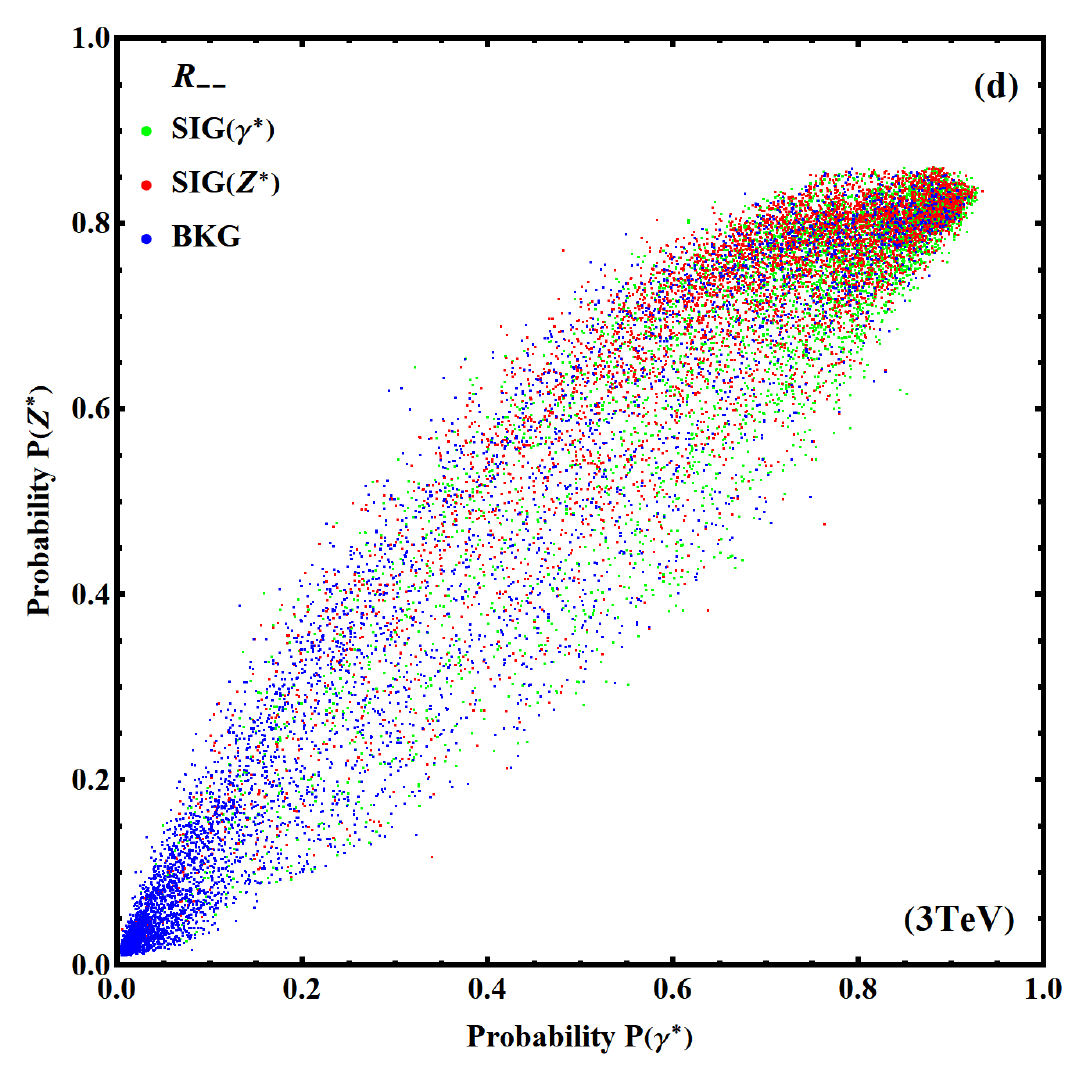}
\vspace*{-2mm}
\caption{\hspace*{-0.5mm}\small 
Distribution of probabilities $P(\gamma^*)$ and $P(Z^*)$ of simulated events belonging to $\gamma^*$ signal or $Z^*$ signal category 
within the 4 regions (plot (a): $R_{++}$, plot (b): $R_{+-}$, plot (c): $R_{-+}$, plot (d): $R_{--}$) of Eq.\eqref{eq:R-ij} for the case of 
$e^+e^-\!\ito V^*\!\ito ZZ\!\ito (\ell\bar{\ell})(d\bar{d})\,$ 
at a collider energy $\sqrt{s\,}\!=\!3$\,TeV
[marked as (3{\hs}TeV) in every plot], 
where $d$ denotes the down-type quarks ($d,s,b$).\
Plots\,(a)-(d) present the probability distributions of events
for the angular regions
$R_{++}$, $R_{+-}$, $R_{-+}$, and $R_{--}$, respectively.\  
The (green,\,red,\,blue) points in each plot represent 
the ($\gamma^*$\!,\,$Z^*$\!,\,SM) contributions respectively.
}
\label{fig:10new}
\end{figure}

We divide the phase space of the final states into 4 regions based on the signs of the angular distribution $f^1_\Omega$ in Eq.\ \eqref{fj} 
or $f^1_{\Omega^\prime}$ in Eq.\eqref{fjn}: 
\begin{subequations}
	\label{eq:R-ij}
	\begin{align}
		R_{++}^{}\!:&\quad \left(f^1_\Omega(\gamma)\hsm\!>\!0,\,f^1_\Omega(Z)\hsm\!>\!0\right)\!, \,~\text{or,}~\, \left(f^1_{\Omega^\prime}(\gamma)\hsm\!>\!0,\,f^1_{\Omega^\prime}(Z)\hsm\!>\!0\right) \!,
		\\
		R_{+-}^{}\!:&\quad \left(f^1_\Omega(\gamma)\hsm\!>\!0,\,f^1_\Omega(Z)\hsm\!<\!0\right)\!, \,~\text{or,}~\, \left(f^1_{\Omega^\prime}(\gamma)\hsm\!>\!0,\,f^1_{\Omega^\prime}(Z)\hsm\!<\!0\right) \!,
		\\
		R_{-+}^{}\!:&\quad \left(f^1_\Omega(\gamma)\hsm\!<\!0,\,f^1_\Omega(Z)\hsm\!>\!0\right)\!, \,~\text{or,}~\, \left(f^1_{\Omega^\prime}(\gamma)\hsm\!<\!0,\,f^1_{\Omega^\prime}(Z)\hsm\!>\!0\right) \!,
		\\
		R_{--}^{}\!:&\quad \left(f^1_\Omega(\gamma)\hsm\!<\!0,\,f^1_\Omega(Z)\hsm\!<\!0\right)\!, \,~\text{or,}~\, \left(f^1_{\Omega^\prime}(\gamma)\hsm\!<\!0,\,f^1_{\Omega^\prime}(Z)\hsm\!<\!0\right) \!.
	\end{align}
\end{subequations}
Within these 4 regions, we can obtain the probability of each event belonging to the
$\gamma^*$ signal or $Z^*$ signal category.\ 
These probabilities $P(\gamma^*)$ and $P(Z^*)$ are provided 
by classifiers trained specifically based on each dataset in the corresponding region.\

In Fig.\,\ref{fig:9}, we present the distribution of probabilities $P(\gamma^*)$ and $P(Z^*)$ of 
the simulated events belonging to the $\gamma^*$ signal 
or $Z^*$ signal category within the 4 regions of 
Eq.\eqref{eq:R-ij} for the case of 
$e^+e^-\!\ito V^*\!\ito ZZ\!\ito (\ell\bar{\ell})(d\bar{d})\,$ at a collider energy  
$\sqrt{s\,}\!=\!250$\,GeV.\ 
For comparison, we present in Fig.\,\ref{fig:10new} an analogous set of plots of $P(\gamma^*)$ and $P(Z^*)$ for a different $e^+e^-$ collider
energy $\sqrt{s\,}\!=\!3$\,TeV.\  
For illustration, we choose in each plot a sample value of the nTGC cutoff scales 
$\Lambda_{3Z}^{}\!=\!\Lambda_{\!\tilde{B}W}^{}\!\!=\!1\hs$TeV, which corresponds to
the input of nTGC form factors $(f_5^{\ga},\hs f_5^{Z})\!=\!(6,\hs 3)\!\times\!10^{-4}$
according to Eqs.\eqref{f5z} and \eqref{f5a}.\ 
As shown in Figs.\,\ref{fig:4} and \ref{fig:6} for the single-observable distributions, 
we see that the differences between the angular distributions (in $\theta$ and $\phi_a^{}$) 
of signals and backgrounds become much larger as the collision energy $\sqrt{s\,}$ increases.\ 
Hence, it is expected that using the machine learning algorithm we can separate the signals (red and green points)
from the SM backgrounds (blue points) much more efficiently at a higher collision energy 
(such as $\sqrt{s\,}\!=\!3$\,TeV) than at a lower collision energy 
(such as $\sqrt{s\,}\!=\!250$\,GeV).\  This enhanced discriminative power of signal-background separation can be
seen by comparing the plots of Figs.\,\ref{fig:9} and \ref{fig:10new}.\

\vs

\begin{figure}[]
\hspace*{6mm}
\includegraphics[width=7.5cm,height=7.3cm]{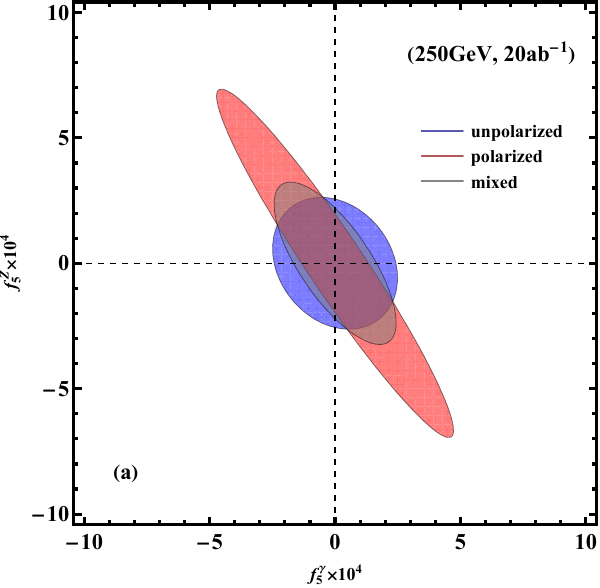}
\hspace*{2mm}
\includegraphics[width=7.5cm,height=7.3cm]{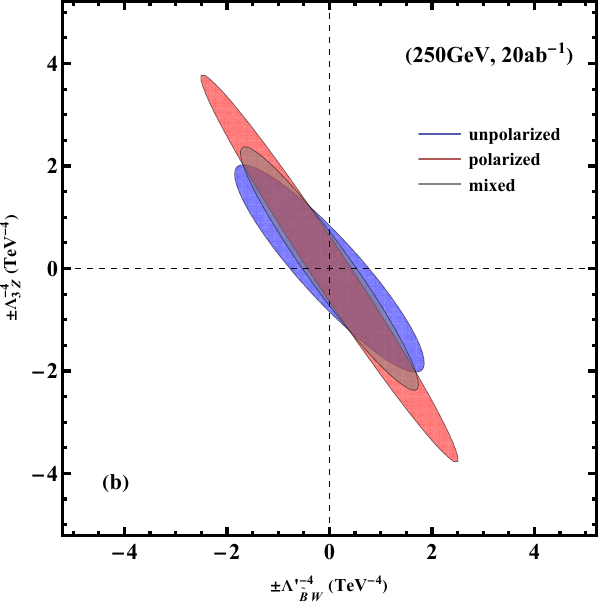}
\\[1.5mm]
\hspace*{6mm}
\includegraphics[width=7.5cm,height=7.3cm]{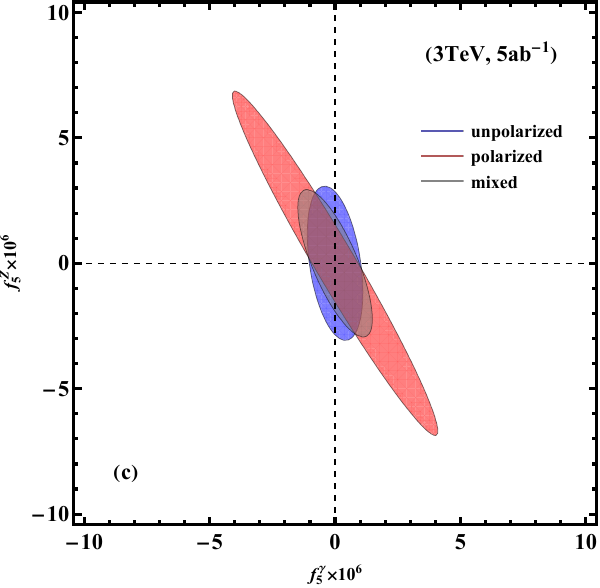}
\hspace*{-1mm}
\includegraphics[width=7.7cm,height=7.3cm]{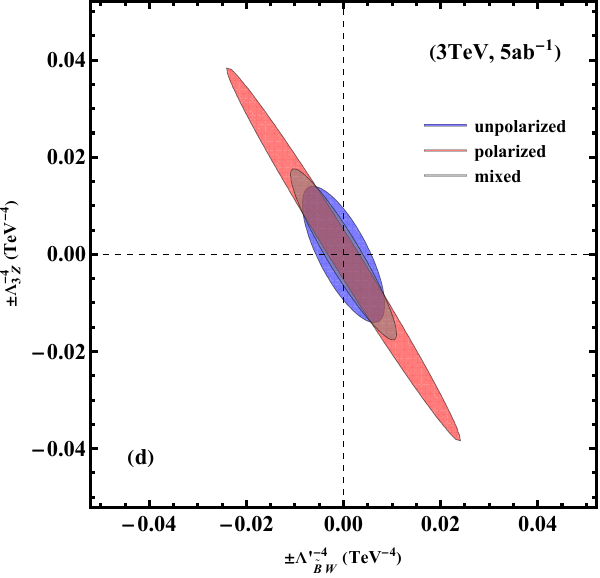}
\vspace*{-2mm}
\caption{\hspace*{-0.5mm}\small 
Bounds on correlations of the nTGC form factors 
and on correlations of the nTGC new physics scales at the $2\sigma$ level.\ 
The plots\,(a)-(b) are for the collision energy $\sqrt{s\,}\!=\!250\hs$GeV 
(with an integrated luminosity 20\,ab$^{-1}$)
and plots\,(c)-(d) are for the collision energy $\sqrt{s\,}\!=\!3\hs$TeV
(with an integrated luminosity 5\,ab$^{-1}$),  
which are marked in the upper-right corner of each plot.\ 
The contours of plots\,(a) and (c) depict the correlations 
between the form factors $(f_5^\gamma,\hs f_5^Z)$, 
whereas the contours of plots\,(b) and (d) present the
correlations between the new physics scales 
$(\cut^{\prime}_{\tilde{B}W}, \cut_{3Z}^{})$.\
In each plot, the correlation bound for the case of unpolarized 
$e^\mp$ beams is shown as the blue contour 
and that of the case for polarized $e^\mp$ beams is given by the pink contour, 
with the $e^{\mp}$ beam polarizations 
$(P_L^e, P_R^{\bar{e}})\!=\!(0.9, 0.65)$; 
whereas the light-blue contour presents the correlation bound for the mixed setup 
in which half of the data comes from unpolarized collisions and the other half 
of the data comes from polarized collisions.} 
\label{contour}
\label{fig:11new}
\end{figure}

We divide events into bins of the probability distribution, whose widths we take as 
$\Delta P(\gamma^*)\!=\hsm 0.1$ and $\Delta P(Z^*)\!=\!0.1\hs$.\ 
Then, we combine the contributions of two nTGC form factors into a global $\chi^2$ function:
\begin{align}
	\chi^2\,=\sum_{\text{region}}\!\sum_{\text{bin}}\!
	\frac{~\big(\sigma_1^\gamma f_5^\gamma\!+\hsm\sigma_1^Z f_5^Z\big)^{\!2}~}{\sigma_0^{}}
	\!\!\times\!\mathcal{L}\!\times\!\epsilon\,.
\end{align}
Minimizing the $\chi^2$ function, we derive correlated constraints on each pair of nTGC form factors and
each pair of nTGC cutoff scales.\
We present these correlated constraints in Fig.\,\ref{contour} as $2\sigma$ contours, 
where the contours of plots\,(a) and (c) depict the correlations 
between the form factors $(f_5^\gamma,\hs f_5^Z)$, whereas the contours of plots\,(b) and (d) demonstrate 
the correlations between the new physics scales $(\cut^{\prime}_{\tilde{B}W}, \cut_{3Z}^{})$.\ 

In each plot of Fig.\,\ref{contour}, 
we compare the correlation bounds for the case of 
unpolarized $e^\mp$ beams (shown as the blue contour) and 
the case of polarized $e^\mp$ beams (shown as the red contour).\ 
For the polarized case, we choose the $e^\mp$ beam polarizations 
$(P_L^e,{\hs}P_R^{\bar{e}})\!=\!(0.9,{\hs}0.65)$.\ 
Figure\,\ref{contour} shows that using the beam polarizations can 
significantly tighten each correlation contour along the 
minor axis of the ellipse by about 50\%, 
but makes the correlation bound much weaker 
along the major axis of the ellipse 
(which is orientated towards the northwestern direction 
in the parameter plane).\ 
Similar features hold for the correlation bounds 
in the parameter space of the cutoff scales 
$(\cut^{\prime}_{\tilde{B}W}, \cut_{3Z}^{})$.\ 
This can be understood as follows.\ 
The angular distributions of $f_5^Z$ and $f_5^\gamma$ are actually 
the same for purely left-handed (or right-handed) electrons.\ 
So the fully polarized $e_L^-$ (or $e_R^+$) 
beam would make $f_5^Z$ and $f_5^\gamma$ distributions indistinguishable 
and thus the correlation contour would collapse into 
a line with slope $-1\hs$.\ 
This explains why the polarized case makes the correlation in each plot 
poorly constrained along the major axis of the ellipse 
(orientated towards the northwestern direction in the parameter 
space of the $f_5^\gamma\!-\!f_5^Z$ plane).\ 
Hence we can modify the polarization setup and optimize the 
correlation bound along this poorly constrained direction 
by combining the data-taking from both the 
unpolarized operation and polarized operation.\
To demonstrate this point, we present in each plot of 
Fig.\,\ref{contour} the correlation bound for a mixed setup 
in which half of the data comes from operation with unpolarized beams 
and the other half of the data comes from operation with polarized beams, 
as shown by the light-blue contour.\ 
We see that the correlation bound in the mixed setup is substantially
tightened along the major axis of the ellipse and becomes comparable to
that of the unpolarized setup.\ On the other hand, the correlation bounds
for the polarized and mixed setups are much stronger than that of 
the unpolarized setup by about $(40\!-\!50)\%$ along the minor axis
of the ellipse contours.\

\vs 

For Fig.\,\ref{contour}, we further compare the correlation contours 
in the plots\,(a)-(b) for the collision energy
$\sqrt{s\,}\!=\!250$\,GeV with the corresponding contours 
in the plots\,(c)-(d) for the collision energy
$\sqrt{s\,}\!=\!3$\,TeV.\ This shows substantial
enhancements of the constraints by factors of
$O(10^2)$ when the collider energy $\sqrt{s\,}$ 
increases from 250\,GeV to 3\,TeV.\  
This feature can be understood by noting that the
signal significance is determined by $S/\hsm\sqrt{B\,}$,
where the signal $S$ is dominated by the interference term
(which scales with energy as $\sqrt{s\,}\hs$) and
the SM background $B$ scales with energy as $1/s$
as shown by Eq.\eqref{eq:fphi-a012} 
for the differential cross sections.\ 
Hence, the signal significance scales with energy as
$\hs s^1\hs$ and is increased by a factor of $O(10^2)$
when the collider energy rises by a factor
of (3${\hs}$TeV$\!$/250${\hs}$GeV)$\,=\hsm O(10)\hs$.\ 

\vs 

\begin{figure}[t]
\centering
\hspace*{-1mm}
\includegraphics[width=7.5cm,height=7.5cm]{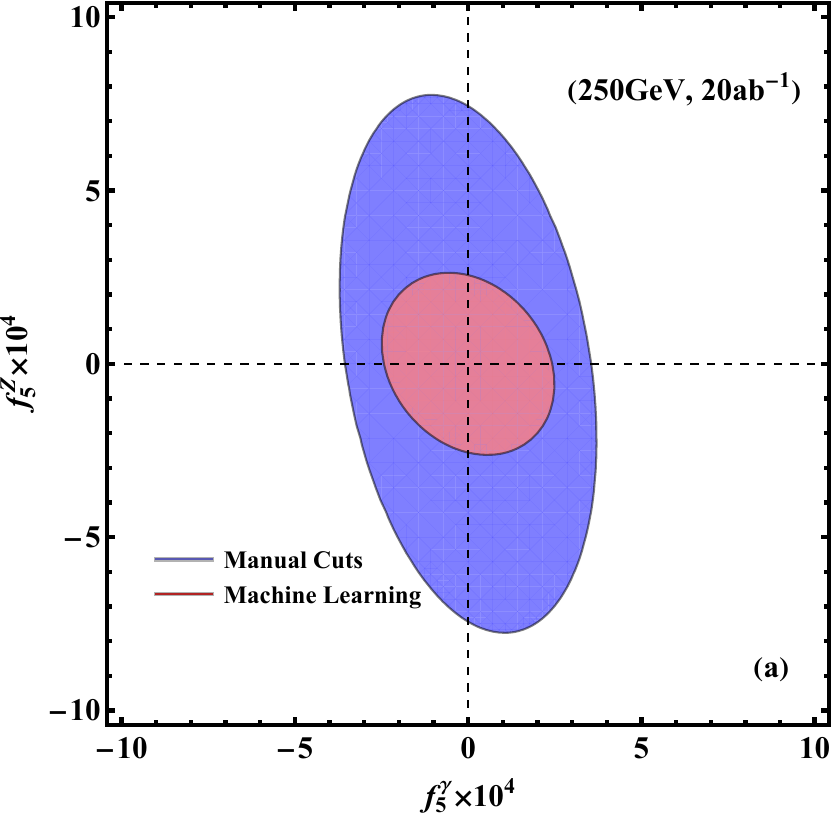}
\includegraphics[width=7.5cm,height=7.5cm]{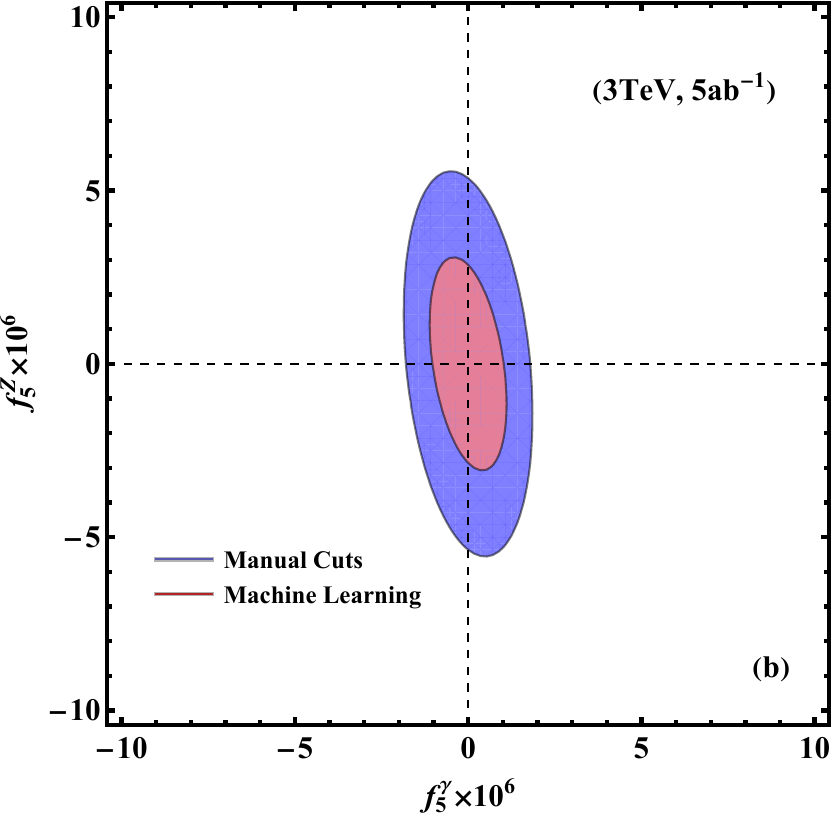}
\vspace*{-2mm}
\caption{\hspace*{-1mm}\small 
Comparison of  2$\sigma$ constraints  on correlations between the nTGC form factors 
$(f_5^{\gamma},{\hs}f_5^Z)$ obtained using conventional Manual Cuts (blue contours)
and by using Machine Learning (pink contours). 
Plot\,(a) shows constraints  on correlations for the $e^-e^+$ collision energy 250\,GeV 
and an integrated luminosity 20\,ab$^{-1}$,  
but without using beam polarization.\ 
Plot\,(b) is similar to plot\,(a) except for the collision energy 3\,TeV
and integrated luminosity 5\,ab$^{-1}$.\
}
\label{fig:13}
\end{figure}

Finally, we compare in Fig.\,\ref{fig:13} the $2\sigma$ constraints on correlations between the nTGC form factors 
$(f_5^{\gamma},\hs f_5^Z)$ obtained 
by using the conventional Manual Cuts 
and by using the Machine Learning (ML) method.\ 
In plot\,(a), we present the constraints on correlations for the $e^-e^+$ collision energy 
250\,GeV and an integrated luminosity 20\,ab$^{-1}$,  
but without using beam polarization.\ 
Plot\,(b) is similar to plot\,(a), but for the collision energy 3\,TeV
and an integrated luminosity 5\,ab$^{-1}$.\
In both plots\,(a) and (b), the blue contours show the constraints
obtained using the conventional manual cuts, whereas the pink contours show 
the constraints obtained using the machine learning method 
[which are taken from the unpolarized contours of Fig.\,\ref{fig:11new}(a)(c)].\  
This comparison shows that machine learning can improve substantially 
the nTGC constraints on $(f_5^{\gamma},{\hs}f_5^Z)$  correlations
by about a factor of 2 to 3 relative to the conventional analysis 
using manual cuts, namely, the nTGC correlation bounds become 
stronger by about (100${\hs}-\hs$200)\%.\ 
This demonstrates that machine learning can provide
a powerful tool for measuring nTGC correlation signals.

\vspace*{2mm}
\subsection{\hspace*{-2.5mm}Comparison with Previous Results}
\label{sec:4.6}
\vspace*{1.5mm}

In this subsection we compare our present sensitivity analysis for probing nTGCs via $ZZ$ final states with those bounds 
extracted from our previous study\,\cite{Ellis:2020ljj}\cite{newformfactor} via the $Z\ga$ final states.\    
%
\begin{table}[]
\vspace*{3mm}
\hspace*{12mm}
\begin{tabular}{c|c||c|c|c|c}
\hline\hline
\multicolumn{2}{c||}{~Energy \& Luminosity~~}
& \multicolumn{2}{|c|}{Previous Results\,\cite{Ellis:2020ljj}} & \multicolumn{2}{|c}{Present Results}\\
\hline
&&&&&\\[-4mm]
$\sqrt{s\,}$\,(TeV) & $\mathcal{L}$
& $\Lambda_{\tilde{B}W}^{\text{tot},2\sigma}$ & $\Lambda_{\tilde{B}W}^{\text{tot},5\sigma}$ & $\Lambda_{\tilde{B}W}^{\text{tot},2\sigma}$ & $\Lambda_{\tilde{B}W}^{\text{tot},5\sigma}$\\
~({collider\,energy})~ & \,(ab$^{-1}$)\,
& ~({unpol},\,{pol})~ & ~({unpol},\,{pol})~ & ~({unpol},\,{pol})~ & ~({unpol},\,{pol})~ \\
\hline\hline
&&&&&\\[-4mm]
0.25 & 20 & (1.4,\,1.5) & (1.2,\,1.2) & (1.3,\,1.6) & (1.1,\,1.3)\\
\hline
0.25 & 5 & (1.2,\,1.3) & (.97,\,1.0) & (1.1,\,1.3) & (.88,\,1.1)\\
\hline
&&&&&\\[-4mm]
0.5 & 5 & (1.8,\,1.9) & (1.4,\,1.4) & (2.0,\,2.1) & (1.6,\,1.6)\\
\hline
&&&&&\\[-4mm]
1 & 5 & (2.6,\,2.6) & (2.0,\,2.1) & (2.9,\,3.0) & (2.3,\,2.4)\\
\hline
&&&&&\\[-4mm]
3 & 5 & (4.3,\,4.5) & (3.5,\,3.6) & (5.2,\,5.3) & (4.1,\,4.2)\\
\hline
&&&&&\\[-4mm]
5 & 5 & (5.7,\,5.9) & (4.5,\,4.7) & (6.7,\,6.9) & (5.4,\,5.5)\\ 
\hline\hline
\end{tabular}
\caption{\hspace*{-1mm}\small Comparison of the previous sensitivity reaches for the nTGC new physics scale $\Lambda_{\tilde{B}W}^{}$
(in TeV) obtained via the reaction $e^+e^-\!\ito Z\ga$ channel\,\cite{Ellis:2020ljj} with the present sensitivity reaches obtained 
via the reaction $e^+e^-\!\ito ZZ\hs$.\  These reaches are presented at $2\sigma$ and $5\sigma$ levels and for different collider energies.\  In each entry the sensitivity bounds correspond to (unpolarized,\,polarized) $e^{\mp}$ beams, with beam polarizations 
$(P_L^e,{\hs}P_R^{\bar{e}})\!=\!(0.9,{\hs}0.65)$.\ 
}
\label{tab:12}
\label{tab:13new}
\end{table}
%

We extract the sensitivity reaches for the nTGC new physics scale $\cut^{}_{\tilde{B}W}$
obtained via $Z\ga$ production (based on Table\,3 of Ref.\,\cite{Ellis:2020ljj}) 
and compare them with the reaches for $\cut^{}_{\tilde{B}W}$ obtained 
via the $ZZ$ production (based on the present analysis),
which are summarized in Table\,\ref{tab:12}.\ 
The previous reaches were derived for the $Z\gamma$ production with $Z\ito q\bar{q}$ \cite{Ellis:2020ljj} 
(shown in the 2nd and 3rd columns of this table),
whereas the present reaches are derived for the $ZZ$ production with $Z\!\ito q\bar{q},\ell\bar{\ell},\nu\bar{\nu}$ 
(shown in the 4th and 5th columns of this table).\  
These reaches are presented at $2\sigma$ and $5\sigma$ levels and for different collider energies.\
In each entry the sensitivity bounds correspond to (unpolarized,\,polarized) $e^{\mp}$ beams, with beam polarizations 
$(P_L^e,{\hs}P_R^{\bar{e}})\!=\!(0.9,{\hs}0.65)$.\  
For illustration, we choose a representative integrated luminosity $\mathcal{L}\!=\!5\,\text{ab}^{-1}$    
and an ideal detection efficiency $\epsilon\!=\!100\%\hs$.\ 
Moreover, for the 250\,GeV collision energy of $e^-e^+$, we choose
another benchmark of integrated luminosity 
$\mathcal{L}\!=\!20\,\text{ab}^{-1}\!$ 
based on the CEPC TDR\,\cite{CEPC-TDR},
which is shown in the third row of Table\,\ref{tab:13new}.\ 

\begin{table}[t]
\vspace*{4mm}
\begin{center}
\begin{tabular}{c|c||c|c|c|c}
\hline\hline
\multicolumn{2}{c||}{~Energy \& Luminosity~~}
& \multicolumn{2}{|c|}{Previous Results\,\cite{newformfactor}} 
& \multicolumn{2}{|c}{Present Results}\\
\hline
&&&&&\\[-4mm]
$\sqrt{s\,}$\,(TeV) & $\mathcal{L}$
& $|h_3^Z|_{2\sigma}$ & $|h_3^Z|_{5\sigma}$ & $|f_5^\gamma|_{2\sigma}$ & $|f_5^\gamma|_{5\sigma}$\\
\,({collider\,energy})\, & \,(ab$^{-1}$)\,
& ({unpol},\,{pol}) & ({unpol},\,{pol}) & ({unpol},\,{pol}) & ({unpol},\,{pol}) \\
\hline\hline
&&&&\\[-4mm]
0.25 & 20 & ~$(1.4,\,1.2)\!\times\!10^{-4}$~ 
& ~$(3.4,\,2.9)\!\times\!10^{-4}$~ 
& ~$(2.0,\,0.96)\!\times\!10^{-4}$~ 
& ~$(4.9,\,2.4)\!\times\!10^{-4}$~ \\
\hline
&&&&&\\[-4mm]
0.25 & 5 & ~$(2.7,\,2.3)\!\times\!10^{-4}$~ 
& ~$(6.8,\,5.8)\!\times\!10^{-4}$~ 
& ~$(3.9,\,1.9)\!\times\!10^{-4}$~ 
& ~$(9.8,\,4.8)\!\times\!10^{-4}$~ \\
\hline
&&&&\\[-4mm]
0.5 & 5 
& $(6.2,\,5.2)\!\times\!10^{-5}$ & $(1.6,\,1.3)\!\times\!10^{-4}$ & $(4.0,\,3.2)\!\times\!10^{-5}$ & $(1.0,\,.81)\!\times\!10^{-4}$\\
\hline
&&&&&\\[-4mm]
1 & 5 & $(1.5,\,1.2)\!\times\!10^{-5}$ & $(3.8,\,3.0)\!\times\!10^{-5}$ & $(8.3,\,7.0)\!\times\!10^{-6}$ & $(2.1,\,1.7)\!\times\!10^{-5}$\\
\hline
&&&&&\\[-4mm]
3 & 5 & $(1.7,\,1.4)\!\times\!10^{-6}$ & $(4.3,\,3.5)\!\times\!10^{-6}$ & $(8.4,\,7.6)\!\times\!10^{-7}$ & $(2.1,\,1.9)\!\times\!10^{-6}$\\
\hline
&&&&&\\[-4mm]
5 & 5 & $(5.8,\,4.9)\!\times\!10^{-7}$ & $(1.5,\,1.2)\!\times\!10^{-6}$ & $(2.9,\,2.6)\!\times\!10^{-7}$ & $(7.3,\,6.4)\!\times\!10^{-7}$\\ 
\hline\hline
\end{tabular}
\caption{\hspace*{-1mm}\small 
Comparison of our previous sensitivity reaches for the nTGC form factors obtained via the $Z\ga$ channel\,\cite{newformfactor} with the present sensitivity reaches obtained obtained via the $ZZ$ channel.\  
These reaches are presented at $2\sigma$ and $5\sigma$ levels and for different collider energies.\
In each entry the sensitivity bounds correspond to (unpolarized,\,polarized) $e^{\mp}$ beams, 
			with beam polarizations $(P_L^e,{\hs}P_R^{\bar{e}})\!=\!(0.9,{\hs}0.65)$.\ 
} 
\label{tab:13}
\label{tab:14new}
\end{center}
\end{table}

In parallel, we can convert the sensitivity reaches for the nTGC new physics scale $\cut^{}_{\tilde{B}W}$ 
obtained via the $Z\ga$ production in Ref.\,\cite{Ellis:2020ljj} to the corresponding reaches 
for the nTGC form factor $h_3^Z$ obtained via the $Z\ga$ production\,\cite{newformfactor}.\ 
Then, in Table\,\ref{tab:13} we compare these bounds 
with the sensitivity reaches for the form factor 
$f_5^\gamma$ [contributed by the same operator $\mathcal{O}^{}_{\!\tilde{B}W}$ 
via Eq.\eqref{f5a}]
obtained through the $ZZ$ production channel of the present study.\ 
The previous reaches were derived for the $Z\gamma$ channel 
with $Z\!\ito q\bar{q}$ (shown in the 2nd and 3rd columns),
whereas the present bounds are derived for the $ZZ$ channel with $Z\!\ito q\bar{q},\ell\bar{\ell},\nu\bar{\nu}$ 
(shown in the 4th and 5th columns).\  These reaches are presented at $2\sigma$ and $5\sigma$ levels and for different collider energies respectively.\
In each entry the reaches correspond to (unpolarized,\,polarized) $e^{\mp}$ beams, with beam polarizations 
$(P_L^e,{\hs}P_R^{\bar{e}})\!=\!(0.9,{\hs}0.65)$.\  
Same as in Table\,\ref{tab:12}, 
we choose a representative integrated luminosity $\mathcal{L}\!=\!5\,\text{ab}^{-1}$    
and an ideal detection efficiency $\epsilon\!=\!100\%$\,.\ 
In addition, for the 250\,GeV collision energy of $e^-e^+$, we choose
another benchmark of integrated luminosity 
$\mathcal{L}\!=\!20\,\text{ab}^{-1}\!$ 
based on the CEPC TDR\,\cite{CEPC-TDR},
as presented in the third row of Table\,\ref{tab:14new}.\ 

\vs

As shown in Table\,\ref{tab:12}, 
it is evident that the sensitivity reaches for the nTGC new physics scales 
estimated in the present study are stronger than those reported in 
Ref.\,\cite{Ellis:2020ljj}.\ 
The current improvement is primarily attributable to the full use of the differential angular distribution $f^1_\Omega$ 
incorporating all angular variables $(\theta,\,\theta_a,\,\phi_a,\,\theta_b^{},\phi_b^{})$, 
combined with machine learning classification algorithms 
that efficiently distinguish between the nTGC signals and SM backgrounds.\ 
For instance, the sensitivity reaches for the new physics scale $\Lambda_{\!\tilde{B}W}^{}$ estimated in the present study 
exhibit an enhancement up to 21\% (18\%) as compared to the previous bounds 
of Ref.\,\cite{Ellis:2020ljj} for the unpolarized (polarized) 
$e^\mp$ beams.\ 
The largest improvement of 21\% (18\%) is achieved for 
$\Lambda^{\text{tot},2\sigma}_{\!\tilde{B}W}$ 
in the unpolarized (polarized) case 
at the collision energy $\sqrt{s\,}\!=\!3\,$TeV.

As shown in Table\,\ref{tab:13}, 
the present sensitivity reaches on the nTGC form factors also  
surpass the previous reaches found in Ref.\,\cite{newformfactor}.\ 
For instance, we find that the sensitivity reaches 
on the nTGC form factors exhibit enhancements of 
about $(17\!-\!51)\%$ as compared to the reaches found in Ref.\,\cite{newformfactor}.\ 
The minimal improvement of 17\% is realized for the sensitivity  bounds
on $|f_5^\gamma|$ in the polarized case at the collision energy $\sqrt{s\,}\!=\!250\,$GeV, 
whereas the most significant enhancement of 51\% is achieved for  
$|f_5^\gamma|$ in the unpolarized case at the collision energy $\sqrt{s\,}\!=\!3\,$TeV.\ 
We also see that using the polarized $e^\mp$ beams the sensitivity reaches
on $f_5^{\ga}$ ($h_3^Z$) can be increased by about 51\%\,(15\%) 
at the collision energy $\sqrt{s\,}\!=\!250\,$GeV, 
and by about 10\%\,(18\%) at the collision energy $\sqrt{s\,}\!=\!3\,$TeV.

\vspace*{1mm}
\section{\hspace*{-2.5mm}Conclusions}
\label{sec:5}

Neutral triple gauge couplings (nTGCs) do not exist within the Standard Model (SM), nor do they receive contributions 
from any dimension-6 SMEFT operators.\ Hence, the nTGCs provide an important window for probing new physics
beyond the SM that can be used to probe directly new physics encoded by dimension-8 SMEFT operators.\ 
The previous phenomenological studies\,\cite{Ellis:2025ghl}-\cite{Degrande:2013kka} 
and experimental measurements\,\cite{Atlas2018nTGC-FF}\cite{ATLAS2025-ZA} mainly focused on the $Z\gamma V^*$ nTGCs
(with $V\!=\!Z,\ga$) and did not probe the triple $Z$ boson coupling $ZZZ^*$.\

\begin{figure}[t]
	\centering
	\includegraphics[width=12.5cm,height=6.5cm]{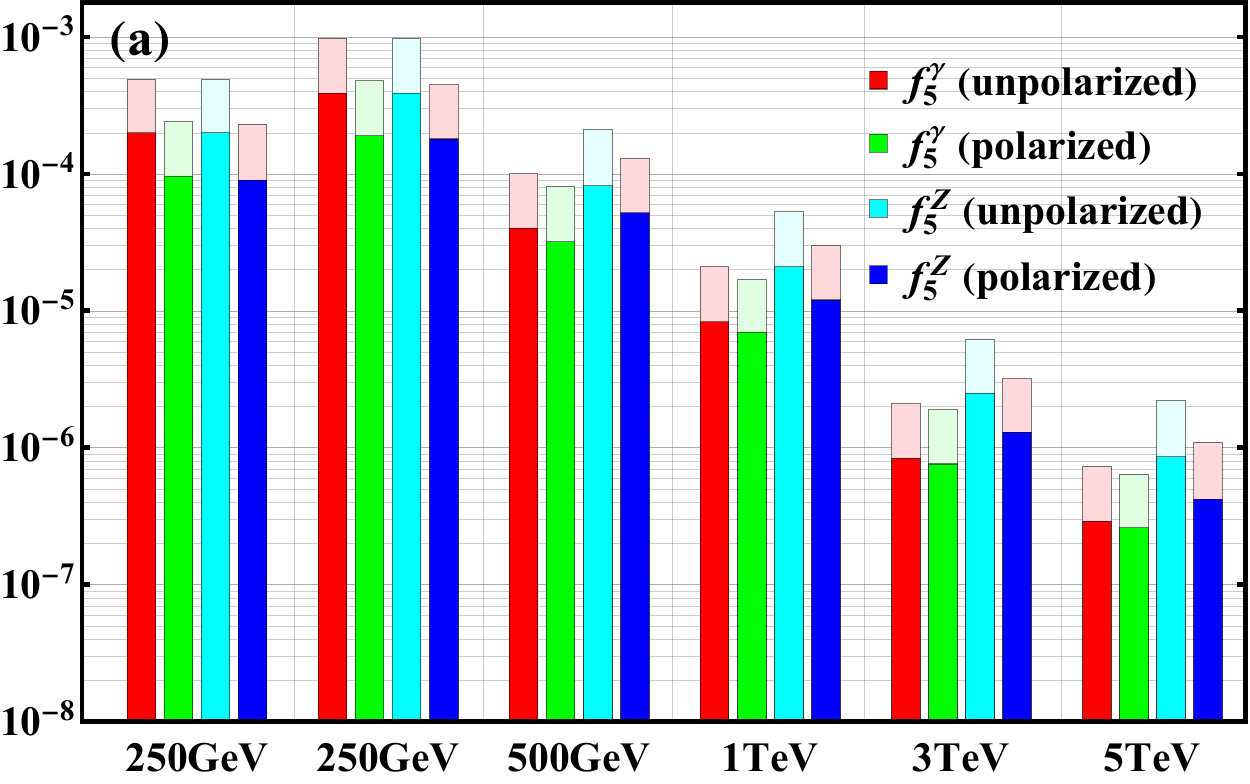}
	\\[2mm]
	\hspace*{-1mm}
	\includegraphics[width=12.5cm,height=7.6cm]{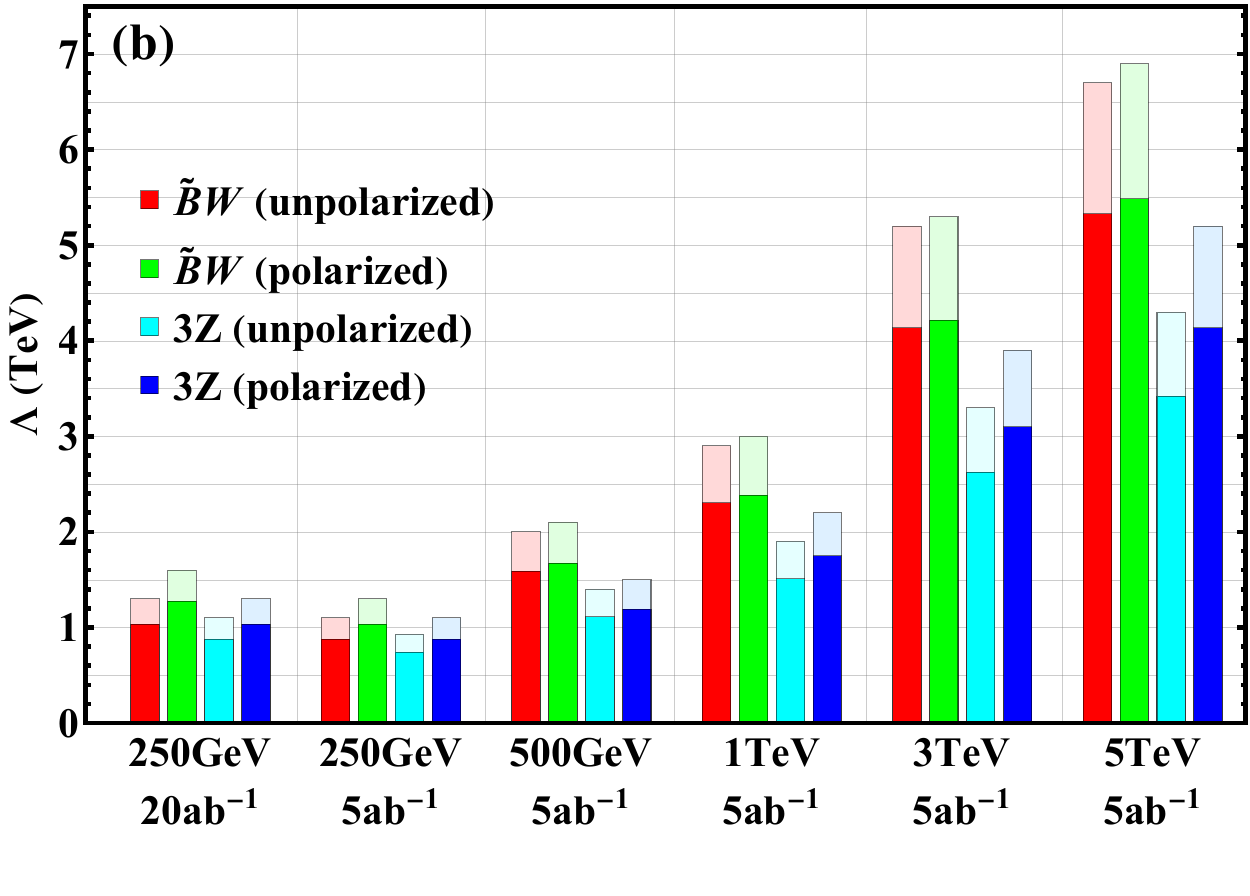}
	\vspace*{-7mm}
	\caption{\hspace*{-1mm}\small {Sensitivity reaches for the nTGC form factors in plot\,(a) and for the nTGC new physics scales in plot\,(b) 
			for various $e^-e^+$ collision energies and integrated luminosities.\ 
			In this figure, plot\,(a) presents the $(2\sigma,\,5\sigma)$ sensitivity reaches in (heavy,\,light) colors, 
			whereas plot\,(b) shows the $(2\sigma,\,5\sigma)$ sensitivity reaches in (light, heavy) colors.}}
	\label{bar}
\label{fig:12}
\label{fig:12new}
\end{figure}

In this work, we formulated the nTGC form factors of $ZZV^*$ ($V\!\!=\!Z,\gamma$) that are compatible with the spontaneous breaking
of the SU(2)$\otimes$U(1) electroweak gauge symmetry and consistently match the dimension-8 nTGC operators in the broken phase.\  
Using machine learning, 
we have systematically studied the sensitivities of probing 
both the $ZZV^*$ form factors and new physics scales through the reaction process  
$e^-e^+\ito ZZ \ito f_a\bar{f}_af_b^{}\bar{f}_b^{}$ (with $f_a\!=\!\ell,q$ and $f_b^{}\!=\!\ell,q,\nu)$ 
at high energy $e^+e^-$ colliders, including CEPC, FCC-ee, LCF, ILC and CLIC.\
We have analyzed the signal significance and derived sensitivity reaches 
for probing the nTGC form factors $(f_5^{\ga},\hs f_5^Z)$ and 
the corresponding nTGC new physics scales $(\cut_{3Z}^{},\cut_{}^{\tilde{B}W})$.\ 
We present a systematic summary of these sensitivity bounds in Figs.\,\ref{fig:12}(a) and \ref{fig:12}(b), 
allowing one nTGC parameter nonzero at a time.\

\vs 

In Section\,\ref{sec:2} we studied how the nTGC couplings of $ZZV^*$ can arise from 
the relevant dimension-8 SMEFT operators as shown in Eq.\eqref{eq:O8-ZZV*}
and Table\,\ref{tab:1}.\  In particular,  we constructed the dimension-8 nTGC operator $\mO_{3Z}^{}$
in Eq.\eqref{eq:O3Z} that contributes to the $ZZZ^*$ vertex only, but not to the $ZZ\gamma^*$ vertex,
as shown in Table\,\ref{tab:1} and Eq.\eqref{f5z}.\ 
This is in constrast with the other operator $\mO_{\!\tilde BW}^{}$ that contributes to
the $ZZ\ga^*$ vertex alone, as shown in Table\,\ref{tab:1} and Eq.\eqref{f5a}.\ 
We presented the consistent formulation of the nTGC form factors that are compatible with 
the full SM gauge symmetry 
$\text{SU(2)}\!\otimes\!\text{U(1)}$ with spontaneous electroweak symmetry breaking.\ 
The unitarity constraints on the nTGCs were derived in Table\,\ref{tab:2} of Section \ref{sec:2.2} 
and are much weaker than the sensitivity bounds derived for the collider probes in Section\,\ref{sec:4}.

In Section\,\ref{sec:3}, we derived the scattering amplitudes, cross sections, and angular distributions 
for the reaction process $e^-e^+\!\ito ZZ\hs$.\ We analyzed systematically the contributions from both the SM and nTGCs,
as given analytically in Eqs.\eqref{eq:f-theta-012}, \eqref{eq:f-theta-a012} and \eqref{eq:fphi-a012}.\  
With these we demonstrated the differences in their differential angular distributions, as shown in 
Figs.\,\ref{fig:4}$-$\ref{fig:6}.\ These are important for our phenomenological analysis of probing the nTGCs in the 
following Section\,\ref{sec:4}.

In Section\,\ref{sec:4}, we presented a systematic analysis of the $ZZ$ production in different fermionic decay channels 
at various $e^+e^-$ colliders.\ We computed the signal significance using the differential angular distributions of 
the cross section, considering both the visible leptonic and hadronic decay channels and the invisible decay channel 
($Z\ito\nu\bar{\nu}\hs$).\  The estimated sensitivity reaches for probing the nTGC form factors are shown in 
Tables\,\ref{tab:3},\,\ref{tab:5},\,\ref{tab:6} and for probing the nTGC new physics scales are shown in 
Tables\,\ref{tab:4},\,\ref{tab:7}.\ These sensitivity limits are presented for the leptonic and hadronic $Z$-decay channels 
in Sections\,\ref{sec:4.1} and for the invisible $Z$-decay channels in Section\,\ref{sec:4.2}.\ 
We studied the significant improvements of the sensitivity reaches on the nTGC form factors 
and on the nTGC new physics scales 
(as well as their correlations) achieved through the use of machine learning algorithms 
in Sections\,\ref{sec:4.3} and \ref{sec:4.5}.\ 
We further derived the enhanced sensitivity bounds on probing the nTGCs by using the polarized $e^\pm$ beamsin in Section\,\ref{sec:4.4}.\
A comprehensive summary of these nTGC sensitivity bounds for both the unpolarized and polarized $e^\mp$ beams is presented 
in Figs.\,\ref{fig:12}(a) and (b).\ 
Moreover, we studied the correlation bounds for the nTGC form factors and for the nTGC new physics scales, which were presented in Figs.\,\ref{fig:11new}(a)(c) and Figs.\,\ref{fig:11new}(b)(d)
respectively.\  
We found that the optimal sensitivity bounds on the nTGC correlations are given 
by the mixed setting including the unpolarized operation 
and a follow-up polarized operation.\
We further demonstrated in Fig.\,\ref{fig:13} that machine learning  
can substantially improve the constraints on nTGC correlations (pink contours) 
over the bounds derived by using conventional manual cuts (blue contours).\   
Finally, Section\,\ref{sec:4.6} compared the sensitivity reaches 
on probing the nTGCs as obtained in this work 
through $ZZ$ production with the previous studies via $Z\gamma$ production\,\cite{newformfactor}\cite{Ellis:2020ljj}, 
which are summarized in Tables\,\ref{tab:12} and \ref{tab:13}.\ 
We have demonstrated that the sensitivity bounds on probing the nTGC form factors and the nTGC new physics scales  
as given in this work are stronger, primarily due to the full use of the differential angular distributions 
based on the machine learning algorithms.


\vspace*{6mm}
\noindent 
{\Large\bf Acknowledgements}
\\
We thank Shu Li, Kun Wang and Haijun Yang 
for useful discussions.\ 
The work of J.E.\ was supported in part by the United Kingdom STFC Grant ST/T000759/1.\ 
The works of H.J.H., R.Q.X.\ and S.P.Z.\ were supported in part 
by the National Natural Science Foundation of China 
(under Grants 12435005 and 12175136), 
by the Shenzhen Science and Technology Program (under Grant JCYJ20240813150911015), 
by the State Key Laboratory of Dark Matter Physics,
by the Key Laboratory for Particle Astrophysics and Cosmology (MOE), and 
by the Shanghai Key Laboratory for Particle Physics and Cosmology.\ 
R.Q.X.\ was also supported in part by the NSFC under Grants 
12505119, 12175006, 12188102 and 12061141002, 
by the Ministry of Science and Technology of China under Grant 2023YFA1605800, 
and by the State Key Laboratory of Nuclear Physics and Technology.

\vspace*{5mm}

\end{document}